\documentclass[aps,prc,twocolumn,superscriptaddress,showpacs,10pt,dvipsnames]{revtex4-1}

\usepackage[utf8]{inputenc}
\usepackage{amsfonts}
\usepackage{amsmath}
\usepackage{amssymb}
\usepackage{bm}
\usepackage[pdftex]{graphicx}
\usepackage{xcolor}
\usepackage[normalem]{ulem}
\usepackage{comment}

\begin{document}

\newcommand{\tj}[6]{ \begin{pmatrix}
  #1 & #2 & #3 \\
  #4 & #5 & #6
 \end{pmatrix}}
 
\newcommand{\js}[1]{\textcolor{red}{\it #1}}
\newcommand{\jsout}[1]{\textcolor{red}{\sout{#1}}}


\title{Electric dipole response of low-lying excitations in the two-neutron halo nucleus $\boldsymbol{^{29}}$F}




\author{J. Casal}
\email{casal@pd.infn.it}
\affiliation{Dipartimento di Fisica e Astronomia ``G.Galilei'', Università degli Studi di Padova, via Marzolo 8, Padova, I-35131, Italy}
\affiliation{INFN-Sezione di Padova, via Marzolo 8, Padova, I-35131, Italy}
\author{Jagjit Singh}
\affiliation{Research Center for Nuclear Physics (RCNP), Osaka University, Ibaraki 567-0047, Japan}
\author{L. Fortunato}
\affiliation{Dipartimento di Fisica e Astronomia ``G.Galilei'', Università degli Studi di Padova, via Marzolo 8, Padova, I-35131, Italy}
\affiliation{INFN-Sezione di Padova, via Marzolo 8, Padova, I-35131, Italy}
\author{W. Horiuchi}
\affiliation{Department of Physics, Hokkaido University, Sapporo 060-0810, Japan}
\author{A. Vitturi}
\affiliation{Dipartimento di Fisica e Astronomia ``G.Galilei'', Università degli Studi di Padova, via Marzolo 8, Padova, I-35131, Italy}
\affiliation{INFN-Sezione di Padova, via Marzolo 8, Padova, I-35131, Italy}


\date{\today}

\begin{abstract}
 \begin{description}
  \item[Background] 
  The neutron-rich $^{28,29}$F isotopes have been recently studied via knockout and interaction cross-section measurements. The two-neutron halo in $^{29}$F has been linked to the occupancy of $pf$ intruder configurations. 
  \item[Purpose] 
  Investigate the bound spectrum and continuum states in $^{29}$F, focusing on the electric dipole ($E1$) response of low-lying excitations and the effect of dipole couplings on nuclear reactions.
  \item[Method] $^{29}\text{F}$ ($^{27}\text{F}+n+n$) wave functions are built within the hyperspherical harmonics expansion formalism, and total reaction cross sections are calculated using the Glauber theory. Continuum states and $B(E1)$ transition probabilities are described in a pseudostate approach using the analytical transformed harmonic oscillator basis. The corresponding structure form factors are used in continuum-discretized coupled-channels (CDCC) calculations to describe low-energy scattering.
  \item[Results] Parity inversion in $^{28}$F leads to a $^{29}$F ground state characterized by 57.5\% of $(p_{3/2})^2$ intruder components, a strong dineutron configuration, and an increase of the matter radius with respect to the core radius of $\Delta R=0.20$ fm. Glauber-model calculations for a carbon target at 240 MeV/nucleon provide a total reaction cross section of 1370 mb, in agreement with recent data. The model produces also a barely bound excited state corresponding to a quadrupole excitation. $B(E1)$ calculations into the continuum yield a total strength of 1.59 e$^2$fm$^2$ up to 6 MeV, and the $E1$ distribution exhibits a resonance at $\approx$ 0.85 MeV. Results using a standard shell-model order for $^{28}$F lead to a considerable reduction of the $B(E1)$ distribution. The four-body CDCC calculations for $^{29}\text{F}+^{120}\text{Sn}$ around the Coulomb barrier are dominated by dipole couplings, which totally cancel the Fresnel peak in the elastic scattering cross section.
  \item[Conclusions] Our three-body calculations for $^{29}$F, using the most recent experimental information on $^{28}$F, are consistent with a two-neutron halo. Our predictions show the low-lying enhancement of the $E1$ response expected for halo nuclei and the relevance of dipole couplings for low-energy reactions on heavy targets. These findings may guide future experimental campaigns.
 \end{description} 
\end{abstract}


\maketitle


\section{Introduction}
\label{sec:introduction}

Nuclei lying away from the stability valley typically show exotic properties which allow us to study shell evolution. Halo nuclei, in particular, severely test our nuclear structure knowledge and have motivated extensive experimental and theoretical developments~\cite{tanihata13}. These systems exhibit a diffuse density distribution, with one or more weakly bound neutrons exploring distances far from a more compact core, and giving rise to an abnormally large matter radius. As a consequence, reactions with halo nuclei are characterized by large interaction cross sections~\cite{tanihata85,Hansen87}. The case of two-neutron halo nuclei is especially interesting, since they are Borromean systems~\cite{Zhukov93}. With the corresponding $\text{core}+n$ subsystems being unbound, the strong correlations between the valence neutrons are key in binding two-neutron halos~\cite{kikuchi16,hagino05}. Text-book examples of two-neutron halo nuclei are $^{6}$He, $^{11}$Li, or $^{14}$Be~\cite{tanihata85,tanihata88}. More recently, heavier two-neutron halos, such as $^{22}$C~\cite{tanaka10,togano16}, have also been explored. In all cases, the low-lying spectra of the unbound $\text{core}+n$ subsystems are known to play an important role in shaping the properties of these nuclei~\cite{Aksyutina13,casalplb17}. 

Together with the extended matter distribution and strong dineutron correlations, one of the most salient features of halo nuclei is an enhancement of the low-lying $E1$ (electric dipole) strength into the continuum~\cite{aumann19}. For different two-neutron halo nuclei ($^{6}$He, $^{11}$Li, $^{14}$Be), this near-threshold enhancement has been observed via invariant mass spectroscopy in Coulomb Dissociation (CD) experiments~\cite{aumann99,labiche01,nakamura06} and is seen as a signature of the halo wave function. Very recently, a high-energy CD measurement for $^{19}$B into $^{17}\text{B}+n+n$ have reported this enhanced $E1$ strength, providing the first evidence that this exotic nucleus has a prominent two-neutron halo~\cite{Cook2020}. Various theoretical investigations within three-body models have been found to describe reasonably well these features in light nuclei (see, for instance, Refs.~\cite{Danilin98,Myo01,MRoGa05,Horiuchi06,Horiuchi07,hagino09,RdDiego10,JCasal13,JPFernandezGarcia13,Cook2020}). Similar studies for heavier systems could help in assessing the limits of halo formation when neutrons occupy higher shells.

In the particular case of $^{11}$Li, the presence of intruder $2s_{1/2}$ components is crucial in developing the two-neutron halo wave function, which is linked to the low-lying spectrum of the unbound $^{10}$Li showing an $s$-wave ground state~\cite{Zhukov93,Riisager94,Jep06,sanetullaev2016}. This is due to the fact that lower $\ell$ orbitals have more extended distributions, which favor the formation of a diffuse halo~\cite{Hamamoto07}. In this context, neutron-rich fluorine isotopes have received special attention recently, since $^{28\text{-}31}$F lay at the southern border of the so-called island of inversion~\cite{WAR91,Nakamura09,Motobayashi1995}. With $^{28}$F and $^{30}$F being unbound, $^{29}$F and $^{31}$F nuclei are Borromean systems~\cite{Gaudefroy2012,Ahn2019}. Thus, parity inversion in the low-lying spectrum of the corresponding $\text{core}+n$ subsystems could lead towards halo formation~\cite{Masui2020,Michel2020}. While $^{27}$F is assumed to be an $sd$-shell nucleus~\cite{Jurado07}, with $pf$ intruder configurations playing a role only for its excited states~\cite{Elekes04}, the situation for fluorine isotopes with $A>27$ has been long debated~\cite{CHRIS1,CHRIS2,DOOR2017}. In Refs.~\cite{CHRIS1,CHRIS2}, a two-resonance structure was found just above the $^{27}\text{F}+n$ threshold. No robust spin-parity assignment could be drawn, but shell-model calculations suggested that $pf$ intruder components should play a minor role in the ground state of $^{28}$F. However, Ref.~\cite{DOOR2017} reported the measurement of a bound excited state in $^{29}$F, and the energy gap with respect to the ground state could be reproduced only by considering $pf$ components.

In Ref.~\cite{Singh2020}, we performed three-body calculations for $^{29}$F considering different scenarios for the corresponding $^{27}\text{F}+n$ subsystem. Our results show that the enhancement of the matter radius, as well as the strong dineutron configuration of the possible halo, is linked to the degree of mixing between standard $1d_{3/2}$ components with $2p_{3/2}$ intruder configurations. Recently, the relative-energy spectra and momentum distributions extracted from high precision nucleon-knockout data showed that the ground-state resonance of $^{28}$F at 0.199(6) MeV is dominated by $\ell=1$ components, thus extending the island of inversion~\cite{Revel2020}. Several other states were measured, including an $\ell=2$ dominated resonance around 0.966 MeV. Even more recently, interaction cross section measurements for $^{27,29}$F observed a significant increase of the matter radius, 0.35$\pm$0.08 fm~\cite{Bagchi2020}, with new shell model calculations also supporting a two-neutron halo wave function characterized by a large occupancy of the $2p_{3/2}$ orbital. These findings are consistent with our predictions in Ref.~\cite{Singh2020}. 

The halo structure in $^{29}$F should have strong implications for direct reactions, such as the high-energy CD discussed above, but also for low-energy elastic scattering and breakup. For instance, in the case of $^{6}$He impinging on heavy targets, continuum-discretized coupled channel (CDCC) calculations showed that dipole Couplings produce a strong reduction of the elastic scattering cross section at near-barrier energies~\cite{MRoGa08}, leading to a cancellation of the typical Fresnel peak. This feature is associated to the large $B(E1)$ strength at low energies, so it provides a direct link between the halo structure and the reaction mechanism. Interestingly, this effect is even larger in the case of dipole resonances, as observed for $^{11}$Li~\cite{Cubero12,JPFernandezGarcia13}.

In view of the recent developments and experimental results, it is the goal of this paper to refine our previous three-body model for $^{29}$F and to make predictions for its low-energy continuum, in particular dipole and quadrupole excitations and their role in nuclear reactions. Some preliminary results were briefly discussed in a recent perspective article~\cite{Fortunato2020}. Here we present a more exhaustive study including additional calculations. In Sec.~\ref{sec:formalism}, we briefly introduce the theoretical framework employed in this work, which is based on the hyperspherical harmonics expansion method. Then, in Sec.~\ref{sec:gs} we discuss on the halo wave function of the $^{29}$F ground state and search for additional bound states. We also provide Glauber-model calculations for total reaction cross sections. In Sec.~\ref{sec:cont} we study electromagnetic transitions into continuum states and present three-body calculations for the $B(E1)$ distribution of this nucleus, and in Sec.~\ref{sec:4b} we give also CDCC predictions for its scattering around the Coulomb barrier on a heavy target, discussing on the role of dipole couplings. Finally, in Sec.~\ref{sec:conclusions} we summarize the main results of the present work, and we highlight how these theoretical predictions could guide further experimental studies.

\section{Three-body model for $\boldsymbol{^{29}}$F}
\label{sec:formalism}
\subsection{Hyperspherical formalism}
\label{sec:HH}
\begin{figure}[t]
\centering
\includegraphics[width=0.7\linewidth]{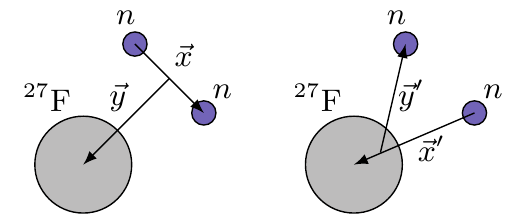}
\caption{Jacobi T (left) and Y (right) coordinates for the $^{29}$F nucleus described as $^{27}\text{F}+n+n$.}
\label{fig:jac}
\end{figure}

In the present work, as in Ref.~\cite{Singh2020}, we describe three-body states by using the hyperspherical framework~\cite{Zhukov93,Nielsen01}. For simplicity, here we focus only on the case of a three-body system composed by a compact core and two valence neutrons. The corresponding Jacobi coordinates $\{\boldsymbol{x},\boldsymbol{y}\}$  are shown in Figure~\ref{fig:jac}, where two distinct sets of coordinates can be identified: i) the Jacobi-T set, in which the valence neutrons are related by the $x$ coordinate, and ii) the Jacobi-Y set, where the coordinate $x'$ connects the core with one of the neutrons. Note that a third set, analogous to the Y system, is obtained by switching the neutron connected to the core. For the particular case under study, and in the Jacobi-T representation, the relations between the scaled Jacobi coordinates ($\boldsymbol{x},\boldsymbol{y}$) and the actual physical distances ($\boldsymbol{r}_x$,$\boldsymbol{r}_y$) between the particles are:
\begin{equation}
\boldsymbol{x}=\boldsymbol{r}_{x}\sqrt{\frac{1}{2}},
\label{eq:coor1}
\end{equation}
\begin{equation}
\boldsymbol{y}=\boldsymbol{r}_{y}\sqrt{\frac{2A}{A+2}},
\label{eq:coor2}
\end{equation}
where $A=27$ is the mass number of the core. From Jacobi coordinates, we introduce the hyperspherical coordinates $\{\rho,\alpha,\widehat{x},\widehat{y}\}$, using the usual definitions for the hyperradius $\rho$ and the hyperangle $\alpha$,
\begin{align}
    \rho& =\sqrt{x^2+y^2},\\
    \alpha&=\arctan{\left(\frac{y}{x}\right)}.
\end{align}
Note that transformation between the T and Y coordinate systems preserve the hyperradius, while the hyperangle is changed~\cite{IJThompson04}. 

In principle, the three-body problem can be solved equivalently in any Jacobi representation. However, the Jacobi-T set is the most convenient one for $\text{core}+n+n$ systems, since the wave function must be antisymmetric under exchange of the valence neutrons related by the $x$ coordinate. The states for a given total angular momentum $j$ are expanded as:
\begin{equation}
  \psi^{j\mu}(\rho,\Omega) = \rho^{-5/2}\sum_{\beta}\chi_{\beta}^{j}(\rho)\mathcal{Y}_{\beta}^{j\mu}(\Omega),
  \label{eq:3bwf}
\end{equation}
where $\Omega\equiv\{\alpha,\widehat{x},\widehat{y}\}$ and $\beta$ represents a set of quantum numbers coupled to $j$. Our coupling scheme for the angular functions $\mathcal{Y}_{\beta}^{j\mu}(\Omega)$ is given in terms of hyperspherical harmonics $\Upsilon_{Klm_l}^{l_xl_y}(\Omega)$, eigenstates of the hypermomentum operator $\widehat{K}$, as
\begin{equation}
\mathcal{Y}_{\beta}^{j\mu}(\Omega)=\left\{\left[\Upsilon_{Kl}^{l_xl_y}(\Omega)\otimes\phi_{S_x}\right]_{j_{ab}}\otimes\kappa_I\right\}_{j\mu}.
\label{eq:Upsilon}
\end{equation}
Here, $\boldsymbol{l}=\boldsymbol{l}_x+\boldsymbol{l}_y$, and the hyperspherical harmonic for a given $(l_x,l_y)_l$ configuration is given by
\begin{equation}
\Upsilon_{Klm_l}^{l_xl_y}(\Omega)=\varphi_K^{l_xl_y}(\alpha)\left[Y_{l_x}(\boldsymbol{x})\otimes Y_{l_y}(\boldsymbol{y})\right]_{lm_l},
\label{eq:HH}
\end{equation}
\begin{equation}
\begin{split}
\varphi_K^{l_xl_y}(\alpha) & = N_{K}^{l_xl_y}\left(\sin\alpha\right)^{l_x}\left(\cos\alpha\right)^{l_y}\\&\times P_n^{l_x+\frac{1}{2},l_y+\frac{1}{2}}\left(\cos 2\alpha\right),
\label{eq:varphi}
\end{split}
\end{equation}
where $P_n^{a,b}$ is a Jacobi polynomial of order $n=(K-l_x-l_y)/2$ and $N_K^{l_xl_y}$ a normalization constant. The spin $S_x$ gives the coupled spin of the two particles related by the $x$ coordinate, $\boldsymbol{j}_{ab}=\boldsymbol{l}+\boldsymbol{S}_x$, and $I$ represents the spin of the core nucleus, which in this work is assumed to be fixed. The spin of $^{27}$F is finite due to the odd number of protons, but for simplicity we neglect it. This allows us to reduce the previous expression by using $j_{ab}=j$ ($I=0$), 
\begin{equation}
\mathcal{Y}_{\beta}^{j\mu}(\Omega)=\left[\Upsilon_{Kl}^{l_xl_y}(\Omega)\otimes\kappa_{S_x}\right]_{j\mu}.
\label{eq:Upsilon0}
\end{equation}
This amounts to considering only neutron degrees of freedom, simplifying the choice of the corresponding $\text{core}+n$ potential and drastically reducing the number of components needed in Eq.~(\ref{eq:3bwf}). Note that a similar approach has been employed in the past, with great success, for other odd-even $\text{core}+n+n$ nuclei such as $^{11}$Li~\cite{Cubero12,Matsumoto19}. It is also worth noting that, in the Jacobi T set, $S_x$ comes from the coupling of two $s=1/2$ neutrons, so only $S_x=0$ or 1 are allowed. Moreover, the Pauli principle for two identical neutrons (with isospin $T=1$) imposes that $S_x+l_x$ must be an even number.

The hyperradial functions in Eq.~(\ref{eq:3bwf}) are solutions of the set of coupled differential equations
\begin{equation}
\begin{split}
&\left[-\frac{\hbar^2}{2m}\left(\frac{d^2}{d\rho^2}-\frac{15/4+K(K+4)}{\rho^2}\right)-\varepsilon\right]\chi_{\beta}^{j}(\rho)\\&+\sum_{\beta'}V_{\beta'\beta}^{j\mu}(\rho) \chi_{\beta'}^{j}(\rho)=0,
\end{split}
\label{eq:coupled}
\end{equation}
where it is clear that $K$ defines an effective three-body barrier, and $V_{\beta'\beta}^{j\mu}(\rho)$ are the so-called coupling potentials defined as
\begin{equation}
V_{\beta'\beta}^{j\mu}(\rho)=\left\langle \mathcal{Y}_{\beta }^{j\mu}(\Omega)\Big|V_{12}+V_{13}+V_{23} \Big|\mathcal{Y}_{\beta'}^{ j\mu}(\Omega) \right\rangle.
\label{eq:3bcoup}
\end{equation}
Here, $V_{ij}$ are the two-body potentials between each pair of particles, which will be described in the next section. In this work, the radial functions are expanded in a discrete basis,
\begin{equation}
    \chi_\beta^j(\rho)=\sum_{i} C_{i\beta}^j U_{i\beta}(\rho),
    \label{eq:PS}
\end{equation}
where the coefficients $C_{i\beta}^j$ can be easily obtained  by diagonalizing the three-body Hamiltonian for $i=0,\dots,N$ basis functions. Eigenstates corresponding to negative-energy eigenvalues describe bound states, while positive-energy states provide a discrete representation of the continuum. This approach is referred to in the literature as the pseudostate (PS) method~\cite{Tolstikhin97}.

Different bases can be used within the PS approach (see, e.g., Refs.~\cite{Desc03,Matsumoto04,MRoGa05}). Our choice is the analytical Transformed Harmonic Oscillator (THO) basis~\cite{JCasal13,Karataglidis}, obtained from the harmonic oscillator (HO) functions in hyperspherical coordinates as
\begin{equation}
  U_{i\beta}^{\text{THO}}(\rho)=\sqrt{\frac{ds}{d\rho}}U_{iK}^{\text{HO}}[s(\rho)],
\label{eq:R}
\end{equation}
where
\begin{equation}
  U_{iK}^{\text{HO}}(s)=D_{iK}s^{K+5/2}L_i^{K+2}(s)\exp{\left(-s^2/2\right)},
\label{eq:fHO}
\end{equation}
the functions $L_i^{K+2}(s)$ are the generalized Laguerre polynomials, and $D_{iK}$ is just a normalization constant. The transformation in Eq.~(\ref{eq:R}) is given by
\begin{equation}
s(\rho) = \frac{1}{\sqrt{2}b}\left[\frac{1}{\left(\frac{1}{\rho}\right)^{4} +
\left(\frac{1}{\gamma\sqrt{\rho}}\right)^4}\right]^{\frac{1}{4}},
\label{eq:LST}
\end{equation}
and depends on two parameters ($b,\gamma$), which change the Gaussian asymptotic behavior of the HO functions ($e^{-\rho^2/2})$ into a simple exponential ($e^{-\gamma^2\rho/2b^2})$. This improves the convergence of the calculations with respect to the number of basis functions $N$. Moreover, the ratio $\gamma/b$ governs the hyperradial extension of the basis. As discussed in Refs.~\cite{JCasal13,JCasal18}, this feature controls the PS density after diagonalization. For instance, a small $\gamma$ parameter, keeping the oscillator length $b$ fixed, provides a larger number of PSs just above the breakup threshold, which are suitable to compute electromagnetic transitions into the continuum. More details will be given in Section~\ref{sec:cont}.

    \subsection{Binary potentials}
    \label{sec:potentials}

The present three-body calculations require appropriate two-body potentials as an input in Eq.~(\ref{eq:3bcoup}). For the $nn$ interaction, we use the Gogny-Pires-Tourreil (GPT) potential~\cite{GPT}, which includes central, spin-orbit and tensor terms. This potential describes $nn$ scattering data reasonably, and it has been used with success in several other three-body calculations for $\text{core}+n+n$ nuclei~\cite{Zhukov93,IJThompson04,JCasal13}. Other choices are available in the literature (e.g.,~the Gaussian interaction in Ref.~\cite{Garrido04}, or more sophisticated interactions such as AV18~\cite{av18}). We have checked that the use of a different $nn$ interaction introduces only minor differences, provided they all reproduce $nn$ scattering data. 

\begin{figure}
\centering
    \includegraphics[width=0.85\linewidth]{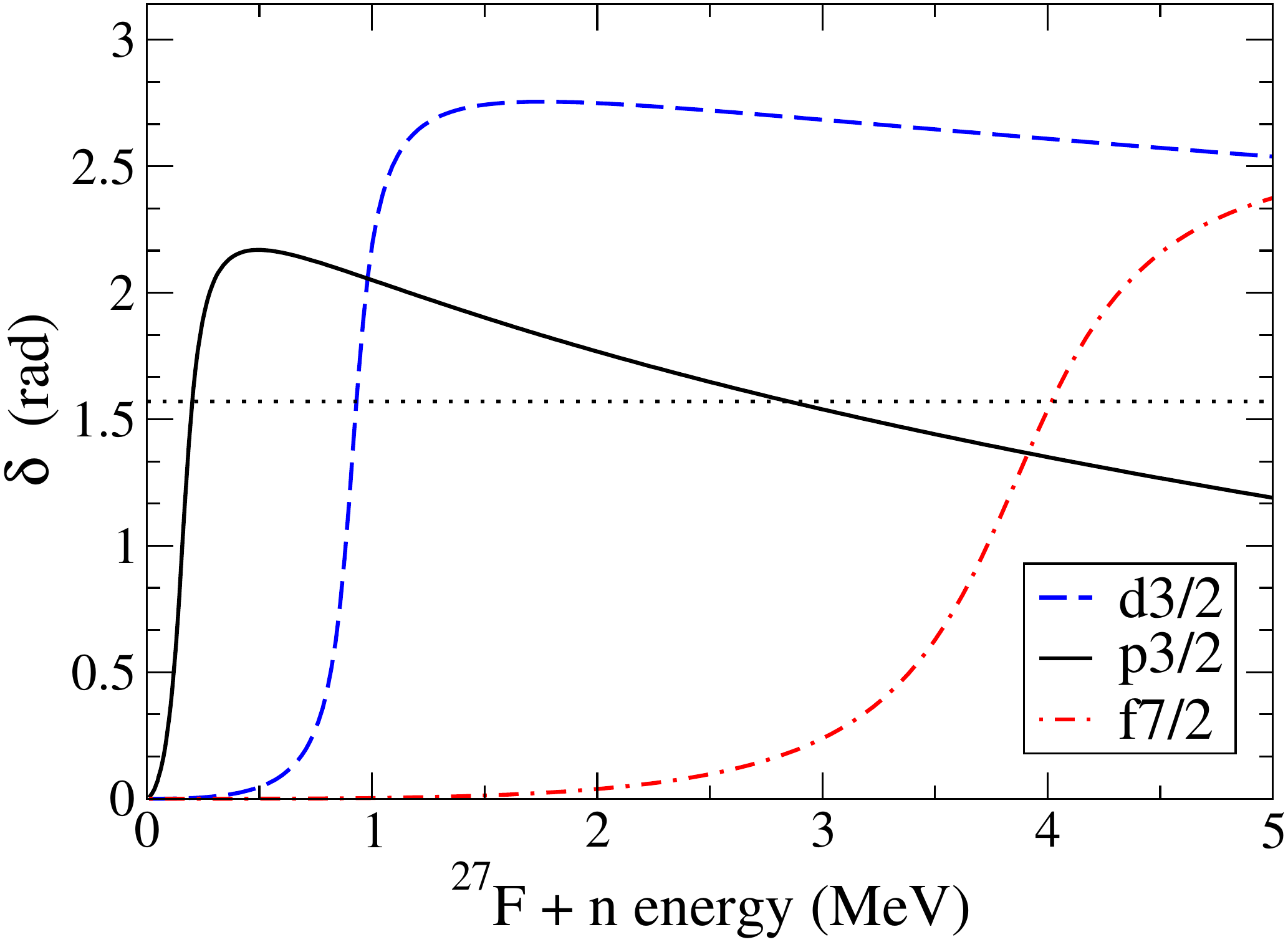}
    \caption{Phase shifts for the $^{27}\text{F}+n$ system in the present work. The dotted line corresponds to $\delta=\pi/2$.}
    \label{fig:ph}
\end{figure}

For the $^{27}$F-$n$ interaction, we fix a Woods-Saxon potential to the available experimental information for $^{28}$F, as presented in Sec.~\ref{sec:introduction}. Including only central and spin-orbit terms, we use the form
\begin{equation}
V_{^{27}\text{F}\text{-}n} = \left(-V_0+V_{ls} \vec{l}\cdot\vec{s}\frac{1}{r}\frac{d}{dr}\right)\frac{1}{1+\exp\left(\frac{r-R}{a}\right)},
\label{vls}
\end{equation}
with $R=r_0A^{1/3}$, the standard values $r_0=1.25$ fm and $a=0.75$ fm adopted from Ref.~\cite{Horiuchi10}, and $V_{ls}=34.666$ MeVfm$^2$ from systematics~\cite{BOHR}. Then, $V_0=V_0^{(\ell)}$ is assumed to be $\ell$ dependent, allowing for effects beyond the simple $\text{inert core}+n$ picture. In our previous work~\cite{Singh2020}, we fixed this depth by considering different scenarios from a standard shell-model picture to an extreme inversion of the $1d_{3/2}$ and $2p_{3/2}$ single-particle levels. By using the recent results of Ref.~\cite{Revel2020}, we adjust $V_0^{(\ell)}$ to produce a $2p_{3/2}$ ground-state resonance around 0.199(6) MeV above the 1$n$ separation threshold and a $1d_{3/2}$ state around 0.996(13) MeV, corresponding to the first two $^{28}$F levels measured in the $-1n$ channel from $^{29}$F. This was the prescription used also in our previous perspective article~\cite{Fortunato2020}, and provides $V_0^{(1)}=46.78$ MeV and $V_0^{(2)}=37.68$ MeV. For other $\ell$ partial waves, we adopt the same value as for $\ell=2$. In this way we obtain also an $f$-wave resonance at about the same energy of one of the states reported in Ref.~\cite{Revel2020}. The corresponding phase shifts for this new potential are shown in Fig.~\ref{fig:ph}. Note that this potential gives rise to $1s_{1/2}$, $1p_{3/2}$, $1p_{1/2}$, $1d_{5/2}$ and $2s_{1/2}$ bound states which represent Pauli forbidden states. As discussed in our previous work~\cite{Singh2020}, in order to perform three-body calculations using this two-body interaction we remove these bound states by constructing phase-equivalent shallow potentials within a supersymmetric transformation~\cite{Baye287}. In our scheme, the $d_{3/2}$ shell is fully available for the valence neutrons, which is a limitation of inert-core models. A particular treatment of this issue might require additional studies.

\vspace{-5pt}

\section{Ground-state properties}
\label{sec:gs}

In addition to the $nn$ and $\text{core}+n$ potentials, it is customary to introduce in Eq.~(\ref{eq:3bcoup}) a phenomenological three-body force to account for possible effects not explicitly included in our three-body description~\cite{IJThompson04,MRoGa05,RdDiego10,JCasal13}. For simplicity, this additional term can be assumed to be diagonal in the HH expansion, $\delta_{\beta,\beta'}V_{3b}(\rho)$, with its hyperradial dependence given by a Gaussian form,
\begin{equation}
    V_{3b}(\rho)=v_{3b}e^{-{\left(\rho/\rho_o\right)}^2},
    \label{eq:3bforce}
\end{equation}
where $\rho_o=6$ fm and the depth $v_{3b}$ is adjusted to recover the known energy of the states, if available. In the case under study, this parameter is fixed to yield a $^{29}$F ground state characterized by the experimental two-neutron separation energy of $S_{2n}=1.44$ MeV~\cite{Gaudefroy2012}. This provides $v_{3b}=-0.35$ MeV, that we keep fixed to generate all the states considered in the present work unless otherwise mentioned. It was checked that the specific shape of the three-body force in Eq.~(\ref{eq:3bforce}) is not crucial. For instance, a less confined potential such as an exponential leads essentially to the same ground-state properties. Note that this choice to correct the three-body spectrum is not unique, and a similar solution can be obtained by incorporating some scaling factors in the binary potentials~\cite{Desc03,Singh16}.

With all these ingredients, we focus first on the $^{29}$F ground-state properties. As discussed in Sec.~\ref{sec:formalism}, we neglect the spin of the core, so that our ground state is characterized by $j^\pi=0^+$. Note that, due to the unpaired proton in the $1d_{5/2}$ shell, the actual spin of $^{29}$F is assumed to be 5/2$^+$~\cite{DOOR2017}. 
We consider $0^+$ states including wave-function components up to a maximum hypermomentum $K_{max}$ (Eq.~(\ref{eq:3bwf})), which restrict also the possible orbital angular momenta with the condition that $l_x+l_y \le K$~\cite{Zhukov93}. To that end, the three-body Hamiltonian is diagonalized in a THO basis with $i=0,\dots,N$ functions (Eq.~(\ref{eq:PS})).

\vspace{-10pt}

    \subsection{Convergence of the ground state}
    \label{sec:conv}
    
\vspace{-5pt}

\begin{figure}[ht]
\centering
\includegraphics[width=0.78\linewidth]{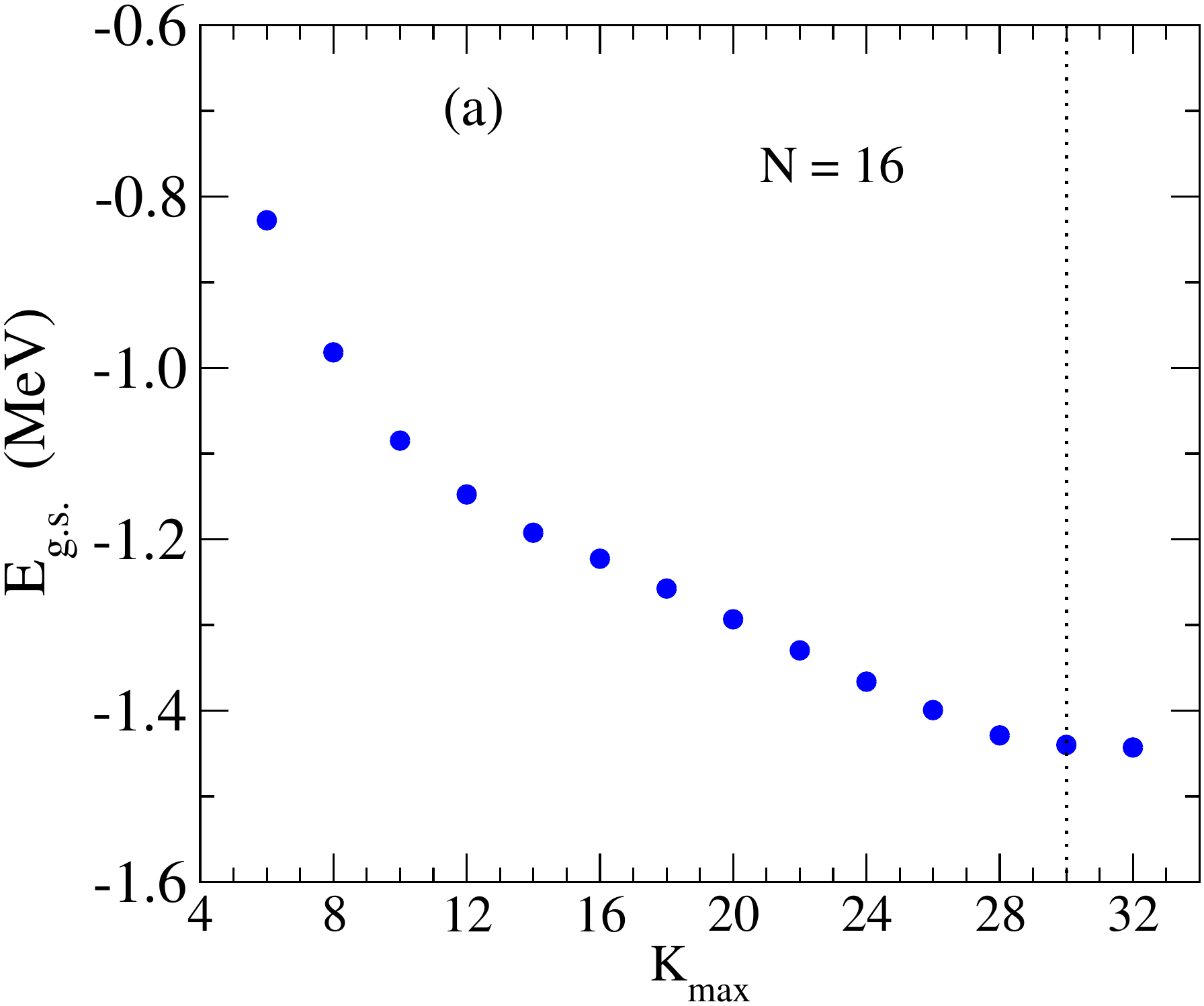}
\vspace{5pt}

\includegraphics[width=0.8\linewidth]{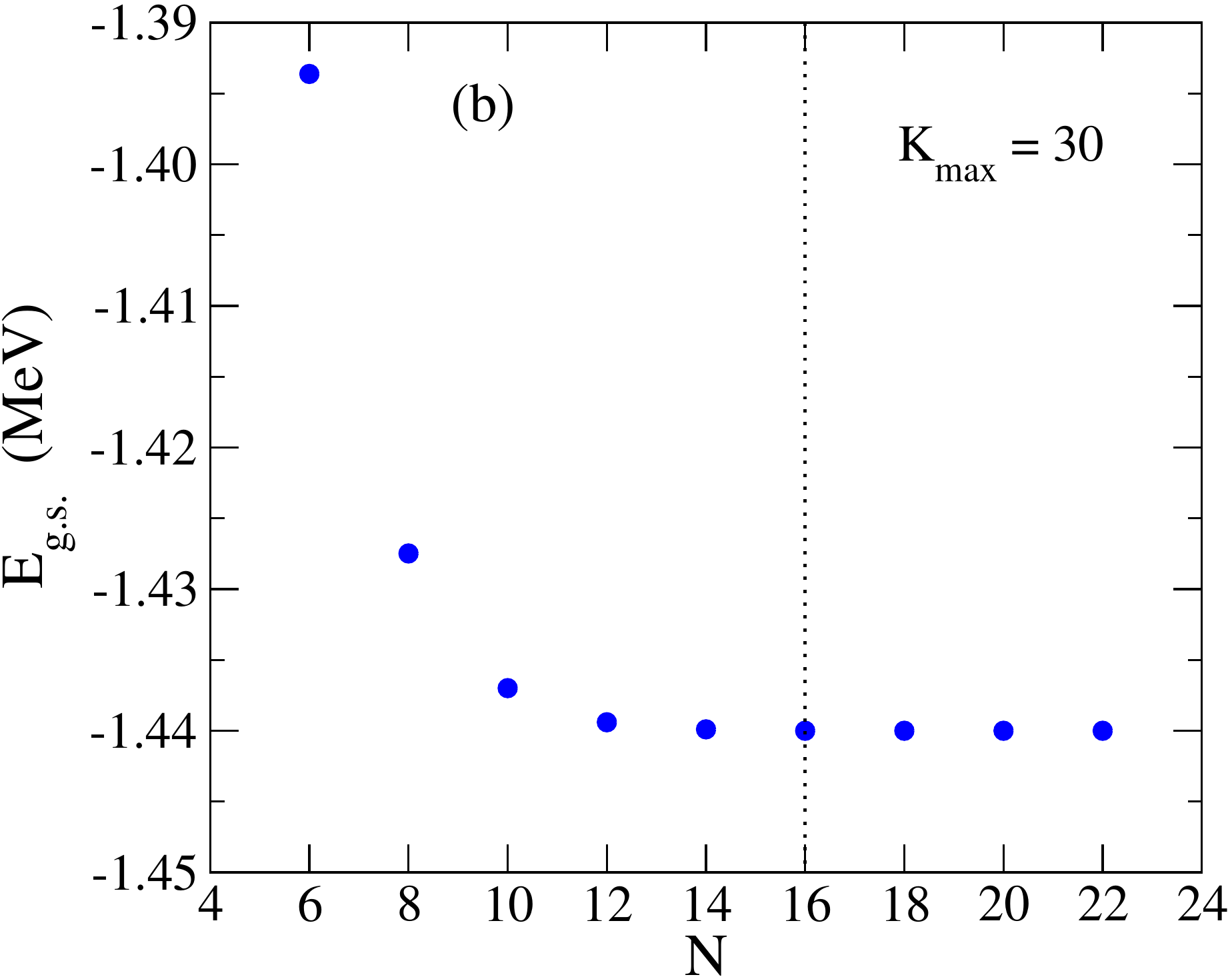}
\caption{Convergence of the ground-state energy with (a) $K_{max}$ and (b) $N$, keeping the other parameter fixed. Vertical dotted lines represent the adopted values at which the three-body force has been adjusted to reproduce $S_{2n} = 1.44$ MeV.}
\label{fig:conv}
\end{figure}

We first analyze the convergence of the ground state. Figure~\ref{fig:conv} shows the behavior of the ground-state energy as a function of $K_{max}$ and $N$. In both cases, the other parameter has been fixed to a sufficiently large number, and calculations correspond to a THO basis characterized by $b=0.7$ fm and $\gamma = 1.4$ fm$^{1/2}$. The position of the ground state evolves rather slowly with the size of the model space, and we need $K_{max}=30$ to achieve a stable solution. The convergence with respect to the number of basis functions $N$ is much faster, and $N=16$ is found to provide converged results. It is worth noting that this is a consequence of the basis choice.

\begin{figure}[t]
\centering
\includegraphics[width=0.90\linewidth]{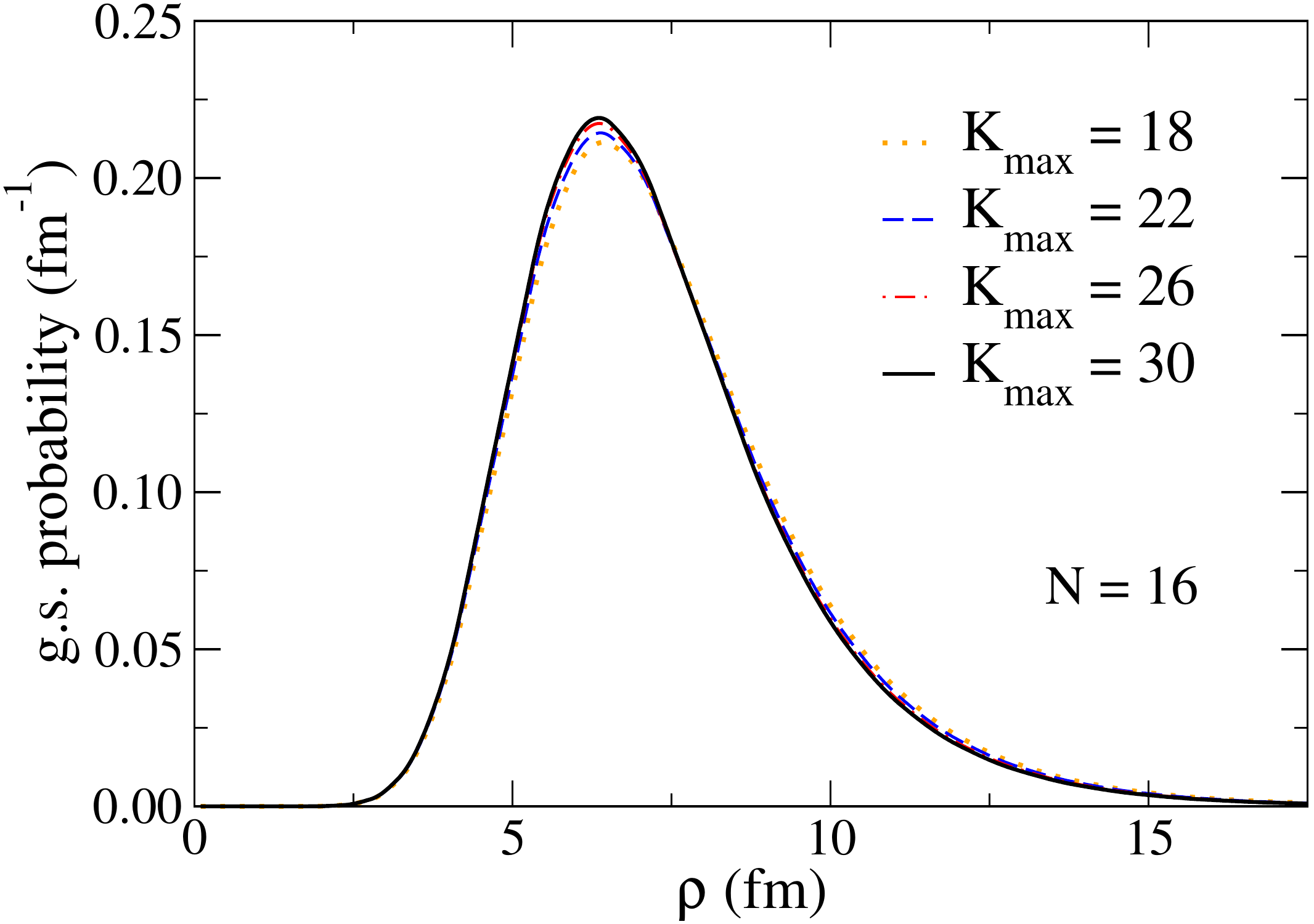}
\caption{Convergence of the ground-state density $P(\rho)$ with $K_{max}$. Calculations correspond to $N=16$.}
\label{fig:prob-conv}
\end{figure}

In Fig.~\ref{fig:prob-conv} we present the convergence of the ground-state hyperradial density, 
\begin{equation}
    P(\rho) = \rho^5 \int d\Omega \left|\psi(\rho,\Omega)\right|^2 = \sum_{\beta}\left|\chi_{\beta}(\rho)\right|^2.
    \label{eq:probrho}
\end{equation}
It is clear that $K_{max}=30$ is more than enough to achieve a robust description of the wave function. The distribution presents a maximum around 6.4 fm and vanishes beyond 15 fm. From the corresponding root-mean-square hyperradius, the matter radius is defined as
\begin{equation}
 R_m = \sqrt{\frac{1}{A+2}\left(R_c^2A + \langle \rho^2\rangle\right)}, 
 \label{eq:radius}
\end{equation}
where $R_c$ is the radius of the core, which is required as an input parameter. 
In Ref.~\cite{Bagchi2020}, a new experimental value ($R_c^{\rm exp} = 3.15(4)$ fm) and several theoretical estimations ($R_c^{\rm th} = 3.12\text{-}3.23$ fm) are presented. In our previous work~\cite{Singh2020}, we adopted $R_c=3.21$ fm, obtained from an auxiliary $^{25}$F$+n+n$ calculation using the experimentally known $^{25}$F radius. 
Here we adopt the value $R_c=3.16$ fm, which is consistent with Ref.~\cite{Bagchi2020}, and analyze the resulting $^{29}$F matter radius in relative terms. 
We obtain $\Delta R = R_m - R_c = 0.20$ fm, well beyond the increase expected by scaling the radius using the standard $A^{1/3}$ formula. As already discussed in Ref.~\cite{Fortunato2020}, this is an indication of the extended matter distribution associated to halo nuclei, but our result is smaller than the value reported in Ref.~\cite{Bagchi2020}, $\Delta R^{\rm exp} = 0.35(8)$ fm. Nonetheless, we note that the calculated value is sensitive to the two-neutron separation energy of $^{29}$F, which entails large uncertainties. The experimental $S_{2n}$ value is 1.443(436) MeV~\cite{Gaudefroy2012}. If we consider its lower limit, by adjusting the three-body force to yield a shallower ground state at $\approx 1$ MeV below the three-body threshold, we obtain instead $\Delta R =0.25$ fm, which is closer to the experimental value. A more precise knowledge of the $S_{2n}$ value could help in bridging the gap between our theoretical estimation of $\Delta R$ and the experiment.

To evaluate the reliability of our wave function, in this work we also investigate the total reaction cross section using the standard Glauber theory~\cite{Glauber}. We adopt the nucleon-target formalism~\cite{AbuIbrahim00} with the nucleon-nucleon profile function given in Ref.~\cite{AbuIbrahim08,*AbuIbrahim09E,*AbuIbrahim10E}. The other theoretical inputs to this reaction model are the density distributions of projectile and target nuclei. The validity of this approach has been confirmed in many examples of high-energy nucleus-nucleus collisions involving unstable nuclei, e.g., the isotope dependence of the total reaction cross sections was reproduced fairly well when appropriate density distributions were employed~\cite{Horiuchi12,Horiuchi15jps}. First, we generate a harmonic-oscillator (HO) type density distributions for $^{27}$F by assuming the lowest shell model filling, whose size parameter is fixed so as to reproduce the interaction cross section data on a carbon target at 240 MeV/nucleon~\cite{Bagchi2020}. The resulting rms matter radius (total reaction cross section) is 3.16 fm (1240 mb), which is consistent with the value $3.15\pm0.04$ fm ($1243\pm 14$ mb) extracted in Ref.~\cite{Bagchi2020}. The density distribution of $^{29}$F is simply constructed by adding the two-neutron density distribution obtained from our three-body model to the HO density of $^{27}$F. The calculated total reaction cross sections are 1370 mb if we assume $S_{2n}=1.44$ MeV, and 1390 mb if we take the lower limit ($S_{2n}\approx 1$ MeV), which are in good agreement with the observed interaction cross section $1396\pm 28$ mb~\cite{Bagchi2020}. 

    \subsection{Mixing and dineutron correlations}
    \label{sec:mix}

\begin{figure}[t]
\centering
\includegraphics[width=0.85\linewidth]{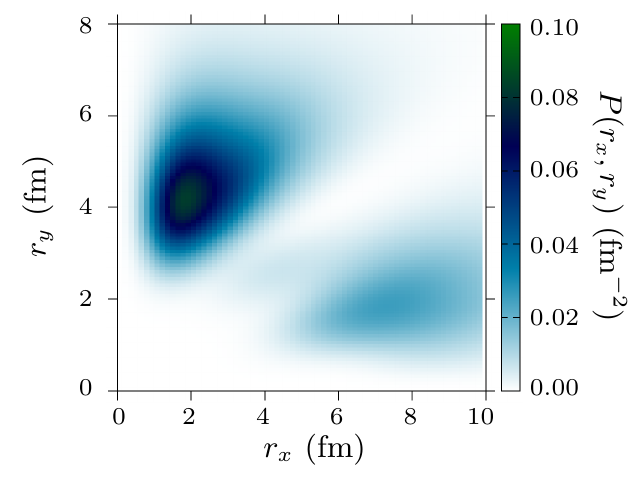}
\caption{Ground-state probability density (in fm$^{-2}$) of $^{29}$F using the present three-body model, as a function of $r_x\equiv r_{nn}$ and $r_y\equiv r_{c\text{-}nn}$.}
\label{fig:xyprob}
\end{figure}

The ground-state probability of two-neutron halo nuclei is generally analyzed in terms of the so-called dineutron and cigar-like configurations~\cite{Zhukov93}. The former corresponds to two neutrons close to each other at some distance from the compact core, while the latter represents the valence neutrons at opposite sides of the core. With the wave function (\ref{eq:3bwf}) written in the Jacobi-T set, the corresponding two-dimensional probability is given by
\begin{equation}
    P(x,y)=x^2y^2\int d\widehat{x}d\widehat{y} \left|\psi(\boldsymbol{x},\boldsymbol{y})\right|^2,
    \label{eq:prob}
\end{equation}
which can be easily converted to physical distances by using Eqs.~(\ref{eq:coor1}) and ~(\ref{eq:coor2}). This probability satisfies
\begin{equation}
    \int P(r_x,r_y) dr_xdr_y = 1.
    \label{eq:normdens}
\end{equation}
In Fig.~\ref{fig:xyprob} we present our results for the ground state of $^{29}$F, as shown in Ref.~\cite{Fortunato2020}, where a dominant dineutron component can be clearly identified. 

Since the mixing between different-parity states in the $\text{core}+n$ system favours dineutron correlations~\cite{Catara84}, the computed distribution can be understood by studying the partial-wave content of the state. This requires a transformation of the wave function into the Jacobi-Y set (see Fig.~\ref{fig:jac}), where the core and a single neutron are related by the $x'$ coordinate. As discussed in Ref.~\cite{Singh2020}, this representation is more akin to a typical shell-model picture and can be numerically obtained by using Reynal-Revai coefficients~\cite{RR70,IJThompson04}. Then, we change the couplings in Eq.~(\ref{eq:Upsilon}) to contain a single-particle angular momentum $\boldsymbol{J}_{x'} = \boldsymbol{l}_{x'}  + \boldsymbol{S}_{x'}$, where $S_{x'}=1/2$ is just the spin of a neutron (provided we neglect the spin of the core). More details on these transformations are given in  Appendix~\ref{app1}.

Our ground state is characterized by 57.5\% of neutron intruder $(p_{3/2})^2$ components, followed by 29.0\% of $(d_{3/2})^2$ and 6.1\% of $(f_{7/2})^2$. These numbers are obtained when the ground-state energy is adjusted to the central value of the experimental $S_{2n}$. By considering instead its lower limit, which corresponds to a more weakly bound system, the $(p_{3/2})^2$ weight increases by 5\%, while the $(d_{3/2})^2$ decreases by the same amount. Both situations are consistent with a dominant $(p_{3/2})^2$ configuration, which clearly leads to the significant increase of the matter radius with respect to the core radius. The large mixing between the $(p_{3/2})^2$ and $(d_{3/2})^2$ components is responsible for the strong dineutron peak observed in Fig.~\ref{fig:xyprob} and plays a key role in the formation of the two-neutron halo in $^{29}$F. We checked also the effect of the small $f_{7/2}$ component by performing test calculations removing the potential for $f$-waves. In that case, the missing $f_{7/2}$ weight goes to $p$-waves, so the mixing between different parity states remains and the resulting density plot is almost identical. Thus, in the present calculations the halo nature and strong dineutron configuration of $^{29}$F is totally driven by the $p_{3/2}$ intruder.

\vspace{-10pt}

    \subsection{Search for additional bound states}
    \label{sec:2+}

In Ref.~\cite{DOOR2017}, a state in $^{29}$F was measured at an excitation energy of $E_{x}=1.080(18)$ MeV. With the adopted value for $S_{2n}$, this corresponds to a bound state at $E=E_x-S_{2n}= -0.363$ MeV with respect to the three-body threshold. The spin-parity of the ground state and this bound excited state are assumed to be 5/2$^+$ and 1/2$^+$, respectively, due to the unpaired proton of the core in the $sd$ shell. In the present three-body model, we neglect the spin of the core, so our ground state is represented by coupling the valence neutrons to $0^+$. Thus, we look for bound excited states corresponding to $2^+$ within our scheme. Note that the inclusion of the spin of the core would lead to a multiplet with several possibilities (from 1/2$^+$ to 9/2$^+$). If these states could be found experimentally and resolved in energy, this splitting could be used to constrain the spin-spin interaction in $^{27}$F-$n$.

\begin{figure}[t]
\centering
\includegraphics[width=0.85\linewidth]{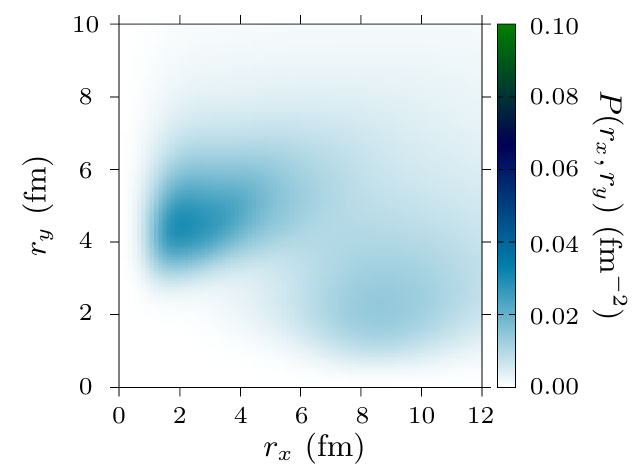}
\caption{As Fig.~\ref{fig:xyprob} but for the 2$^+$ excited bound state predicted in this work. Note that, for a proper comparison, the probability scale is the same in both figures.}
\label{fig:xyprob2p}
\end{figure}

Interestingly, the same Hamiltonian without any fitting parameters (i.e.,~using the same binary potentials and the three-body force adjusted to the 0$^+$ ground-state energy) produces a very weakly bound 2$^+$ state at $E\simeq - 0.1$ MeV. This energy is not quantitatively consistent with the experimental value, but given the simplicity of the model, the qualitative agreement is remarkable. By following the same partial-wave analysis done for the ground state, the leading (valence-neutron) components in the wave function for this excited state are 65.6\% of $(p_{3/2})^2$, 16.5\% of $(p_{3/2})(p_{1/2})$ and 7.1\% of $(d_{3/2})^2$. Since the valence neutrons occupy mostly the $p$-wave orbitals, the mixing between different-parity states is small, and the dineutron configuration for this state is weaker. This can be seen in Fig.~\ref{fig:xyprob2p}, which corresponds to a very diffuse wave function. While some dineutron component is visible, the relative probability of finding the two valence neutrons close to each other is clearly smaller than that in the ground state. Nevertheless, the presence of this state very close to the breakup threshold will have an impact on low-energy reactions involving $^{29}$F, as will be discussed in Sec.~\ref{sec:4b}.

It is worth noting that, with the same Hamiltonian, no bound $1^-$ states were found.

\section{$\boldsymbol{E1}$ transitions}
\label{sec:cont}

We are interested in studying the $E1$ strength function from the ground state of $^{29}$F into the continuum. In general, the reduced transition probability for an $E\lambda$ excitation from the ground state is given by
\begin{equation}
    B(E\lambda) = |\langle n_0j_0 ||\widehat{O}_{E\lambda}||nj\rangle|^2.
    \label{eq:BE1}
\end{equation}
Here, $|n_0j_0\rangle$ represents the ground state, $|nj\rangle$ stands for states with total angular momentum $j$, and $\widehat{O}_{E\lambda}$ is the electric operator of order $\lambda$. In the case of a $\text{core}+n+n$ system described in the Jacobi-T coordinates, this operator can be written as
\begin{equation}
    \widehat{O}_{E\lambda} = Zer_c^\lambda Y_{\lambda M_{\lambda}}(\widehat{y}).
    \label{eq:op}
\end{equation}
where $Z$ is the charge of the core, and $r_c$ its relative distance from the center of mass of the three-body system,
\begin{equation}
    r_c=\sqrt{\frac{2}{A(A+2)}}y.
    \label{eq:cm}
\end{equation}
By using the expansion~(\ref{eq:3bwf}) and Eq.~(\ref{eq:Upsilon0}), the matrix elements for the particular case of $\lambda=1$ transitions between the $0^+$ ground state and $1^-$ states are given by~\cite{JCasal13}
\begin{equation}
    \begin{split}
        & \langle \text{g.s.}||\widehat{Q}_1||n1\rangle = \sqrt{3}Ze\sqrt{\frac{2}{A(A+2)}} \\
        & \times \sum_{\beta,\beta'} \delta_{l_xl_x'} \delta_{S_xS_x'}  (-)^{l_x+S_x} \hat{l}_y\hat{l}_y'\hat{l}\hat{l}' \\ 
        & \times W(ll'l_yl_y';1l_x) W(01ll';1S_x) \tj{l_y}{1}{l_y'}{0}{0}{0} \\ 
        & \times \sum_{ii'} C_\text{g.s.}^{i\beta 0} C_\text{n}^{i'\beta' 1} I_{i\beta,i'\beta'},
        \end{split}
    \label{eq:meQ}
\end{equation}
where we use the common notation $\hat{\ell}=\sqrt{2\ell+1}$, $W$ are Racah coefficients, and $I_{i\beta,i'\beta'}$ represents the double integrals (in $\rho$ and $\alpha$) of $y=\rho\cos\alpha$ between hyperradial basis functions $U_{i\beta}(\rho)$ and hyperangular functions $\varphi_{K}^{l_xl_y}(\alpha)$. Note that the index $n$ labels the final states in a discrete representation, i.e., corresponds to the different 1$^-$ pseudostates obtained upon diagonalization of the three-body Hamiltonian. This means that, within the present formalism, we obtain discrete $B(E1)[\text{g.s.}\rightarrow n]$ values even if the $1^-$ states lay in the continuum. In order to construct an energy distribution from the discrete $B(E1)$ values, we may perform a standard convolution with Gaussian or Poisson distributions (see, e.g., Refs.~\cite{MRoGa05,Pinilla2010,JCasal14,Singh16,Descouvemont20}), which preserves the total strength. 

In order to achieve a detailed description of the $B(E1)$ distribution within our pseudostate representation of the continuum, it is convenient to diagonalize the Hamiltonian for $1^-$ states using a THO basis characterized by a large hyperradial extension, as discussed in Sec.~\ref{sec:formalism}. This enables a larger concentration of pseudostates at low relative energies. It is also convenient to work with large basis sets (i.e., large $N$ values). Then, the completeness of the basis can be checked by comparing the calculated transition probabilities with the corresponding cluster sum rule for electric dipole transitions~\cite{RdDiego08}. In the case under study, the non-energy-weighted sum rule reads
\begin{align}
    S_T(E1) & = \sum_{n}B(E1)[\text{g.s.}\rightarrow n] = \sum_n \left|\langle \text{g.s.}||\widehat{Q}_1||n1\rangle\right|^2 \nonumber \\
    & = \frac{3}{4\pi}\frac{2Ze^2}{A(A+2)} \langle \text{g.s.}|y^2|\text{g.s.}\rangle,
    \label{eq:sumrule}
\end{align}
which depends solely on the ground-state properties.

    \subsection{$\boldsymbol{B(E1)}$ distribution for $\boldsymbol{^{29}}$F}
    \label{sec:E1}

\begin{figure}[t]
\centering
\includegraphics[width=0.85\linewidth]{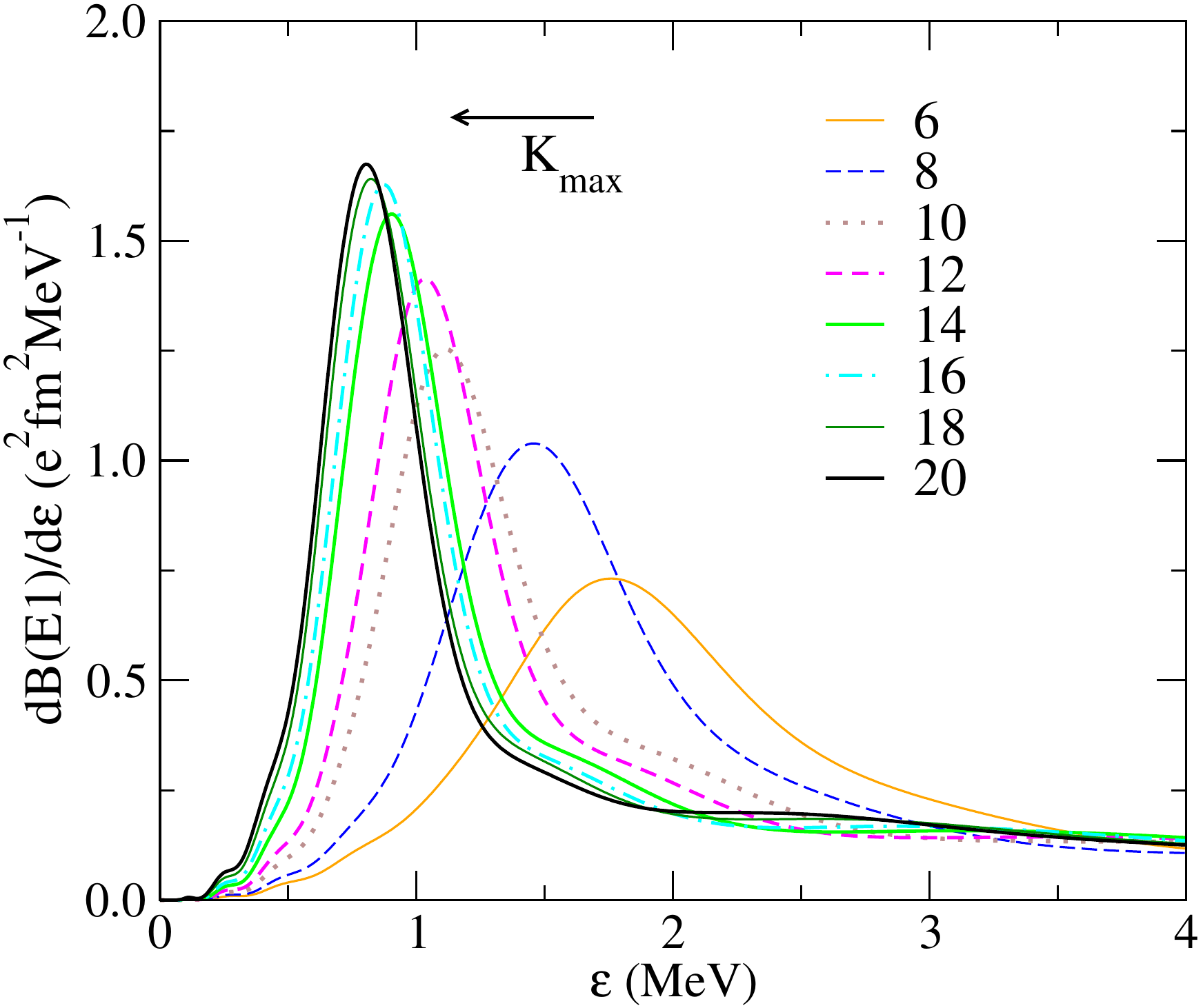}
\caption{Convergence of the $B(E1)$ distribution for $^{29}$F with respect to $K_{max}$.}
\label{fig:BE1conv}
\end{figure}

\begin{figure}[t]
\centering
\includegraphics[width=0.95\linewidth]{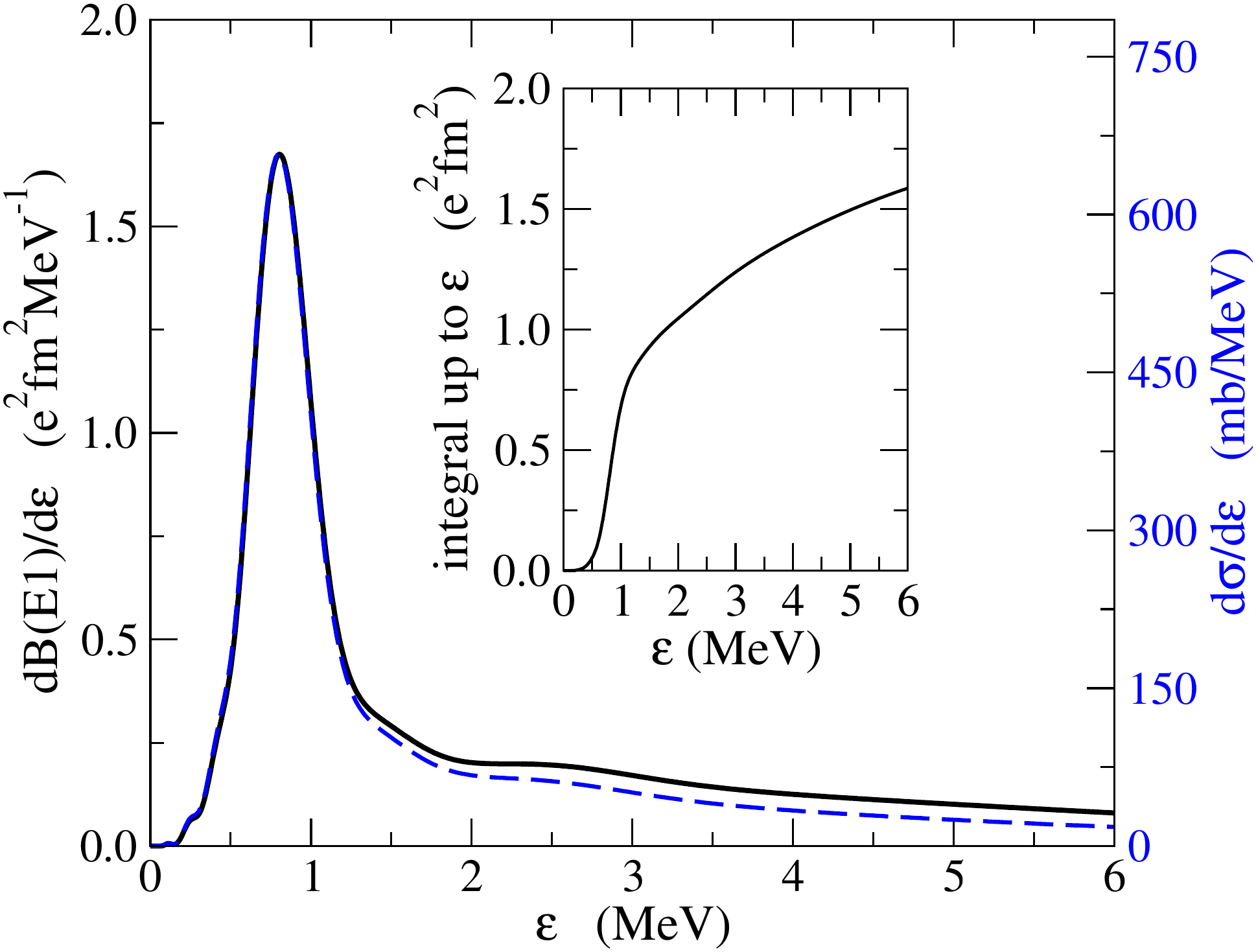}
\caption{$B(E1)$ distribution for $^{29}$F as a function of the continuum energy. The inset shows the cumulative integral up to $\varepsilon$. The dashed blue line corresponds to the RCE cross section of $^{29}$F at 235 MeV/u on a lead target. Note the different scales.}
\label{fig:BE1}
\end{figure}

\begin{figure}[t]
\centering
\includegraphics[width=0.85\linewidth]{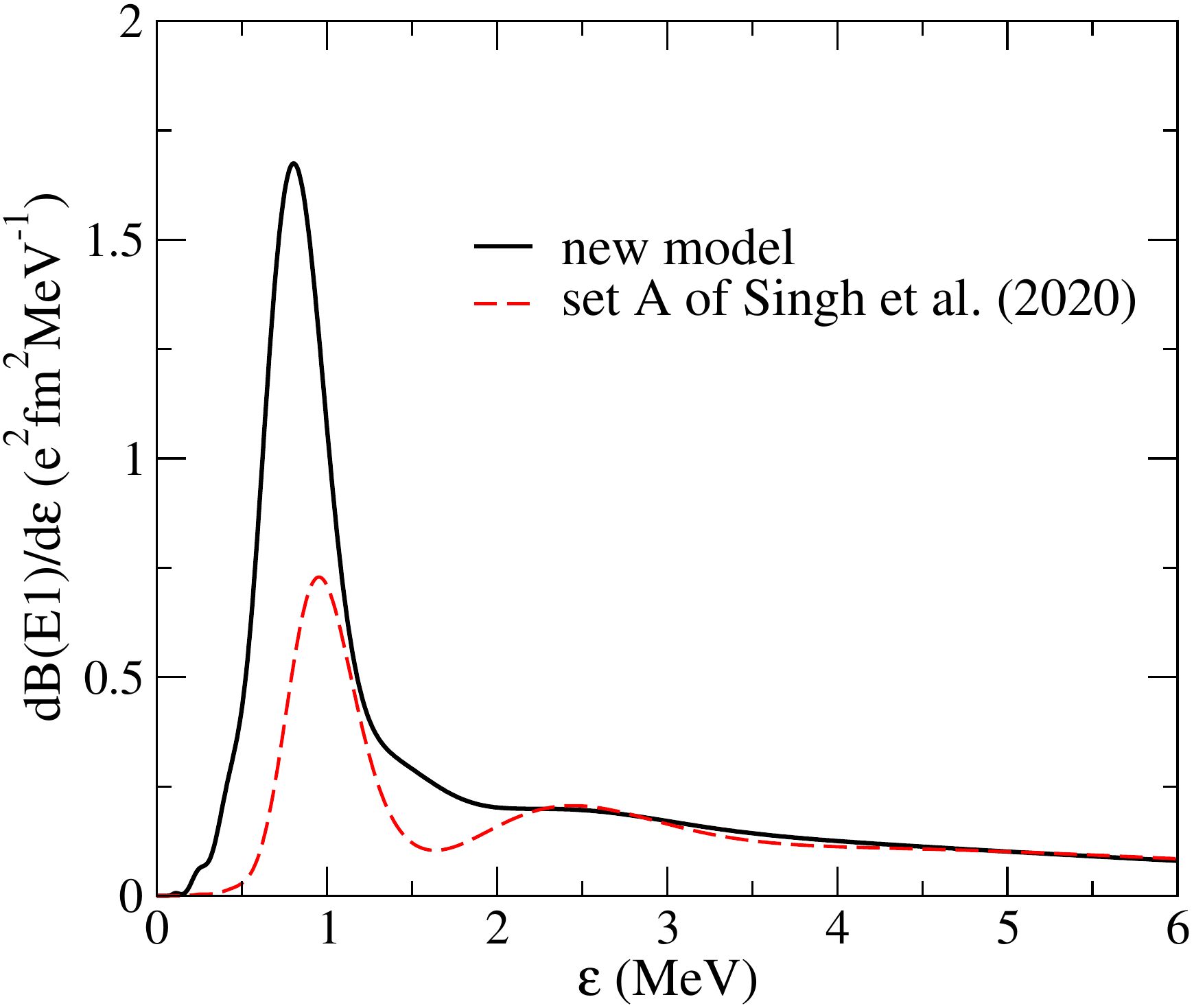}
\caption{$B(E1)$ distribution obtained in the present work (solid black line) compared with the results using the potential set A of Ref.~\cite{Singh2020} (dashed red line), which corresponds to a standard shell-model order.}
\label{fig:BE1comp}
\end{figure}

We generate the $1^-$ states using a THO basis with $b=0.7$ fm and $\gamma=1.0$ fm$^{1/2}$, $N=20$, and including partial waves up to $K_{max}=20$. No bound states are obtained, so ensuring convergence of the calculations with respect to $K_{max}$ and $N$ is not straightforward. In this case, we looked instead for a stable $B(E1)$ distribution. With the adopted values, our discretized $1^-$ continuum contains 1578 positive-energy pseudostates from 0.1 up to 10 MeV. As in the case of the 2$^+$ bound excited state discussed in Sec.~\ref{sec:2+}, here we used the same three-body force employed to fix the 0$^+$ ground state to the experimental $S_{2n}$ value. In that sense, once convergence is achieved, our $B(E1)$ calculations involve no parameter fitting. Our results for the $B(E1)$ distribution are shown in Figs.~\ref{fig:BE1conv}, \ref{fig:BE1} and ~\ref{fig:BE1comp}.

Figure \ref{fig:BE1conv} presents the convergence of the calculations with respect to the maximum hypermomentum $K_{max}$ used to generate the dipole continuum. The curves have been obtained by smoothing the discrete $B(E1)$ values with Poisson functions. Note that one should pay more attention to this issue in case that experimental data become available, in particular to take into account the corresponding energy resolution. It is shown that the distribution converges to a localized peak at low excitation energies. Calculations with $K_{max}>20$ were found to be indistinguishable from that with $K_{max}=20$. 

The converged distribution, up to a continuum energy of 6 MeV, is singled out in Fig.~\ref{fig:BE1}. The calculations are characterized by a clear maximum around 0.85 MeV, as we already discussed in Ref.~\cite{Fortunato2020}. The inset shows the cumulative integral up to a given continuum energy, which gives 1.59 e$^2$fm$^2$ up to 6 MeV. This number is a large fraction (83\%) of the dipole cluster sum rule given by Eq.~(\ref{eq:sumrule}), that yields $S_{T}(E1)=1.92$ e$^2$fm$^2$. If we integrate the $B(E1)$ distribution up to 15 MeV, the result approaches the exact value for the sum rule. As discussed in Sec.~\ref{sec:introduction}, such a large integrated $B(E1)$ value at low continuum energies is a signature of neutron halo nuclei. Note that a similar value was measured recently for the case of $^{19}$B~\cite{Cook2020}, which was claimed to present a two-neutron halo, and it is also consistent with the reported values for other halo nuclei~\cite{Nakamura94,nakamura06}. 

\begin{figure}[t]
\centering
\includegraphics[width=0.85\linewidth]{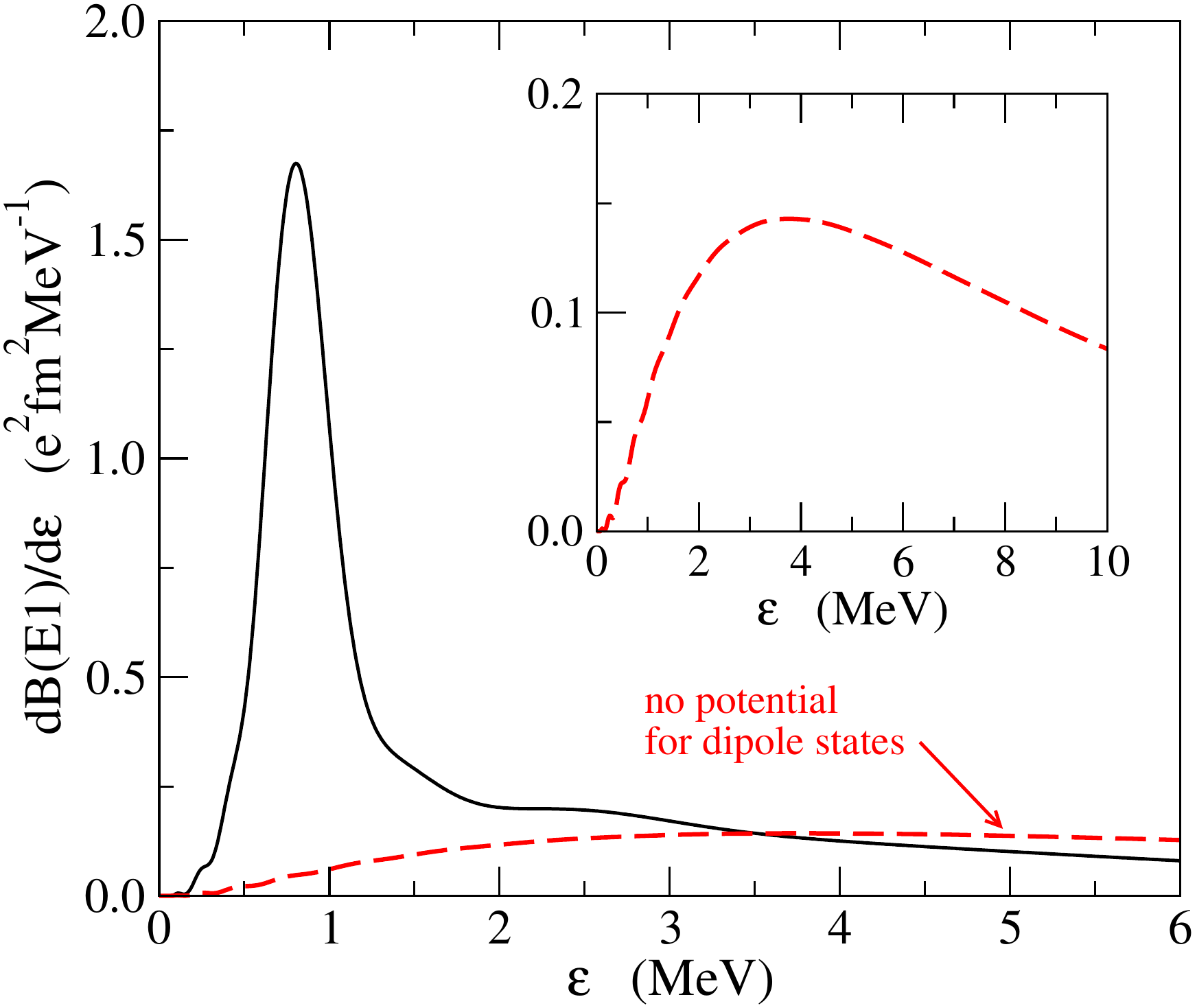}
\caption{$B(E1)$ distribution compared with the results setting the potential for dipole states to zero (dashed red line). The inset shows this unrealistic calculation with a more appropriate scale and up to 10 MeV in the continuum.}
\label{fig:BE1-pw}
\end{figure}

To assess whether the peak in the $B(E1)$ distributions corresponds to a resonant state, we have checked the stability of its position with changes in the ground-state energy. If we move the ground state by changing the three-body force, the total $E1$ strength changes, but the maximum remains at the same energy. This implies that the peak may arise from a dipole resonance. We explicitly checked that the maximum is not due to a threshold effect (c.f., the low-energy enhancement discussed, for instance, in Ref.~\cite{Nagarajan05}) by performing a calculation with plane waves for the dipole states, i.e., setting the potential to zero. The result is shown in Fig.~\ref{fig:BE1-pw}, where the narrow peak observed in Figs.~\ref{fig:BE1conv} and \ref{fig:BE1} disappears and the E1 strength spreads towards higher energies in the continuum. 
To gain more insight into this structure, we did additional calculations using the identification method proposed by one of us in Ref.~\cite{JCasal19}, and we confirmed the presence of a dipole resonance at the peak energy. We also computed the three-body eigenphases using the \textsc{sturmxx} code~\cite{sturmxx}, finding the same resonant state. The situation for the dipole continuum in $^{29}$F resembles that of $^{11}$Li, where several works indicate the presence of a low-energy dipole resonance~\cite{Cubero12,JPFernandezGarcia13,aumann19}. 
As shown in Fig.~\ref{fig:BE1}, the resonance extinguishes a substantial portion of the total strength. By studying the partial-wave content of a single 1$^-$ pseudostate around the peak, we find that it corresponds to 73\% of $(2p_{3/2})(1d_{3/2})$ configurations, so the computed $E1$ distribution is governed by $(2p_{3/2})^2\rightarrow (2p_{3/2})(1d_{3/2})$ and $(1d_{3/2})^2\rightarrow (2p_{3/2})(1d_{3/2})$ transitions from the ground state. These predictions for the low-lying $B(E1)$ strength in $^{29}$F, including its resonance character, require an experimental confirmation in Coulomb dissociation (CD) experiments.

In Fig.~\ref{fig:BE1} we show also the Relativistic Coulomb Excitation (RCE) cross section obtained by using the well-known Alder \& Winther theory~\cite{WintherAdler79,BertulaniEPM}. Within this semiclassical approach, the cross section is proportional to the electromagnetic strength function. In the case of $E1$ transitions, this is given by
\begin{equation}
    \left.\frac{d\sigma}{d\varepsilon}\right|_{E1} =(\alpha Z_t)^2 \sum_{\mu} \frac{dB(E1)}{d\varepsilon}\mid G_{E1\mu}(\beta^{-1})\mid^2g_\mu(\xi),
    \label{eq:RCE}
\end{equation}
where $Z_t$ is the charge of the target, $\alpha$ is the fine structure constant, and $G_{E1\mu}$ are relativistic Winther-Alder functions for dipole excitations, which depend on the bombarding energy. Note that, although the previous expression does not contain an explicit dependence on the continuum energy, the function $g$ has a decreasing dependence on the energy through the so-called adiabaticity parameter $\xi$. This modulates the cross section, flattering its tail. By considering the case of a $^{29}$F beam impinging at 235 MeV/u on a $^{208}$Pb target, we get an estimate for the Coulomb dissociation cross section of $\approx$ 550 mb up to 6 MeV. By using again the Glauber theory, we provide new estimations for the nuclear breakup contribution simply by taking the difference between the total reaction cross sections of $^{29}$F and $^{27}$F. The values are 300 mb for $S_{2n}=1.44$ MeV and 390 mb for $S_{2n}=1.00$ MeV, which are comparable to the CD cross section. This estimation may guide future CD experiments to assess the halo structure of the exotic $^{29}$F nucleus. Note that the higher-order contributions might play a role, and this requires further investigations.

As a final comment on the $B(E1)$ distribution, we note that this observable (and the corresponding cross section) is highly sensitive to the ground-state radius and configuration mixing. In particular, calculations using the model A (standard shell-model order) of Ref.~\cite{Singh2020} leads to a reduction of the total  strength by $\approx 40$\%. This is illustrated in Fig.~\ref{fig:BE1comp}. The present calculations are built on the recent experimental results for the unbound $^{28}$F subsystem~\cite{Revel2020} and are in agreement with the interpretations in Ref.~\cite{Bagchi2020} that $^{29}$F is a two-neutron halo nucleus linked to the occupancy of the intruder $2p_{3/2}$ orbital. Therefore, a measurement of the $B(E1)$ distribution for this nucleus could provide a natural confirmation of these findings.

\section{Four-body reaction calculations}
\label{sec:4b}

The coupling to breakup channels plays a key role in low-energy reactions involving halo nuclei (see, for instance, the studies on $^{6}$He or $^{11}$Li~\cite{MRoGa08,Cubero12,JPFernandezGarcia13}).
While such data for $^{29}$F are not yet available, the low-energy dynamics involving this nucleus could also help in assessing its halo nature, complementing the interaction cross section measurement and the high-energy Coulomb dissociation discussed in the previous sections.

\begin{figure}[t]
\centering
\includegraphics[width=0.55\linewidth]{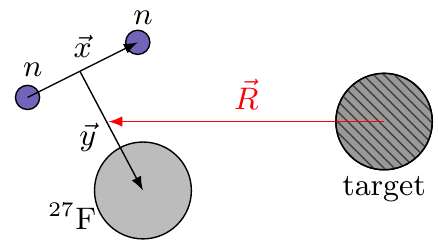}
\caption{Coordinates in the four-body CDCC framework, considering a three-body projectile impinging on a structureless target.}
\label{fig:4b}
\end{figure}

Formally, these effects can be studied within the Continuum-Discretized Coupled-Channels (CDCC) formalism~\cite{Austern87,Yahiro86}, by expanding the total projectile-target scattering wave functions in (bound and unbound) states of the projectile, which can be generated in a given few-body model. In the case of three-body projectiles, this is typically referred to as four-body CDCC~\cite{MRoGa08,Matsumoto04-2}. Assuming a structureless target, and introducing the relative coordinate $\boldsymbol{R}$ in Fig.~\ref{fig:4b}, the scattering wave function is expanded as
\begin{equation}
    \Psi^{JM} = \sum_{c}\frac{i^L}{R}u_c^J(R)\Phi_c^{JM}(\widehat{R},\boldsymbol{x},\boldsymbol{y}),
    \label{eq:4bwf}
\end{equation}
where $c\equiv\{L(nj)\}$ stands for a channel wave function corresponding to a projectile state with labels $(nj)$ and a relative projectile-target angular momentum $L$, 
\begin{equation}
    \Phi_c^{JM}(\widehat{R},\boldsymbol{x},\boldsymbol{y}) = \left[Y_{L}(\widehat{R})\otimes\psi_n^j(\boldsymbol{x},\boldsymbol{y})\right]_{JM}.
    \label{eq:channelwf}
\end{equation}
As in Sec.~\ref{sec:E1}, here index $n$ labels the different pseudostates for a given total angular momentum $j$ of the three-body nucleus. We consider the Hamiltonian of the projectile-target system,
\begin{equation}
    H(\boldsymbol{R},\boldsymbol{x},\boldsymbol{y}) = h_p(\boldsymbol{x},\boldsymbol{y}) + T_R + V_{pt}(\boldsymbol{R},\boldsymbol{x},\boldsymbol{y}),
    \label{eq:reacH}
\end{equation}
where $h_p$ is the internal three-body Hamiltonian of the projectile (used to generate the states $\psi^j_n$ in Sec.~\ref{sec:formalism}), $T_R$ is the kinetic-energy term associated to the relative motion, and $V_{pt}$ represents the interaction between projectile and target. Thus, the radial functions in Eq.~(\ref{eq:4bwf}) are obtained from the coupled equations
\begin{equation}
\begin{split}
&\left[-\frac{\hbar^2}{2m_r}\left(\frac{d^2}{dR^2}-\frac{L(L+1)}{R^2}\right)+E_{nj}-E\right]u_c^{J}(R)\\&+\sum_{c'}V_{c,c'}^{J}(R) u_{c'}^{J}(R)=0,
\end{split}
\label{eq:4beq}
\end{equation}
where $m_r$ is the reduced mass, and the scattering coupling potentials are 
\begin{equation}
    V_{c,c'}^J(R) = \langle \Phi_c^{JM} | V_{pt}|\Phi_{c'}^{JM} \rangle,
    \label{eq:4bcoup}
\end{equation}
integrated over $\widehat{R}$ and \{$\boldsymbol{x},\boldsymbol{y}\}$. Here, we assume that $V_{pt}=V_1+V_2+V_3$, and $V_i$ are optical potentials between each particle and the target. In this case, we need suitable core-target and $n$-target optical potentials. Then, the coupling potentials in the previous expression are generated by performing a multipole expansion of the projectile-target interaction. Details on how to compute these couplings can be found in Refs.~\cite{MRoGa08,CasalTh} and are summarized in Appendix~\ref{app2}. Once the radial equations are solved with scattering boundary conditions, we can obtain the corresponding scattering matrices and cross sections. In the present work, this is done by inserting the computed coupling potentials in the code \textsc{fresco}~\cite{fresco}.

It is worth noting that, in the CDCC framework, only inelastic excitations and breakup are treated explicitly. Thus, possible target excitations and/or the effect of other reaction channels not explicitly included in the coupling form factors are considered implicitly through the imaginary part of the optical potentials employed. For the present calculations, we use the Koning-Delaroche ($n$-target)~\cite{KD} and the S\~ao Paulo (core-target)~\cite{SPP} global optical potentials at the appropriate energy per nucleon. 

In this work, we present predictions for the elastic scattering of $^{29}$F on $^{120}$Sn at $E_{\rm lab}=84$ MeV, i.e., slightly above the Coulomb barrier for this system. We choose this situation in order to illustrate the qualitative effect of dipole couplings due to the strong Coulomb field generated by the target. Note that the calculations for heavier targets, such as $^{208}$Pb, are computationally more demanding due to convergence issues (see, e.g., Ref.~\cite{MRoGa08}), so here we study the reaction on $^{120}$Sn. We consider $j=0^+,1^-,2^+$ excitations of the $^{29}$F projectile, including bound and continuum states generated up to a given maximum energy ($\varepsilon_{max}$) within our THO basis. Since the number of continuum states to be coupled depends not only on the maximum energy but also on the size of the model space and basis choice, one needs to ensure the convergence of the elastic cross section also with respect to $K_{max}$ and $N$, in a similar fashion as the convergence analysis presented in Sec.~\ref{sec:gs} for the ground-state properties. From such analysis, we set $K_{max}=14$, $N=8$ and $\varepsilon_{max}=8$ MeV. Note that these $K_{max}$ and $N$ values are well below the ones needed to ensure convergence of the ground state. Nevertheless, we have checked that the final elastic distributions are converged within a 5-10\% difference, which is sufficient for the present discussion. This required a small change in the three-body force ($v_{3b}$) so that the ground-state energy matches that of the full calculations. 

\begin{figure}[t]
\centering
\includegraphics[width=0.75\linewidth]{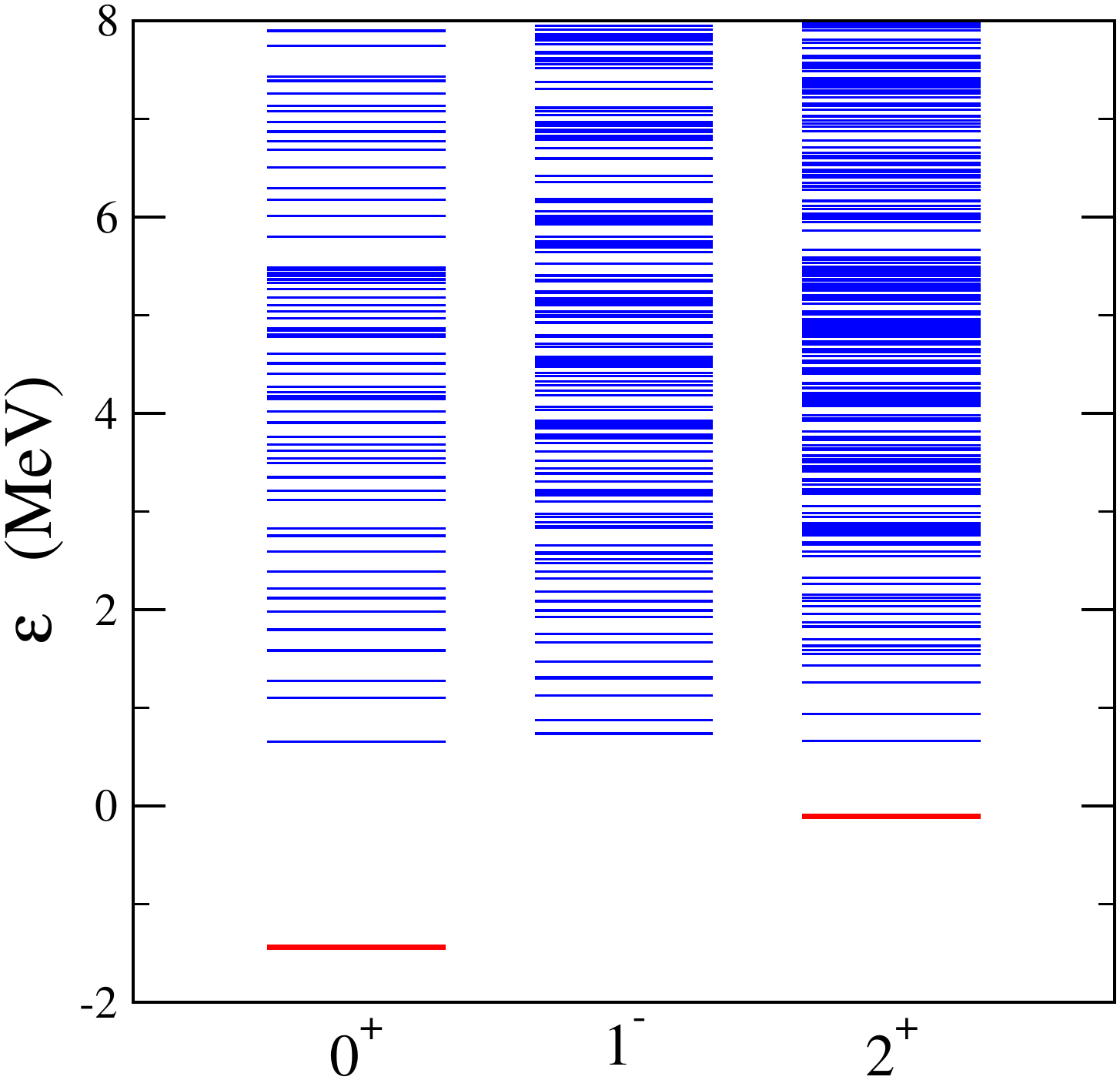}
\caption{Energies of the 0$^+$, 1$^-$ and 2$^+$ states of $^{29}$F included in the present four-body coupled-channels calculations. Note there are two bound states (thick red lines).}
\label{fig:spectraCDCC}
\end{figure}

\begin{figure}[t]
\centering
\includegraphics[width=0.9\linewidth]{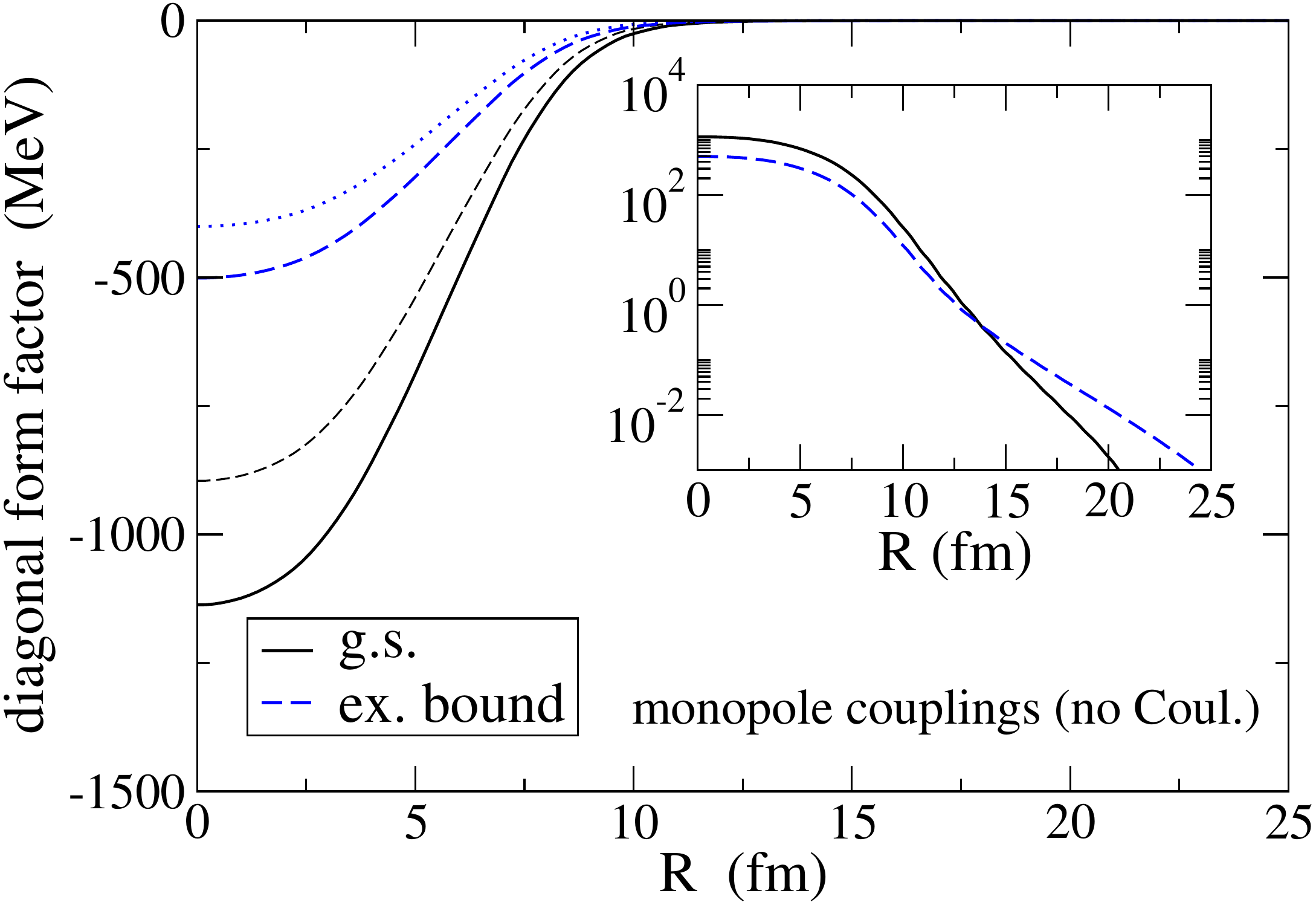}
\caption{Diagonal real form factors (nuclear monopole) for the ground state (solid black line) and the only excited bound state (dashed blue line) considered in the $^{29}\text{F}+^{120}\text{Sn}$ reaction at $E_{\rm lab}=84$ MeV. The thin dashed and dotted lines correspond to the imaginary parts. The inset shows the real part (absolute value) in logarithmic scale.}
\label{fig:ff_diag}
\end{figure}

\begin{figure}[t]
\centering
\includegraphics[width=0.85\linewidth]{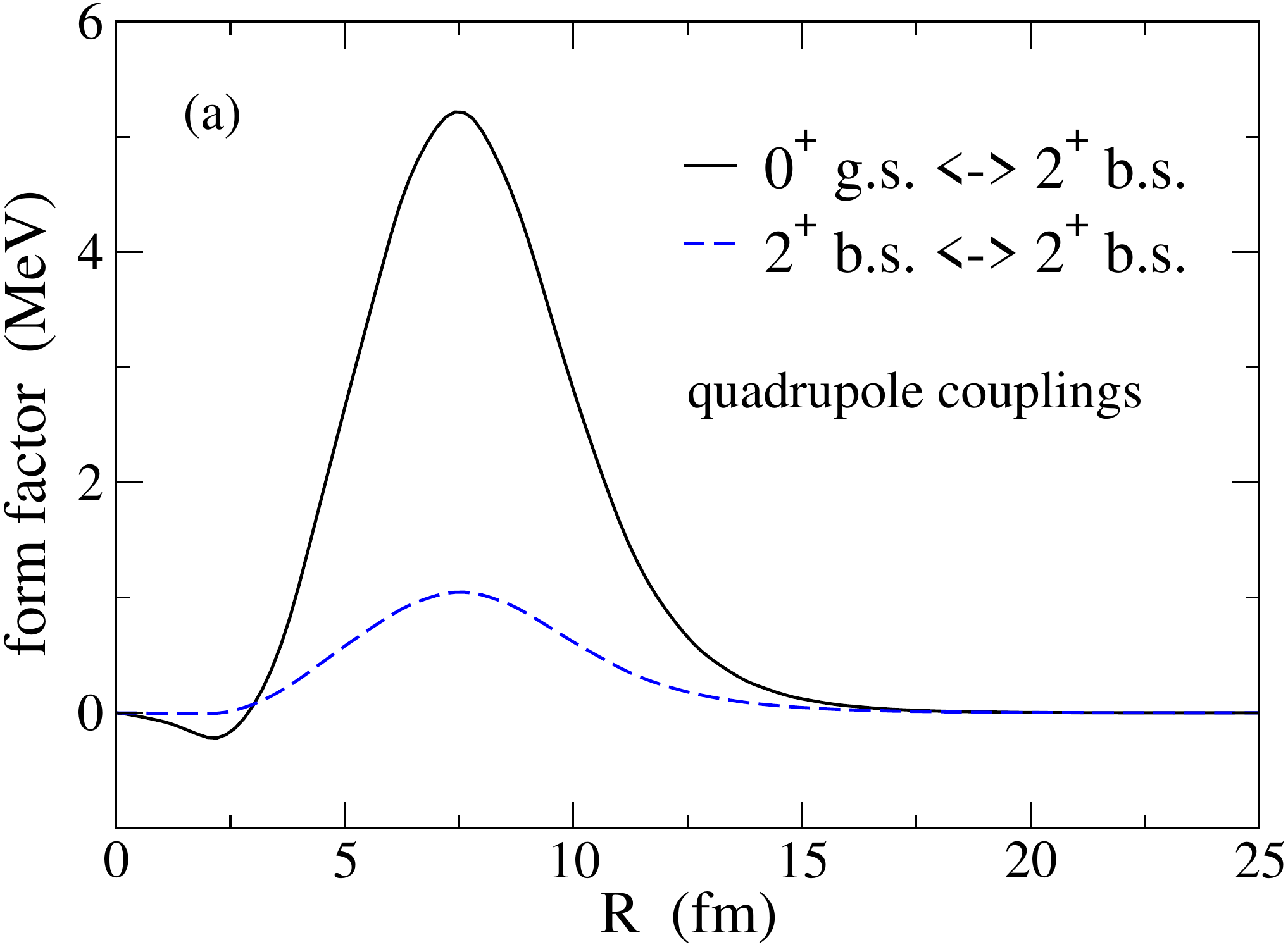}
\vspace{5pt}

\includegraphics[width=0.85\linewidth]{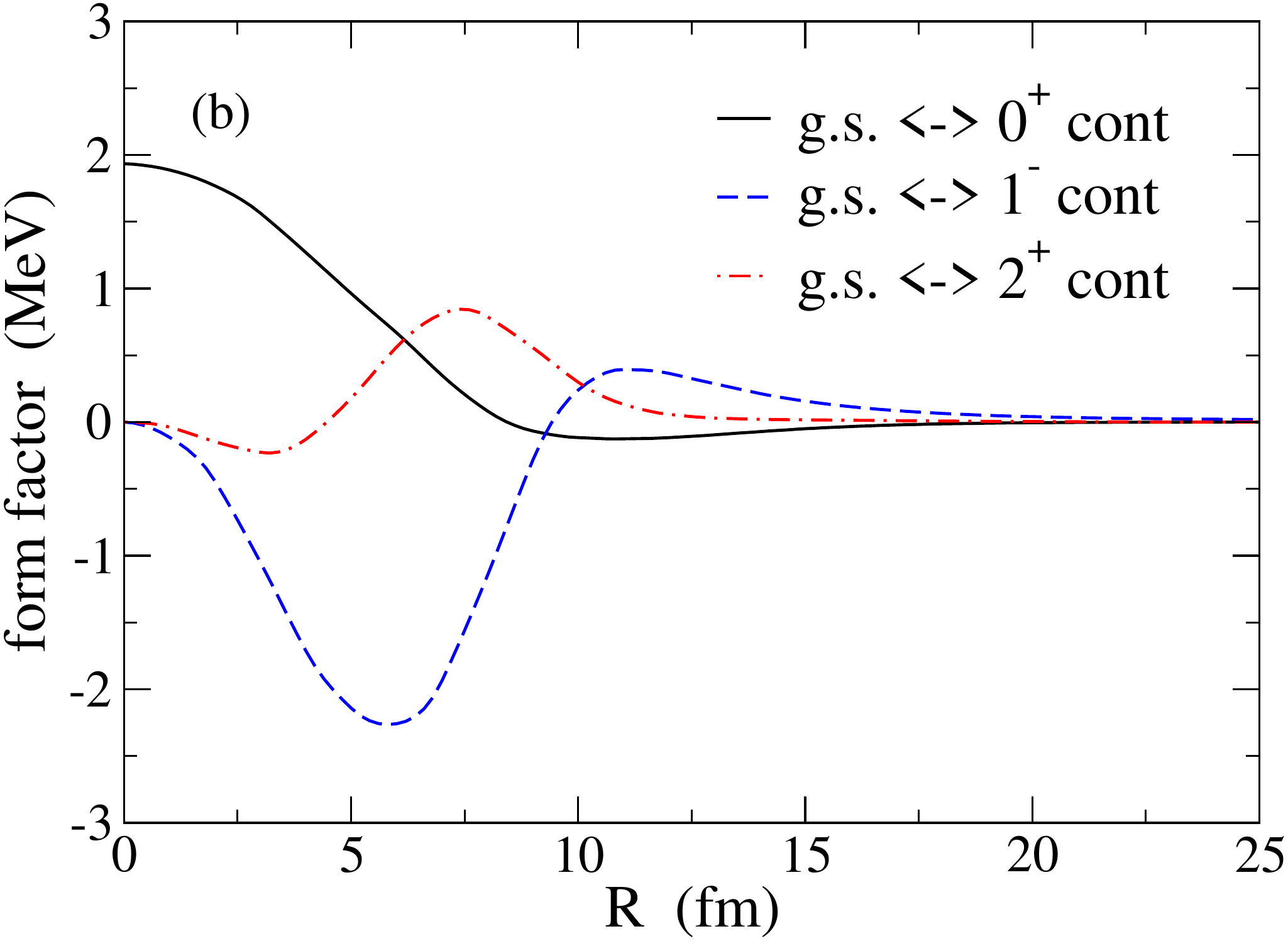}
\caption{Form factors (real part only) for the $^{29}\text{F}+{^{120}\text{Sn}}$ reaction at $E_{\rm lab}=84$ MeV. (a) Quadrupole couplings involving the bound states, i.e., $0^+\Leftrightarrow 2^+$ (solid black line) and $2^+\Leftrightarrow 2^+$ (dashed blue line). (b) Monopole (solid black line), dipole (dashed blue line) and quadrupole (dot-dashed red line) couplings connecting the ground state with continuum pseudostates at $\varepsilon\approx 1$ MeV.}
\label{fig:ff}
\end{figure}

Since the optical potentials and the states describing the structure part are fixed, our four-body CDCC calculations involve no parameter fitting. The states we couple in the calculations are shown in Fig.~\ref{fig:spectraCDCC}, which include the ground state of $^{29}$F and the only excited bound state in the present model, as well as our pseudostate representation of the continuum. In Fig.~\ref{fig:ff_diag} we show the monopole diagonal form factors for the two bound states. Note the slightly longer extension of the form factor corresponding to the excited bound state, which is highlighted in the inset. In Fig.~\ref{fig:ff}, we present some higher-order couplings. The top panel shows the quadrupole form factors involving the two bound states, while the bottom panel illustrates the couplings between the ground state and continuum states for monopole (only nuclear), dipole and quadrupole excitations. The latter panel has been obtained by considering, in the three cases, a pseudostate around 1 MeV above the $^{27}\text{F}+n+n$ threshold. From this figure, it is clear that low-energy dipole couplings have a longer range than their quadrupole counterpart, so they are expected to play a significant role in for the low-energy reaction under consideration.

\begin{figure}[t]
\centering
\includegraphics[width=0.9\linewidth]{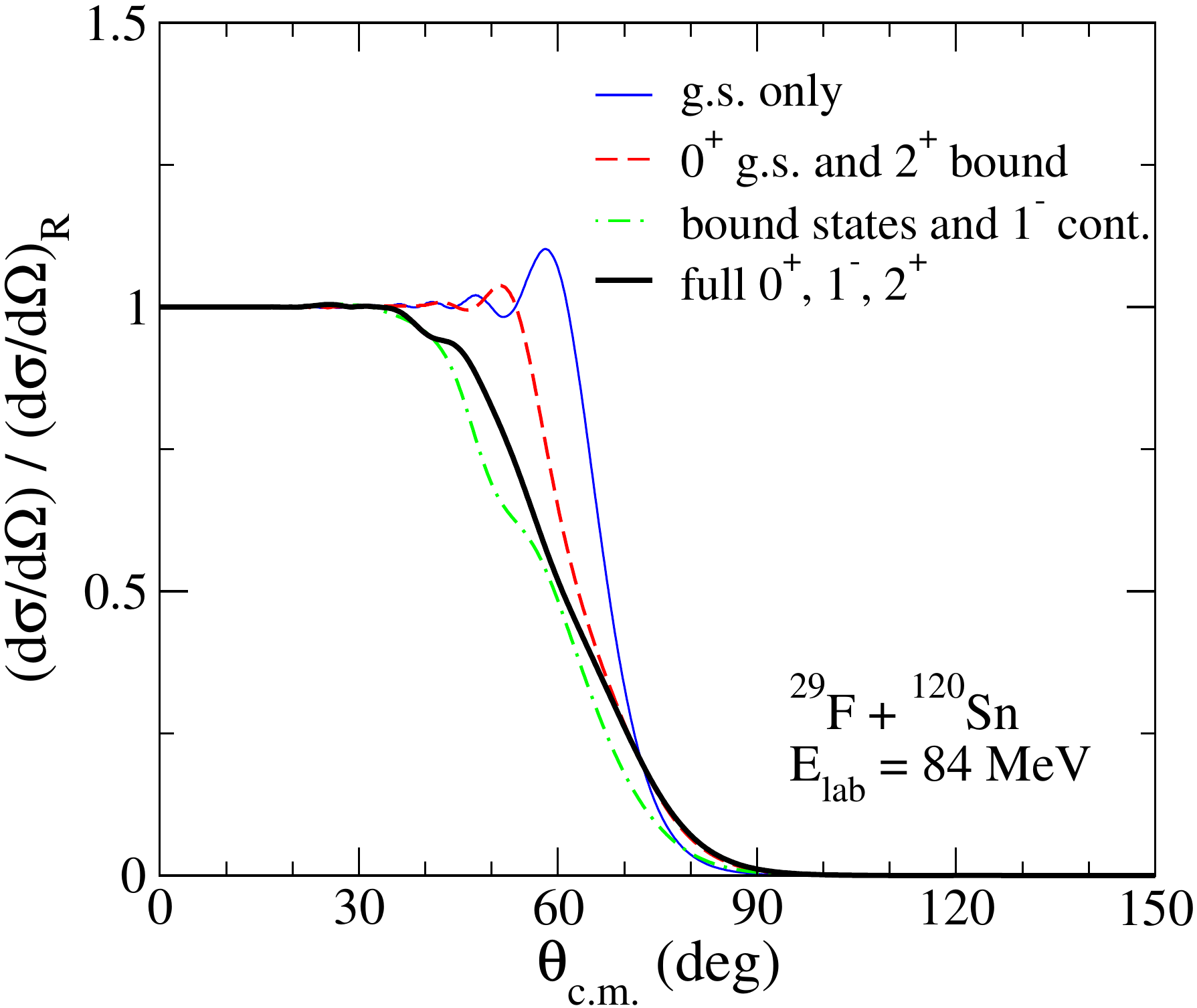}
\caption{Angular distribution of the elastic scattering cross section (relative to Rutherford) for the $^{29}\text{F}+{^{120}\text{Sn}}$ reaction at $E_{\rm lab}=84$ MeV. The thick solid line corresponds to the full CDCC calculations, while the others contain a restricted set of $^{29}$F states. See the text for details.}
\label{fig:elas_84}
\end{figure}

The elastic cross section, relative to the Rutherford cross section, is presented in Fig.~\ref{fig:elas_84}. This figure shows different calculations to illustrate the effect of an increasing number of coupled states. The thin blue line is the result by considering only the ground state of $^{29}$F, which has a clear Fresnel peak around 60$^\circ$. The dashed red line corresponds to a calculation including also the bound excited state (i.e., includes the inelastic excitation of $^{29}$F). In that case, the quadrupole coupling between bound states shifts the maximum to smaller angles and slighly increases the cross section at backward angles. The dot-dashed green line is the result considering the two bound states and the $1^-$ continuum. It is clear that the inclusion of dipole coupling leads to a strong reduction of the elastic scattering cross section and cancels the Fresnel peak, in total analogy with the behavior observed for other two-neutron halo nuclei at energies slightly above the Coulomb barrier~\cite{MRoGa08,Acosta11,Cubero12}, and at variance with the no-halo case (e.g., Ref.~\cite{AdiPietro04} for a comparison between $^4$He and $^6$He). Lastly, the thick black line is the results when all 0$^+$, 1$^-$ and 2$^+$ states are included. While this contains a fair number of continuum-continuum couplings, the final result is not very different from the calculations including only the dipole states. This is another signal that dipole couplings dominate the reaction mechanism.

Our CDCC calculations provide also the inelastic, breakup and total reaction cross sections. In this particular case, we get $\sigma_{inel}(2^+)=60.5$ mb, which is just a small fraction of the total reaction cross section, $\sigma_{reac}=2.37$ b. For the total breakup cross section, we obtain $\sigma_{BU}=440$ mb. It is worth noting that this cross section corresponds to the so-called elastic breakup only, i.e., $^{29}$F breaks into its $^{27}\text{F}+n+n$ constituents without exciting either the core or the target nucleus. The experimental counterpart for that quantity will require exclusive measurements.

\vspace{-10pt}

\section{Summary and conclusions}
\label{sec:conclusions}

We reported three-body ($^{27}\text{F}+n+n$) calculations for the $^{29}$F nucleus taking into account recent experimental information on the unbound $^{28}$F system. We fixed a new $^{27}\text{F}+n$ effective potential, which is consistent with parity inversion and gives rise to a $p_{3/2}$ ground-state resonance (0.199 MeV) and an excited $d_{3/2}$ resonance (0.966 MeV). As in our previous work, we neglected the spin of the core and solved the three-body problem within the hyperspherical harmonics expansion method, describing the radial functions using the analytical THO basis. We provide a more exhaustive analysis with respect to our results in Ref.~\cite{Fortunato2020}, including additional details regarding the theoretical formalism, the ground state properties and the $B(E1)$ distribution, the search for additional bound states, Glauber-model calculations for high-energy reactions, and four-body CDCC calculations at low energy.

With the updated model, and after fixing the two-neutron separation energy of $^{29}$F to the experimental value of 1.44 MeV by using a small three-body force, our ground state consists of two valence neutrons occupying mostly $(p_{3/2})^2$ intruder configurations (57.5\%). Glauber model calculations of the total reaction cross section on a carbon target using our ground-state density are in good agreement with the available data. The present model yields a relative increase of the matter radius of $\Delta R = R_m-R(^{27}\text{F})=0.20$ fm (which goes up to 0.25 fm if the lower limit of $S_{2n}$ is considered). The large mixing with the  $(d_{3/2})^2$ configuration (29.0\%) favors dineutron correlations in the ground state and reinforces the interpretation of $^{29}$F as the heaviest two-neutron halo nucleus up to date. These results are in reasonable agreement with recent estimations from interaction cross section measurements and shell-model calculations. Our model produces also a bound excited state at $\approx$ 0.10 MeV below the three-body threshold and consistent with a quadrupole excitation of the valence neutrons.

Our calculations for the continuum in a pseudostate representation yield a large $E1$ strength at low energies, as expected for a halo nucleus. The computed $B(E1)$ distribution exhibits a clear peak around 0.85 MeV above the $^{27}\text{F}+n+n$ threshold. This maximum is consistent with a dipole resonance. The integrated strength up to 6 MeV is 1.59 e$^2$fm$^2$, which is similar to the total $B(E1)$ reported for other two-neutron halo nuclei. We found that this strength is reduced by $\approx$ 40\% if one considers a standard shell-model order for the spectrum of the $^{28}$F subsystem. Therefore, a measurement of the $B(E1)$ distribution for $^{29}$F could provide an definitive confirmation of its halo structure. We also computed, using the Winther \& Alder theory, the RCE cross section at 235 MeV/u on a lead target to be $\approx$ 550 mb (up to 6 MeV). A Glauber model estimation for the nuclear contribution yields $\approx 300$ mb, which is comparable to the Coulomb contribution. These results may guide future Coulomb dissociation experiments via invariant mass spectroscopy. 

Since the near-barrier dynamics of halo nuclei are known to be strongly driven by the coupling to breakup channels, we performed also the first four-body CDCC calculations with $^{29}$F as a projectile. We considered the reaction on $^{120}$Sn at $E_{\rm lab}=84$ MeV, i.e., slightly above the Coulomb barrier. Dipole couplings involving the low-lying continuum in $^{29}$F were found to play a key role, leading to a complete cancellation of the Fresnel peak in the elastic cross section. This resembles what has already been observed for other ``more standard'' two-neutron halos. Thus, low-energy reactions may provide a complementary way to probe the halo structure in $^{29}$F and to assess our prediction of a large $E1$ strength. 

It is worth noting that the present calculations do not consider the finite spin of the core or possible core excitations, which may affect the structure of $^{29}$F. This could be constrained by a more detailed study of the level structure in $^{28}$F. Our three-body wave functions, including bound and unbound states, could also be employed to study other processes involving this nucleus, e.g., knockout or transfer. Work along these lines will require further investigations.

\begin{acknowledgments}
This work has been partially supported by SID funds 2019 (Università degli Studi di Padova, Italy) under project No.~CASA\_SID19\_01, by the Ministerio de Ciencia e Innovación and FEDER funds (Spain) under project No.~FIS2017-88410-P, by the European Union's Horizon 2020 research and innovation program under grant agreement No.~654002, and by JSPS KAKENHI Grants No.~18K03635, 18H04569, and 19H05140. WH acknowledges the collaborative research program 2020, Information Initiative Center, Hokkaido University.
\end{acknowledgments}

\appendix
\section{Wave-function transformations}
\label{app1}

Within the hyperspherical formalism presented in Sec.~\ref{sec:formalism}, wave function in the Jacobi-Y system (see Fig.~\ref{fig:jac}) can be obtained from that in the Jacobi-T set as
\begin{align}
    \nonumber \psi_{Y}^{j\mu}(\rho,\Omega') & = \rho^{-5/2}\sum_{\beta'} \left(\sum_{\beta}N_{\beta,\beta'}\chi_{\beta}^{j}(\rho)\right) \mathcal{Y}_{\beta'}^{j\mu}(\Omega')\\
    & = \rho^{-5/2}\sum_{\beta'} \eta_{\beta'}^j(\rho) \mathcal{Y}_{\beta'}^{j\mu}(\Omega'),
    \label{eq:wfY}
\end{align}
where $N_{\beta,\beta'}\equiv \langle Y,\beta'|T,\beta \rangle$ are the angular basis overlaps between the $T$ and $Y$ Jacobi representations. These overlaps are analytical and can be easily obtained from the Reynal-Revai coefficients~\cite{RR70} (see, e.g., Eqs.~(13) and (14) of Ref.~\cite{IJThompson04}). Note that the transformation preserves $K$ and $l$, so the angular functions $\mathcal{Y}_{\beta'}^{j\mu}$ in the previous expression follow the coupling order
\begin{equation}
\mathcal{Y}_{\beta'}^{j\mu}(\Omega')=\left\{\left[\Upsilon_{Kl}^{l_x'l_y'}(\Omega)\otimes\phi_{S_x'}\right]_{j_{ab}'}\otimes\kappa_I'\right\}_{j\mu}.
    \label{eq:Ycoup}
\end{equation}
For a $\text{core}+n+n$ system, $S_x'=I'=s=1/2$ is the spin of a single neutron. 

We are interested in expressing the wave function in terms of a single-particle angular momentum of the neutrons with respect to the core. First, we can express the wave function~(\ref{eq:wfY}) in Jacobi-Y coordinates as
\begin{equation}
   \psi_{Y}^{j\mu} = \frac{1}{xy} \sum_{\gamma} w_{\gamma}^j(x',y') |Y, \gamma\rangle,
    \label{eq:wfYxy}
\end{equation}
where $\gamma$ is a set of quantum numbers without the hypermomentum $K$, so that $\beta'\equiv\{K,\gamma\}$, the radial functions are
\begin{equation}
    w^j_\gamma(x',y')=\rho^{-1/2}\sum_{K} \eta_{K,\gamma}^j(\rho)\varphi_{K}^{l_x'l_y'}(\alpha'),
    \label{eq:xyrad}
\end{equation}
and $\rho=\sqrt{x'^2+y'^2}$, $\alpha'=\arctan{(x'/y')}$. The ket in Eq.~(\ref{eq:wfYxy}) represents
\begin{equation}
    |Y, \gamma\rangle = |\left[(l_x'l_y')l,s\right]j_{ab}',s \rangle^j,
    \label{eq:ketgamma}
\end{equation}
where the orbital angular momenta in the $x'$ and $y'$ coordinates are coupled to $l$ before adding up the spins. We can decouple these quantum numbers in order to explicitly separate all the $\boldsymbol{x}'$ and $\boldsymbol{y}'$ dependence. After working out the algebra, this leads to
\begin{align}
    \nonumber & |\left[(l_x'l_y')l,s\right]j_{ab}',s \rangle^j = (-)^{l-l_x'-l_y'}\hat{l}\hat{j}_{ab}' \\ \nonumber &  \times \sum_{j_1}  (-)^{j_{ab}'-l_y'-j_1} \hat{j}_1 W(l_y'l_x'j_{ab}'s;lj_1) \\ &\times \sum_{j_2} \hat{j}_2 W(j_1l_y'js;j_{ab}'j_2) |(l_x's)j_1,(l_y's)j_2 \rangle^j.
    \label{eq:changecoup}
\end{align}
In this expression, $j_1$ and $j_2$ are associated to the $x'$ and $y'$ coordinate, respectively, and $\boldsymbol{j}_1+\boldsymbol{j}_2=\boldsymbol{j}$. By inserting Eq.~(\ref{eq:changecoup}) in Eq.~(\ref{eq:wfYxy}), and replacing the sums in $j_{ab}'$ and $l$ by the corresponding ones for $j_1$ and $j_2$, we get
\begin{equation}
   \psi_{Y}^{j\mu} = \frac{1}{xy} \sum_{\gamma'} f_{\gamma'}^j(x',y') |Y, \gamma'\rangle,
    \label{eq:wfYxych}
\end{equation}
where the radial functions are given by
\begin{align}
    \nonumber & f_{\gamma'}^j(x',y')=\hat{j}_1\hat{j}_2 \sum_{j_{ab}'} (-)^{j_{ab}-l_y-j_1} \hat{j}_{ab}' W(j_1l_y'js;j_{ab}'j_2) \\& \times \sum_{l} (-)^{l-l_x'-l_y'} \hat{l} W(l_y'l_x'j_{ab}'s;lj_1) \eta_{\gamma}^j(x',y').
    \label{eq:xyradch}
\end{align}
and $\gamma'$ represents the new set of quantum numbers including the shell-model-like angular momenta $j_1$, $j_2$, i.e., 
\begin{equation}
|Y, \gamma'\rangle = |(l_x's)j_1,(l_y's)j_2 \rangle^j.
\end{equation}
Once the wave function has been transformed and follows Eq.~(\ref{eq:wfYxych}), it is straightforward to extract the configurations corresponding to the valence neutrons in a given single-particle orbital.

\section{Four-body coupling form factors}
\label{app2}

To generate the coupling potentials needed in the four-body CDCC calculations (Eq.~(\ref{eq:4bcoup})), we consider a sum over the three projectile constituents. Let us define $\{\boldsymbol{R}_q\}$ as the position vectors of each particle to the target nucleus. In general, these vectors can be expressed in terms of $\boldsymbol{R}$ and the Jacobi coordinates $\{\boldsymbol{x},\boldsymbol{y}\}$ (see Fig.~\ref{fig:4b}). For simplicity, each contribution can be computed after rotating the projectile wave function to a Jacobi set in which $\boldsymbol{R}_q$ depends only on $\boldsymbol{R}$ and $\boldsymbol{y}$, and we use the label $q$ to represent those coordinates. Assuming central potentials, we consider a multipole expansion,
\begin{equation}
    V_q(R_q) = \sum_Q (2Q+1)\mathcal{V}_Q^q(R,y_q)P_Q(z_q),
    \label{eq:Q}
\end{equation}
where $Q$ denotes the multipole order, $P_Q(z_q)$ is a Legendre polynomial, and $z_q\equiv \widehat{y}_q\cdot\widehat{R}$ is the cosine of the angle between $\boldsymbol{y}_q$ and $\boldsymbol{R}$. The coefficients in this expansion are given by
\begin{equation}
    \mathcal{V}^q_Q(R,y_q) = \int_{-1}^{+1} dz_qV_q(R_q)P_Q(z_q).
    \label{eq:Qcoeff}
\end{equation}
Using these definitions and the hyperspherical harmonic expansion of the states, it is possible to write the coupling potentials~(\ref{eq:4bcoup}) as
\begin{align}
    \nonumber V^{J}_{Lnj,L'n'j'}(R) & = \sum_Q (-)^{J-j}\hat{L}\hat{L}'(2Q+1) \tj{L}{Q}{L'}{0}{0}{0}\\
    & \times W(LL'jj';QJ) F^Q_{nj,n'j'}(R),
    \label{eq:Vcc}
\end{align}
Here, all the dependence on the projectile states is factorized in a set of radial form factors $F^Q_{nj,n'j'}$ given by~\cite{CasalTh}
\begin{align}
\nonumber & F_{nj,n'j'}^Q(R)  = (-1)^{Q+2j-j'}\hat{j}\hat{j}'\sum_{\beta\beta'}\sum_{q=1}^3\sum_{\beta_q\beta_{q}'} N_{\beta\beta_q} N_{\beta_q\beta_{q}'} \\ 
\nonumber & \times (-1)^{l_{x_q}+S_{x_q}+j_{ab_q}'-j_{ab_q}-I_q} \delta_{l_ql_q'}\delta_{S_{x_q}S_{x_q}'}\hat{l}_{y_q}\hat{l}_{y_q}'\hat{l}_q\hat{l}_q' \hat{j}_{ab_q}\hat{j}_{ab_q}'\\ 
\nonumber & \times\left(\begin{array}{ccc}l_{y_q}&Q&l_{y_q}'\\0&0&0\end{array}\right) W(l_ql_q'l_{y_q}l_{y_q'};Ql_{x_q}) \\
\nonumber & \times W(j_{ab_q}j_{ab_q}'l_ql_q';QS_{x_q}) W(jj'j_{ab_q}j_{ab_q}';QI_q) \\
& \times \sum_{ii'}C_n^{i\beta_qj}C_{n'}^{i'\beta_q'j'} \mathcal{I}_{i\beta_q,i'\beta_q'},
\end{align}
where $\mathcal{I}_{i\beta_q,i'\beta_q'}$ represents the double integrals (in $\rho$ and $\alpha_q$) of $\mathcal{V}^q_Q(R,y_q)$ between hyperradial and hyperangular basis functions. Note that the coefficients $N_{\beta_q,\beta_q'}$ provide the transformations between different sets of Jacobi coordinates and have been introduced in Appendix~\ref{app1}.

\bibliography{bibfile}

\begin{thebibliography}{90}%
\makeatletter
\providecommand \@ifxundefined [1]{%
 \@ifx{#1\undefined}
}%
\providecommand \@ifnum [1]{%
 \ifnum #1\expandafter \@firstoftwo
 \else \expandafter \@secondoftwo
 \fi
}%
\providecommand \@ifx [1]{%
 \ifx #1\expandafter \@firstoftwo
 \else \expandafter \@secondoftwo
 \fi
}%
\providecommand \natexlab [1]{#1}%
\providecommand \enquote  [1]{``#1''}%
\providecommand \bibnamefont  [1]{#1}%
\providecommand \bibfnamefont [1]{#1}%
\providecommand \citenamefont [1]{#1}%
\providecommand \href@noop [0]{\@secondoftwo}%
\providecommand \href [0]{\begingroup \@sanitize@url \@href}%
\providecommand \@href[1]{\@@startlink{#1}\@@href}%
\providecommand \@@href[1]{\endgroup#1\@@endlink}%
\providecommand \@sanitize@url [0]{\catcode `\\12\catcode `\$12\catcode
  `\&12\catcode `\#12\catcode `\^12\catcode `\_12\catcode `\%12\relax}%
\providecommand \@@startlink[1]{}%
\providecommand \@@endlink[0]{}%
\providecommand \url  [0]{\begingroup\@sanitize@url \@url }%
\providecommand \@url [1]{\endgroup\@href {#1}{\urlprefix }}%
\providecommand \urlprefix  [0]{URL }%
\providecommand \Eprint [0]{\href }%
\providecommand \doibase [0]{http://dx.doi.org/}%
\providecommand \selectlanguage [0]{\@gobble}%
\providecommand \bibinfo  [0]{\@secondoftwo}%
\providecommand \bibfield  [0]{\@secondoftwo}%
\providecommand \translation [1]{[#1]}%
\providecommand \BibitemOpen [0]{}%
\providecommand \bibitemStop [0]{}%
\providecommand \bibitemNoStop [0]{.\EOS\space}%
\providecommand \EOS [0]{\spacefactor3000\relax}%
\providecommand \BibitemShut  [1]{\csname bibitem#1\endcsname}%
\let\auto@bib@innerbib\@empty
\bibitem [{\citenamefont {Tanihata}\ \emph {et~al.}(2013)\citenamefont
  {Tanihata}, \citenamefont {Savajols},\ and\ \citenamefont
  {Kanungo}}]{tanihata13}%
  \BibitemOpen
  \bibfield  {author} {\bibinfo {author} {\bibfnamefont {I.}~\bibnamefont
  {Tanihata}}, \bibinfo {author} {\bibfnamefont {H.}~\bibnamefont {Savajols}},
  \ and\ \bibinfo {author} {\bibfnamefont {R.}~\bibnamefont {Kanungo}},\ }\href
  {\doibase https://doi.org/10.1016/j.ppnp.2012.07.001} {\bibfield  {journal}
  {\bibinfo  {journal} {Prog. Part. Nucl. Phys.}\ }\textbf {\bibinfo {volume}
  {68}},\ \bibinfo {pages} {215 } (\bibinfo {year} {2013})}\BibitemShut
  {NoStop}%
\bibitem [{\citenamefont {Tanihata}\ \emph {et~al.}(1985)\citenamefont
  {Tanihata}, \citenamefont {Hamagaki}, \citenamefont {Hashimoto},
  \citenamefont {Shida}, \citenamefont {Yoshikawa}, \citenamefont {Sugimoto},
  \citenamefont {Yamakawa}, \citenamefont {Kobayashi},\ and\ \citenamefont
  {Takahashi}}]{tanihata85}%
  \BibitemOpen
  \bibfield  {author} {\bibinfo {author} {\bibfnamefont {I.}~\bibnamefont
  {Tanihata}}, \bibinfo {author} {\bibfnamefont {H.}~\bibnamefont {Hamagaki}},
  \bibinfo {author} {\bibfnamefont {O.}~\bibnamefont {Hashimoto}}, \bibinfo
  {author} {\bibfnamefont {Y.}~\bibnamefont {Shida}}, \bibinfo {author}
  {\bibfnamefont {N.}~\bibnamefont {Yoshikawa}}, \bibinfo {author}
  {\bibfnamefont {K.}~\bibnamefont {Sugimoto}}, \bibinfo {author}
  {\bibfnamefont {O.}~\bibnamefont {Yamakawa}}, \bibinfo {author}
  {\bibfnamefont {T.}~\bibnamefont {Kobayashi}}, \ and\ \bibinfo {author}
  {\bibfnamefont {N.}~\bibnamefont {Takahashi}},\ }\href {\doibase
  10.1103/PhysRevLett.55.2676} {\bibfield  {journal} {\bibinfo  {journal}
  {Phys. Rev. Lett.}\ }\textbf {\bibinfo {volume} {55}},\ \bibinfo {pages}
  {2676} (\bibinfo {year} {1985})}\BibitemShut {NoStop}%
\bibitem [{\citenamefont {Hansen}\ and\ \citenamefont
  {Jonson}(1987)}]{Hansen87}%
  \BibitemOpen
  \bibfield  {author} {\bibinfo {author} {\bibfnamefont {P.~G.}\ \bibnamefont
  {Hansen}}\ and\ \bibinfo {author} {\bibfnamefont {B.}~\bibnamefont
  {Jonson}},\ }\href {\doibase 10.1209/0295-5075/4/4/005} {\bibfield  {journal}
  {\bibinfo  {journal} {Europhys. Lett.}\ }\textbf {\bibinfo {volume} {4}},\
  \bibinfo {pages} {409} (\bibinfo {year} {1987})}\BibitemShut {NoStop}%
\bibitem [{\citenamefont {Zhukov}\ \emph {et~al.}(1993)\citenamefont {Zhukov},
  \citenamefont {Danilin}, \citenamefont {Fedorov}, \citenamefont {Bang},
  \citenamefont {Thompson},\ and\ \citenamefont {Vaagen}}]{Zhukov93}%
  \BibitemOpen
  \bibfield  {author} {\bibinfo {author} {\bibfnamefont {M.}~\bibnamefont
  {Zhukov}}, \bibinfo {author} {\bibfnamefont {B.}~\bibnamefont {Danilin}},
  \bibinfo {author} {\bibfnamefont {D.}~\bibnamefont {Fedorov}}, \bibinfo
  {author} {\bibfnamefont {J.}~\bibnamefont {Bang}}, \bibinfo {author}
  {\bibfnamefont {I.}~\bibnamefont {Thompson}}, \ and\ \bibinfo {author}
  {\bibfnamefont {J.}~\bibnamefont {Vaagen}},\ }\href {\doibase
  https://doi.org/10.1016/0370-1573(93)90141-Y} {\bibfield  {journal} {\bibinfo
   {journal} {Phys. Rep.}\ }\textbf {\bibinfo {volume} {231}},\ \bibinfo
  {pages} {151 } (\bibinfo {year} {1993})}\BibitemShut {NoStop}%
\bibitem [{\citenamefont {Kikuchi}\ \emph {et~al.}(2016)\citenamefont
  {Kikuchi}, \citenamefont {Ogata}, \citenamefont {Kubota}, \citenamefont
  {Sasano},\ and\ \citenamefont {Uesaka}}]{kikuchi16}%
  \BibitemOpen
  \bibfield  {author} {\bibinfo {author} {\bibfnamefont {Y.}~\bibnamefont
  {Kikuchi}}, \bibinfo {author} {\bibfnamefont {K.}~\bibnamefont {Ogata}},
  \bibinfo {author} {\bibfnamefont {Y.}~\bibnamefont {Kubota}}, \bibinfo
  {author} {\bibfnamefont {M.}~\bibnamefont {Sasano}}, \ and\ \bibinfo {author}
  {\bibfnamefont {T.}~\bibnamefont {Uesaka}},\ }\href {\doibase
  10.1093/ptep/ptw148} {\bibfield  {journal} {\bibinfo  {journal} {Prog. Theor.
  Exp. Phys.}\ }\textbf {\bibinfo {volume} {2016}},\ \bibinfo {pages} {103D03}
  (\bibinfo {year} {2016})}\BibitemShut {NoStop}%
\bibitem [{\citenamefont {Hagino}\ and\ \citenamefont
  {Sagawa}(2005)}]{hagino05}%
  \BibitemOpen
  \bibfield  {author} {\bibinfo {author} {\bibfnamefont {K.}~\bibnamefont
  {Hagino}}\ and\ \bibinfo {author} {\bibfnamefont {H.}~\bibnamefont
  {Sagawa}},\ }\href {\doibase 10.1103/PhysRevC.72.044321} {\bibfield
  {journal} {\bibinfo  {journal} {Phys. Rev. C}\ }\textbf {\bibinfo {volume}
  {72}},\ \bibinfo {pages} {044321} (\bibinfo {year} {2005})}\BibitemShut
  {NoStop}%
\bibitem [{\citenamefont {Tanihata}\ \emph {et~al.}(1988)\citenamefont
  {Tanihata}, \citenamefont {Kobayashi}, \citenamefont {Yamakawa},
  \citenamefont {Shimoura}, \citenamefont {Ekuni}, \citenamefont {Sugimoto},
  \citenamefont {Takahashi}, \citenamefont {Shimoda},\ and\ \citenamefont
  {Sato}}]{tanihata88}%
  \BibitemOpen
  \bibfield  {author} {\bibinfo {author} {\bibfnamefont {I.}~\bibnamefont
  {Tanihata}}, \bibinfo {author} {\bibfnamefont {T.}~\bibnamefont {Kobayashi}},
  \bibinfo {author} {\bibfnamefont {O.}~\bibnamefont {Yamakawa}}, \bibinfo
  {author} {\bibfnamefont {S.}~\bibnamefont {Shimoura}}, \bibinfo {author}
  {\bibfnamefont {K.}~\bibnamefont {Ekuni}}, \bibinfo {author} {\bibfnamefont
  {K.}~\bibnamefont {Sugimoto}}, \bibinfo {author} {\bibfnamefont
  {N.}~\bibnamefont {Takahashi}}, \bibinfo {author} {\bibfnamefont
  {T.}~\bibnamefont {Shimoda}}, \ and\ \bibinfo {author} {\bibfnamefont
  {H.}~\bibnamefont {Sato}},\ }\href {\doibase
  https://doi.org/10.1016/0370-2693(88)90702-2} {\bibfield  {journal} {\bibinfo
   {journal} {Phys. Lett. B}\ }\textbf {\bibinfo {volume} {206}},\ \bibinfo
  {pages} {592 } (\bibinfo {year} {1988})}\BibitemShut {NoStop}%
\bibitem [{\citenamefont {{Tanaka \textsl{et al.}}}(2010)}]{tanaka10}%
  \BibitemOpen
  \bibfield  {author} {\bibinfo {author} {\bibfnamefont {K.}~\bibnamefont
  {{Tanaka \textsl{et al.}}}},\ }\href {\doibase 10.1103/PhysRevC.82.044309}
  {\bibfield  {journal} {\bibinfo  {journal} {Phys. Rev. C}\ }\textbf {\bibinfo
  {volume} {82}},\ \bibinfo {pages} {044309} (\bibinfo {year}
  {2010})}\BibitemShut {NoStop}%
\bibitem [{\citenamefont {Togano}\ \emph {et~al.}(2016)\citenamefont {Togano},
  \citenamefont {Nakamura}, \citenamefont {Kondo}, \citenamefont {Tostevin},
  \citenamefont {Saito}, \citenamefont {Gibelin}, \citenamefont {Orr},
  \citenamefont {Achouri}, \citenamefont {Aumann}, \citenamefont {Baba},
  \citenamefont {Delaunay}, \citenamefont {Doornenbal}, \citenamefont {Fukuda},
  \citenamefont {Hwang}, \citenamefont {Inabe}, \citenamefont {Isobe},
  \citenamefont {Kameda}, \citenamefont {Kanno}, \citenamefont {Kim},
  \citenamefont {Kobayashi}, \citenamefont {Kobayashi}, \citenamefont {Kubo},
  \citenamefont {Leblond}, \citenamefont {Lee}, \citenamefont {Marqués},
  \citenamefont {Minakata}, \citenamefont {Motobayashi}, \citenamefont {Murai},
  \citenamefont {Murakami}, \citenamefont {Muto}, \citenamefont {Nakashima},
  \citenamefont {Nakatsuka}, \citenamefont {Navin}, \citenamefont {Nishi},
  \citenamefont {Ogoshi}, \citenamefont {Otsu}, \citenamefont {Sato},
  \citenamefont {Satou}, \citenamefont {Shimizu}, \citenamefont {Suzuki},
  \citenamefont {Takahashi}, \citenamefont {Takeda}, \citenamefont {Takeuchi},
  \citenamefont {Tanaka}, \citenamefont {Tuff}, \citenamefont {Vandebrouck},\
  and\ \citenamefont {Yoneda}}]{togano16}%
  \BibitemOpen
  \bibfield  {author} {\bibinfo {author} {\bibfnamefont {Y.}~\bibnamefont
  {Togano}}, \bibinfo {author} {\bibfnamefont {T.}~\bibnamefont {Nakamura}},
  \bibinfo {author} {\bibfnamefont {Y.}~\bibnamefont {Kondo}}, \bibinfo
  {author} {\bibfnamefont {J.}~\bibnamefont {Tostevin}}, \bibinfo {author}
  {\bibfnamefont {A.}~\bibnamefont {Saito}}, \bibinfo {author} {\bibfnamefont
  {J.}~\bibnamefont {Gibelin}}, \bibinfo {author} {\bibfnamefont
  {N.}~\bibnamefont {Orr}}, \bibinfo {author} {\bibfnamefont {N.}~\bibnamefont
  {Achouri}}, \bibinfo {author} {\bibfnamefont {T.}~\bibnamefont {Aumann}},
  \bibinfo {author} {\bibfnamefont {H.}~\bibnamefont {Baba}}, \bibinfo {author}
  {\bibfnamefont {F.}~\bibnamefont {Delaunay}}, \bibinfo {author}
  {\bibfnamefont {P.}~\bibnamefont {Doornenbal}}, \bibinfo {author}
  {\bibfnamefont {N.}~\bibnamefont {Fukuda}}, \bibinfo {author} {\bibfnamefont
  {J.}~\bibnamefont {Hwang}}, \bibinfo {author} {\bibfnamefont
  {N.}~\bibnamefont {Inabe}}, \bibinfo {author} {\bibfnamefont
  {T.}~\bibnamefont {Isobe}}, \bibinfo {author} {\bibfnamefont
  {D.}~\bibnamefont {Kameda}}, \bibinfo {author} {\bibfnamefont
  {D.}~\bibnamefont {Kanno}}, \bibinfo {author} {\bibfnamefont
  {S.}~\bibnamefont {Kim}}, \bibinfo {author} {\bibfnamefont {N.}~\bibnamefont
  {Kobayashi}}, \bibinfo {author} {\bibfnamefont {T.}~\bibnamefont
  {Kobayashi}}, \bibinfo {author} {\bibfnamefont {T.}~\bibnamefont {Kubo}},
  \bibinfo {author} {\bibfnamefont {S.}~\bibnamefont {Leblond}}, \bibinfo
  {author} {\bibfnamefont {J.}~\bibnamefont {Lee}}, \bibinfo {author}
  {\bibfnamefont {F.}~\bibnamefont {Marqués}}, \bibinfo {author}
  {\bibfnamefont {R.}~\bibnamefont {Minakata}}, \bibinfo {author}
  {\bibfnamefont {T.}~\bibnamefont {Motobayashi}}, \bibinfo {author}
  {\bibfnamefont {D.}~\bibnamefont {Murai}}, \bibinfo {author} {\bibfnamefont
  {T.}~\bibnamefont {Murakami}}, \bibinfo {author} {\bibfnamefont
  {K.}~\bibnamefont {Muto}}, \bibinfo {author} {\bibfnamefont {T.}~\bibnamefont
  {Nakashima}}, \bibinfo {author} {\bibfnamefont {N.}~\bibnamefont
  {Nakatsuka}}, \bibinfo {author} {\bibfnamefont {A.}~\bibnamefont {Navin}},
  \bibinfo {author} {\bibfnamefont {S.}~\bibnamefont {Nishi}}, \bibinfo
  {author} {\bibfnamefont {S.}~\bibnamefont {Ogoshi}}, \bibinfo {author}
  {\bibfnamefont {H.}~\bibnamefont {Otsu}}, \bibinfo {author} {\bibfnamefont
  {H.}~\bibnamefont {Sato}}, \bibinfo {author} {\bibfnamefont {Y.}~\bibnamefont
  {Satou}}, \bibinfo {author} {\bibfnamefont {Y.}~\bibnamefont {Shimizu}},
  \bibinfo {author} {\bibfnamefont {H.}~\bibnamefont {Suzuki}}, \bibinfo
  {author} {\bibfnamefont {K.}~\bibnamefont {Takahashi}}, \bibinfo {author}
  {\bibfnamefont {H.}~\bibnamefont {Takeda}}, \bibinfo {author} {\bibfnamefont
  {S.}~\bibnamefont {Takeuchi}}, \bibinfo {author} {\bibfnamefont
  {R.}~\bibnamefont {Tanaka}}, \bibinfo {author} {\bibfnamefont
  {A.}~\bibnamefont {Tuff}}, \bibinfo {author} {\bibfnamefont {M.}~\bibnamefont
  {Vandebrouck}}, \ and\ \bibinfo {author} {\bibfnamefont {K.}~\bibnamefont
  {Yoneda}},\ }\href {\doibase https://doi.org/10.1016/j.physletb.2016.08.062}
  {\bibfield  {journal} {\bibinfo  {journal} {Phys. Lett. B}\ }\textbf
  {\bibinfo {volume} {761}},\ \bibinfo {pages} {412 } (\bibinfo {year}
  {2016})}\BibitemShut {NoStop}%
\bibitem [{\citenamefont {Aksyutina}\ \emph {et~al.}(2008)\citenamefont
  {Aksyutina}, \citenamefont {Johansson}, \citenamefont {Adrich}, \citenamefont
  {Aksouh}, \citenamefont {Aumann}, \citenamefont {Boretzky}, \citenamefont
  {Borge}, \citenamefont {Chatillon}, \citenamefont {Chulkov}, \citenamefont
  {Cortina-Gil}, \citenamefont {Pramanik}, \citenamefont {Emling},
  \citenamefont {Forssén}, \citenamefont {Fynbo}, \citenamefont {Geissel},
  \citenamefont {Hellström}, \citenamefont {Ickert}, \citenamefont {Jones},
  \citenamefont {Jonson}, \citenamefont {Kliemkiewicz}, \citenamefont {Kratz},
  \citenamefont {Kulessa}, \citenamefont {Lantz}, \citenamefont {LeBleis},
  \citenamefont {Lindahl}, \citenamefont {Mahata}, \citenamefont {Matos},
  \citenamefont {Meister}, \citenamefont {Münzenberg}, \citenamefont
  {Nilsson}, \citenamefont {Nyman}, \citenamefont {Palit}, \citenamefont
  {Pantea}, \citenamefont {Paschalis}, \citenamefont {Prokopowicz},
  \citenamefont {Reifarth}, \citenamefont {Richter}, \citenamefont {Riisager},
  \citenamefont {Schrieder}, \citenamefont {Simon}, \citenamefont {Sümmerer},
  \citenamefont {Tengblad}, \citenamefont {Walus}, \citenamefont {Weick},\ and\
  \citenamefont {Zhukov}}]{Aksyutina13}%
  \BibitemOpen
  \bibfield  {author} {\bibinfo {author} {\bibfnamefont {Y.}~\bibnamefont
  {Aksyutina}}, \bibinfo {author} {\bibfnamefont {H.}~\bibnamefont
  {Johansson}}, \bibinfo {author} {\bibfnamefont {P.}~\bibnamefont {Adrich}},
  \bibinfo {author} {\bibfnamefont {F.}~\bibnamefont {Aksouh}}, \bibinfo
  {author} {\bibfnamefont {T.}~\bibnamefont {Aumann}}, \bibinfo {author}
  {\bibfnamefont {K.}~\bibnamefont {Boretzky}}, \bibinfo {author}
  {\bibfnamefont {M.}~\bibnamefont {Borge}}, \bibinfo {author} {\bibfnamefont
  {A.}~\bibnamefont {Chatillon}}, \bibinfo {author} {\bibfnamefont
  {L.}~\bibnamefont {Chulkov}}, \bibinfo {author} {\bibfnamefont
  {D.}~\bibnamefont {Cortina-Gil}}, \bibinfo {author} {\bibfnamefont {U.~D.}\
  \bibnamefont {Pramanik}}, \bibinfo {author} {\bibfnamefont {H.}~\bibnamefont
  {Emling}}, \bibinfo {author} {\bibfnamefont {C.}~\bibnamefont {Forssén}},
  \bibinfo {author} {\bibfnamefont {H.}~\bibnamefont {Fynbo}}, \bibinfo
  {author} {\bibfnamefont {H.}~\bibnamefont {Geissel}}, \bibinfo {author}
  {\bibfnamefont {M.}~\bibnamefont {Hellström}}, \bibinfo {author}
  {\bibfnamefont {G.}~\bibnamefont {Ickert}}, \bibinfo {author} {\bibfnamefont
  {K.}~\bibnamefont {Jones}}, \bibinfo {author} {\bibfnamefont
  {B.}~\bibnamefont {Jonson}}, \bibinfo {author} {\bibfnamefont
  {A.}~\bibnamefont {Kliemkiewicz}}, \bibinfo {author} {\bibfnamefont
  {J.}~\bibnamefont {Kratz}}, \bibinfo {author} {\bibfnamefont
  {R.}~\bibnamefont {Kulessa}}, \bibinfo {author} {\bibfnamefont
  {M.}~\bibnamefont {Lantz}}, \bibinfo {author} {\bibfnamefont
  {T.}~\bibnamefont {LeBleis}}, \bibinfo {author} {\bibfnamefont
  {A.}~\bibnamefont {Lindahl}}, \bibinfo {author} {\bibfnamefont
  {K.}~\bibnamefont {Mahata}}, \bibinfo {author} {\bibfnamefont
  {M.}~\bibnamefont {Matos}}, \bibinfo {author} {\bibfnamefont
  {M.}~\bibnamefont {Meister}}, \bibinfo {author} {\bibfnamefont
  {G.}~\bibnamefont {Münzenberg}}, \bibinfo {author} {\bibfnamefont
  {T.}~\bibnamefont {Nilsson}}, \bibinfo {author} {\bibfnamefont
  {G.}~\bibnamefont {Nyman}}, \bibinfo {author} {\bibfnamefont
  {R.}~\bibnamefont {Palit}}, \bibinfo {author} {\bibfnamefont
  {M.}~\bibnamefont {Pantea}}, \bibinfo {author} {\bibfnamefont
  {S.}~\bibnamefont {Paschalis}}, \bibinfo {author} {\bibfnamefont
  {W.}~\bibnamefont {Prokopowicz}}, \bibinfo {author} {\bibfnamefont
  {R.}~\bibnamefont {Reifarth}}, \bibinfo {author} {\bibfnamefont
  {A.}~\bibnamefont {Richter}}, \bibinfo {author} {\bibfnamefont
  {K.}~\bibnamefont {Riisager}}, \bibinfo {author} {\bibfnamefont
  {G.}~\bibnamefont {Schrieder}}, \bibinfo {author} {\bibfnamefont
  {H.}~\bibnamefont {Simon}}, \bibinfo {author} {\bibfnamefont
  {K.}~\bibnamefont {Sümmerer}}, \bibinfo {author} {\bibfnamefont
  {O.}~\bibnamefont {Tengblad}}, \bibinfo {author} {\bibfnamefont
  {W.}~\bibnamefont {Walus}}, \bibinfo {author} {\bibfnamefont
  {H.}~\bibnamefont {Weick}}, \ and\ \bibinfo {author} {\bibfnamefont
  {M.}~\bibnamefont {Zhukov}},\ }\href {\doibase
  https://doi.org/10.1016/j.physletb.2008.07.093} {\bibfield  {journal}
  {\bibinfo  {journal} {Phys. Lett. B}\ }\textbf {\bibinfo {volume} {666}},\
  \bibinfo {pages} {430 } (\bibinfo {year} {2008})}\BibitemShut {NoStop}%
\bibitem [{\citenamefont {Casal}\ \emph {et~al.}(2017)\citenamefont {Casal},
  \citenamefont {Gómez-Ramos},\ and\ \citenamefont {Moro}}]{casalplb17}%
  \BibitemOpen
  \bibfield  {author} {\bibinfo {author} {\bibfnamefont {J.}~\bibnamefont
  {Casal}}, \bibinfo {author} {\bibfnamefont {M.}~\bibnamefont {Gómez-Ramos}},
  \ and\ \bibinfo {author} {\bibfnamefont {A.}~\bibnamefont {Moro}},\ }\href
  {\doibase https://doi.org/10.1016/j.physletb.2017.02.017} {\bibfield
  {journal} {\bibinfo  {journal} {Phys. Lett. B}\ }\textbf {\bibinfo {volume}
  {767}},\ \bibinfo {pages} {307 } (\bibinfo {year} {2017})}\BibitemShut
  {NoStop}%
\bibitem [{\citenamefont {Aumann}(2019)}]{aumann19}%
  \BibitemOpen
  \bibfield  {author} {\bibinfo {author} {\bibfnamefont {T.}~\bibnamefont
  {Aumann}},\ }\href {\doibase 10.1140/epja/i2019-12862-7} {\bibfield
  {journal} {\bibinfo  {journal} {Eur. Phys. J. A}\ }\textbf {\bibinfo {volume}
  {55}},\ \bibinfo {pages} {234} (\bibinfo {year} {2019})}\BibitemShut
  {NoStop}%
\bibitem [{\citenamefont {Aumann}\ \emph {et~al.}(1999)\citenamefont {Aumann},
  \citenamefont {Aleksandrov}, \citenamefont {Axelsson}, \citenamefont
  {Baumann}, \citenamefont {Borge}, \citenamefont {Chulkov}, \citenamefont
  {Cub}, \citenamefont {Dostal}, \citenamefont {Eberlein}, \citenamefont
  {Elze}, \citenamefont {Emling}, \citenamefont {Geissel}, \citenamefont
  {Goldberg}, \citenamefont {Golovkov}, \citenamefont {Gr\"unschlo\ss{}},
  \citenamefont {Hellstr\"om}, \citenamefont {Hencken}, \citenamefont
  {Holeczek}, \citenamefont {Holzmann}, \citenamefont {Jonson}, \citenamefont
  {Korshenninikov}, \citenamefont {Kratz}, \citenamefont {Kraus}, \citenamefont
  {Kulessa}, \citenamefont {Leifels}, \citenamefont {Leistenschneider},
  \citenamefont {Leth}, \citenamefont {Mukha}, \citenamefont {M\"unzenberg},
  \citenamefont {Nickel}, \citenamefont {Nilsson}, \citenamefont {Nyman},
  \citenamefont {Petersen}, \citenamefont {Pf\"utzner}, \citenamefont
  {Richter}, \citenamefont {Riisager}, \citenamefont {Scheidenberger},
  \citenamefont {Schrieder}, \citenamefont {Schwab}, \citenamefont {Simon},
  \citenamefont {Smedberg}, \citenamefont {Steiner}, \citenamefont {Stroth},
  \citenamefont {Surowiec}, \citenamefont {Suzuki}, \citenamefont {Tengblad},\
  and\ \citenamefont {Zhukov}}]{aumann99}%
  \BibitemOpen
  \bibfield  {author} {\bibinfo {author} {\bibfnamefont {T.}~\bibnamefont
  {Aumann}}, \bibinfo {author} {\bibfnamefont {D.}~\bibnamefont {Aleksandrov}},
  \bibinfo {author} {\bibfnamefont {L.}~\bibnamefont {Axelsson}}, \bibinfo
  {author} {\bibfnamefont {T.}~\bibnamefont {Baumann}}, \bibinfo {author}
  {\bibfnamefont {M.~J.~G.}\ \bibnamefont {Borge}}, \bibinfo {author}
  {\bibfnamefont {L.~V.}\ \bibnamefont {Chulkov}}, \bibinfo {author}
  {\bibfnamefont {J.}~\bibnamefont {Cub}}, \bibinfo {author} {\bibfnamefont
  {W.}~\bibnamefont {Dostal}}, \bibinfo {author} {\bibfnamefont
  {B.}~\bibnamefont {Eberlein}}, \bibinfo {author} {\bibfnamefont {T.~W.}\
  \bibnamefont {Elze}}, \bibinfo {author} {\bibfnamefont {H.}~\bibnamefont
  {Emling}}, \bibinfo {author} {\bibfnamefont {H.}~\bibnamefont {Geissel}},
  \bibinfo {author} {\bibfnamefont {V.~Z.}\ \bibnamefont {Goldberg}}, \bibinfo
  {author} {\bibfnamefont {M.}~\bibnamefont {Golovkov}}, \bibinfo {author}
  {\bibfnamefont {A.}~\bibnamefont {Gr\"unschlo\ss{}}}, \bibinfo {author}
  {\bibfnamefont {M.}~\bibnamefont {Hellstr\"om}}, \bibinfo {author}
  {\bibfnamefont {K.}~\bibnamefont {Hencken}}, \bibinfo {author} {\bibfnamefont
  {J.}~\bibnamefont {Holeczek}}, \bibinfo {author} {\bibfnamefont
  {R.}~\bibnamefont {Holzmann}}, \bibinfo {author} {\bibfnamefont
  {B.}~\bibnamefont {Jonson}}, \bibinfo {author} {\bibfnamefont {A.~A.}\
  \bibnamefont {Korshenninikov}}, \bibinfo {author} {\bibfnamefont {J.~V.}\
  \bibnamefont {Kratz}}, \bibinfo {author} {\bibfnamefont {G.}~\bibnamefont
  {Kraus}}, \bibinfo {author} {\bibfnamefont {R.}~\bibnamefont {Kulessa}},
  \bibinfo {author} {\bibfnamefont {Y.}~\bibnamefont {Leifels}}, \bibinfo
  {author} {\bibfnamefont {A.}~\bibnamefont {Leistenschneider}}, \bibinfo
  {author} {\bibfnamefont {T.}~\bibnamefont {Leth}}, \bibinfo {author}
  {\bibfnamefont {I.}~\bibnamefont {Mukha}}, \bibinfo {author} {\bibfnamefont
  {G.}~\bibnamefont {M\"unzenberg}}, \bibinfo {author} {\bibfnamefont
  {F.}~\bibnamefont {Nickel}}, \bibinfo {author} {\bibfnamefont
  {T.}~\bibnamefont {Nilsson}}, \bibinfo {author} {\bibfnamefont
  {G.}~\bibnamefont {Nyman}}, \bibinfo {author} {\bibfnamefont
  {B.}~\bibnamefont {Petersen}}, \bibinfo {author} {\bibfnamefont
  {M.}~\bibnamefont {Pf\"utzner}}, \bibinfo {author} {\bibfnamefont
  {A.}~\bibnamefont {Richter}}, \bibinfo {author} {\bibfnamefont
  {K.}~\bibnamefont {Riisager}}, \bibinfo {author} {\bibfnamefont
  {C.}~\bibnamefont {Scheidenberger}}, \bibinfo {author} {\bibfnamefont
  {G.}~\bibnamefont {Schrieder}}, \bibinfo {author} {\bibfnamefont
  {W.}~\bibnamefont {Schwab}}, \bibinfo {author} {\bibfnamefont
  {H.}~\bibnamefont {Simon}}, \bibinfo {author} {\bibfnamefont {M.~H.}\
  \bibnamefont {Smedberg}}, \bibinfo {author} {\bibfnamefont {M.}~\bibnamefont
  {Steiner}}, \bibinfo {author} {\bibfnamefont {J.}~\bibnamefont {Stroth}},
  \bibinfo {author} {\bibfnamefont {A.}~\bibnamefont {Surowiec}}, \bibinfo
  {author} {\bibfnamefont {T.}~\bibnamefont {Suzuki}}, \bibinfo {author}
  {\bibfnamefont {O.}~\bibnamefont {Tengblad}}, \ and\ \bibinfo {author}
  {\bibfnamefont {M.~V.}\ \bibnamefont {Zhukov}},\ }\href {\doibase
  10.1103/PhysRevC.59.1252} {\bibfield  {journal} {\bibinfo  {journal} {Phys.
  Rev. C}\ }\textbf {\bibinfo {volume} {59}},\ \bibinfo {pages} {1252}
  (\bibinfo {year} {1999})}\BibitemShut {NoStop}%
\bibitem [{\citenamefont {Labiche}\ \emph {et~al.}(2001)\citenamefont
  {Labiche}, \citenamefont {Orr}, \citenamefont {Marqu\'es}, \citenamefont
  {Ang\'elique}, \citenamefont {Axelsson}, \citenamefont {Benoit},
  \citenamefont {Bergmann}, \citenamefont {Borge}, \citenamefont {Catford},
  \citenamefont {Chappell}, \citenamefont {Clarke}, \citenamefont {Costa},
  \citenamefont {Curtis}, \citenamefont {D'Arrigo}, \citenamefont
  {de~G\'oes~Brennand}, \citenamefont {Dorvaux}, \citenamefont {Fazio},
  \citenamefont {Freer}, \citenamefont {Fulton}, \citenamefont {Giardina},
  \citenamefont {Gr\'evy}, \citenamefont {Guillemaud-Mueller}, \citenamefont
  {Hanappe}, \citenamefont {Heusch}, \citenamefont {Jones}, \citenamefont
  {Jonson}, \citenamefont {Le~Brun}, \citenamefont {Leenhardt}, \citenamefont
  {Lewitowicz}, \citenamefont {Lopez}, \citenamefont {Markenroth},
  \citenamefont {Mueller}, \citenamefont {Nilsson}, \citenamefont {Ninane},
  \citenamefont {Nyman}, \citenamefont {de~Oliveira}, \citenamefont {Piqueras},
  \citenamefont {Riisager}, \citenamefont {Saint~Laurent}, \citenamefont
  {Sarazin}, \citenamefont {Singer}, \citenamefont {Sorlin},\ and\
  \citenamefont {Stuttg\'e}}]{labiche01}%
  \BibitemOpen
  \bibfield  {author} {\bibinfo {author} {\bibfnamefont {M.}~\bibnamefont
  {Labiche}}, \bibinfo {author} {\bibfnamefont {N.~A.}\ \bibnamefont {Orr}},
  \bibinfo {author} {\bibfnamefont {F.~M.}\ \bibnamefont {Marqu\'es}}, \bibinfo
  {author} {\bibfnamefont {J.~C.}\ \bibnamefont {Ang\'elique}}, \bibinfo
  {author} {\bibfnamefont {L.}~\bibnamefont {Axelsson}}, \bibinfo {author}
  {\bibfnamefont {B.}~\bibnamefont {Benoit}}, \bibinfo {author} {\bibfnamefont
  {U.~C.}\ \bibnamefont {Bergmann}}, \bibinfo {author} {\bibfnamefont
  {M.~J.~G.}\ \bibnamefont {Borge}}, \bibinfo {author} {\bibfnamefont {W.~N.}\
  \bibnamefont {Catford}}, \bibinfo {author} {\bibfnamefont {S.~P.~G.}\
  \bibnamefont {Chappell}}, \bibinfo {author} {\bibfnamefont {N.~M.}\
  \bibnamefont {Clarke}}, \bibinfo {author} {\bibfnamefont {G.}~\bibnamefont
  {Costa}}, \bibinfo {author} {\bibfnamefont {N.}~\bibnamefont {Curtis}},
  \bibinfo {author} {\bibfnamefont {A.}~\bibnamefont {D'Arrigo}}, \bibinfo
  {author} {\bibfnamefont {E.}~\bibnamefont {de~G\'oes~Brennand}}, \bibinfo
  {author} {\bibfnamefont {O.}~\bibnamefont {Dorvaux}}, \bibinfo {author}
  {\bibfnamefont {G.}~\bibnamefont {Fazio}}, \bibinfo {author} {\bibfnamefont
  {M.}~\bibnamefont {Freer}}, \bibinfo {author} {\bibfnamefont {B.~R.}\
  \bibnamefont {Fulton}}, \bibinfo {author} {\bibfnamefont {G.}~\bibnamefont
  {Giardina}}, \bibinfo {author} {\bibfnamefont {S.}~\bibnamefont {Gr\'evy}},
  \bibinfo {author} {\bibfnamefont {D.}~\bibnamefont {Guillemaud-Mueller}},
  \bibinfo {author} {\bibfnamefont {F.}~\bibnamefont {Hanappe}}, \bibinfo
  {author} {\bibfnamefont {B.}~\bibnamefont {Heusch}}, \bibinfo {author}
  {\bibfnamefont {K.~L.}\ \bibnamefont {Jones}}, \bibinfo {author}
  {\bibfnamefont {B.}~\bibnamefont {Jonson}}, \bibinfo {author} {\bibfnamefont
  {C.}~\bibnamefont {Le~Brun}}, \bibinfo {author} {\bibfnamefont
  {S.}~\bibnamefont {Leenhardt}}, \bibinfo {author} {\bibfnamefont
  {M.}~\bibnamefont {Lewitowicz}}, \bibinfo {author} {\bibfnamefont {M.~J.}\
  \bibnamefont {Lopez}}, \bibinfo {author} {\bibfnamefont {K.}~\bibnamefont
  {Markenroth}}, \bibinfo {author} {\bibfnamefont {A.~C.}\ \bibnamefont
  {Mueller}}, \bibinfo {author} {\bibfnamefont {T.}~\bibnamefont {Nilsson}},
  \bibinfo {author} {\bibfnamefont {A.}~\bibnamefont {Ninane}}, \bibinfo
  {author} {\bibfnamefont {G.}~\bibnamefont {Nyman}}, \bibinfo {author}
  {\bibfnamefont {F.}~\bibnamefont {de~Oliveira}}, \bibinfo {author}
  {\bibfnamefont {I.}~\bibnamefont {Piqueras}}, \bibinfo {author}
  {\bibfnamefont {K.}~\bibnamefont {Riisager}}, \bibinfo {author}
  {\bibfnamefont {M.~G.}\ \bibnamefont {Saint~Laurent}}, \bibinfo {author}
  {\bibfnamefont {F.}~\bibnamefont {Sarazin}}, \bibinfo {author} {\bibfnamefont
  {S.~M.}\ \bibnamefont {Singer}}, \bibinfo {author} {\bibfnamefont
  {O.}~\bibnamefont {Sorlin}}, \ and\ \bibinfo {author} {\bibfnamefont
  {L.}~\bibnamefont {Stuttg\'e}},\ }\href {\doibase 10.1103/PhysRevLett.86.600}
  {\bibfield  {journal} {\bibinfo  {journal} {Phys. Rev. Lett.}\ }\textbf
  {\bibinfo {volume} {86}},\ \bibinfo {pages} {600} (\bibinfo {year}
  {2001})}\BibitemShut {NoStop}%
\bibitem [{\citenamefont {Nakamura}\ \emph {et~al.}(2006)\citenamefont
  {Nakamura}, \citenamefont {Vinodkumar}, \citenamefont {Sugimoto},
  \citenamefont {Aoi}, \citenamefont {Baba}, \citenamefont {Bazin},
  \citenamefont {Fukuda}, \citenamefont {Gomi}, \citenamefont {Hasegawa},
  \citenamefont {Imai}, \citenamefont {Ishihara}, \citenamefont {Kobayashi},
  \citenamefont {Kondo}, \citenamefont {Kubo}, \citenamefont {Miura},
  \citenamefont {Motobayashi}, \citenamefont {Otsu}, \citenamefont {Saito},
  \citenamefont {Sakurai}, \citenamefont {Shimoura}, \citenamefont {Watanabe},
  \citenamefont {Watanabe}, \citenamefont {Yakushiji}, \citenamefont
  {Yanagisawa},\ and\ \citenamefont {Yoneda}}]{nakamura06}%
  \BibitemOpen
  \bibfield  {author} {\bibinfo {author} {\bibfnamefont {T.}~\bibnamefont
  {Nakamura}}, \bibinfo {author} {\bibfnamefont {A.~M.}\ \bibnamefont
  {Vinodkumar}}, \bibinfo {author} {\bibfnamefont {T.}~\bibnamefont
  {Sugimoto}}, \bibinfo {author} {\bibfnamefont {N.}~\bibnamefont {Aoi}},
  \bibinfo {author} {\bibfnamefont {H.}~\bibnamefont {Baba}}, \bibinfo {author}
  {\bibfnamefont {D.}~\bibnamefont {Bazin}}, \bibinfo {author} {\bibfnamefont
  {N.}~\bibnamefont {Fukuda}}, \bibinfo {author} {\bibfnamefont
  {T.}~\bibnamefont {Gomi}}, \bibinfo {author} {\bibfnamefont {H.}~\bibnamefont
  {Hasegawa}}, \bibinfo {author} {\bibfnamefont {N.}~\bibnamefont {Imai}},
  \bibinfo {author} {\bibfnamefont {M.}~\bibnamefont {Ishihara}}, \bibinfo
  {author} {\bibfnamefont {T.}~\bibnamefont {Kobayashi}}, \bibinfo {author}
  {\bibfnamefont {Y.}~\bibnamefont {Kondo}}, \bibinfo {author} {\bibfnamefont
  {T.}~\bibnamefont {Kubo}}, \bibinfo {author} {\bibfnamefont {M.}~\bibnamefont
  {Miura}}, \bibinfo {author} {\bibfnamefont {T.}~\bibnamefont {Motobayashi}},
  \bibinfo {author} {\bibfnamefont {H.}~\bibnamefont {Otsu}}, \bibinfo {author}
  {\bibfnamefont {A.}~\bibnamefont {Saito}}, \bibinfo {author} {\bibfnamefont
  {H.}~\bibnamefont {Sakurai}}, \bibinfo {author} {\bibfnamefont
  {S.}~\bibnamefont {Shimoura}}, \bibinfo {author} {\bibfnamefont
  {K.}~\bibnamefont {Watanabe}}, \bibinfo {author} {\bibfnamefont {Y.~X.}\
  \bibnamefont {Watanabe}}, \bibinfo {author} {\bibfnamefont {T.}~\bibnamefont
  {Yakushiji}}, \bibinfo {author} {\bibfnamefont {Y.}~\bibnamefont
  {Yanagisawa}}, \ and\ \bibinfo {author} {\bibfnamefont {K.}~\bibnamefont
  {Yoneda}},\ }\href {\doibase 10.1103/PhysRevLett.96.252502} {\bibfield
  {journal} {\bibinfo  {journal} {Phys. Rev. Lett.}\ }\textbf {\bibinfo
  {volume} {96}},\ \bibinfo {pages} {252502} (\bibinfo {year}
  {2006})}\BibitemShut {NoStop}%
\bibitem [{\citenamefont {Cook}\ \emph {et~al.}(2020)\citenamefont {Cook},
  \citenamefont {Nakamura}, \citenamefont {Kondo}, \citenamefont {Hagino},
  \citenamefont {Ogata}, \citenamefont {Saito}, \citenamefont {Achouri},
  \citenamefont {Aumann}, \citenamefont {Baba}, \citenamefont {Delaunay},
  \citenamefont {Deshayes}, \citenamefont {Doornenbal}, \citenamefont {Fukuda},
  \citenamefont {Gibelin}, \citenamefont {Hwang}, \citenamefont {Inabe},
  \citenamefont {Isobe}, \citenamefont {Kameda}, \citenamefont {Kanno},
  \citenamefont {Kim}, \citenamefont {Kobayashi}, \citenamefont {Kobayashi},
  \citenamefont {Kubo}, \citenamefont {Leblond}, \citenamefont {Lee},
  \citenamefont {Marqu\'es}, \citenamefont {Minakata}, \citenamefont
  {Motobayashi}, \citenamefont {Muto}, \citenamefont {Murakami}, \citenamefont
  {Murai}, \citenamefont {Nakashima}, \citenamefont {Nakatsuka}, \citenamefont
  {Navin}, \citenamefont {Nishi}, \citenamefont {Ogoshi}, \citenamefont {Orr},
  \citenamefont {Otsu}, \citenamefont {Sato}, \citenamefont {Satou},
  \citenamefont {Shimizu}, \citenamefont {Suzuki}, \citenamefont {Takahashi},
  \citenamefont {Takeda}, \citenamefont {Takeuchi}, \citenamefont {Tanaka},
  \citenamefont {Togano}, \citenamefont {Tsubota}, \citenamefont {Tuff},
  \citenamefont {Vandebrouck},\ and\ \citenamefont {Yoneda}}]{Cook2020}%
  \BibitemOpen
  \bibfield  {author} {\bibinfo {author} {\bibfnamefont {K.~J.}\ \bibnamefont
  {Cook}}, \bibinfo {author} {\bibfnamefont {T.}~\bibnamefont {Nakamura}},
  \bibinfo {author} {\bibfnamefont {Y.}~\bibnamefont {Kondo}}, \bibinfo
  {author} {\bibfnamefont {K.}~\bibnamefont {Hagino}}, \bibinfo {author}
  {\bibfnamefont {K.}~\bibnamefont {Ogata}}, \bibinfo {author} {\bibfnamefont
  {A.~T.}\ \bibnamefont {Saito}}, \bibinfo {author} {\bibfnamefont {N.~L.}\
  \bibnamefont {Achouri}}, \bibinfo {author} {\bibfnamefont {T.}~\bibnamefont
  {Aumann}}, \bibinfo {author} {\bibfnamefont {H.}~\bibnamefont {Baba}},
  \bibinfo {author} {\bibfnamefont {F.}~\bibnamefont {Delaunay}}, \bibinfo
  {author} {\bibfnamefont {Q.}~\bibnamefont {Deshayes}}, \bibinfo {author}
  {\bibfnamefont {P.}~\bibnamefont {Doornenbal}}, \bibinfo {author}
  {\bibfnamefont {N.}~\bibnamefont {Fukuda}}, \bibinfo {author} {\bibfnamefont
  {J.}~\bibnamefont {Gibelin}}, \bibinfo {author} {\bibfnamefont {J.~W.}\
  \bibnamefont {Hwang}}, \bibinfo {author} {\bibfnamefont {N.}~\bibnamefont
  {Inabe}}, \bibinfo {author} {\bibfnamefont {T.}~\bibnamefont {Isobe}},
  \bibinfo {author} {\bibfnamefont {D.}~\bibnamefont {Kameda}}, \bibinfo
  {author} {\bibfnamefont {D.}~\bibnamefont {Kanno}}, \bibinfo {author}
  {\bibfnamefont {S.}~\bibnamefont {Kim}}, \bibinfo {author} {\bibfnamefont
  {N.}~\bibnamefont {Kobayashi}}, \bibinfo {author} {\bibfnamefont
  {T.}~\bibnamefont {Kobayashi}}, \bibinfo {author} {\bibfnamefont
  {T.}~\bibnamefont {Kubo}}, \bibinfo {author} {\bibfnamefont {S.}~\bibnamefont
  {Leblond}}, \bibinfo {author} {\bibfnamefont {J.}~\bibnamefont {Lee}},
  \bibinfo {author} {\bibfnamefont {F.~M.}\ \bibnamefont {Marqu\'es}}, \bibinfo
  {author} {\bibfnamefont {R.}~\bibnamefont {Minakata}}, \bibinfo {author}
  {\bibfnamefont {T.}~\bibnamefont {Motobayashi}}, \bibinfo {author}
  {\bibfnamefont {K.}~\bibnamefont {Muto}}, \bibinfo {author} {\bibfnamefont
  {T.}~\bibnamefont {Murakami}}, \bibinfo {author} {\bibfnamefont
  {D.}~\bibnamefont {Murai}}, \bibinfo {author} {\bibfnamefont
  {T.}~\bibnamefont {Nakashima}}, \bibinfo {author} {\bibfnamefont
  {N.}~\bibnamefont {Nakatsuka}}, \bibinfo {author} {\bibfnamefont
  {A.}~\bibnamefont {Navin}}, \bibinfo {author} {\bibfnamefont
  {S.}~\bibnamefont {Nishi}}, \bibinfo {author} {\bibfnamefont
  {S.}~\bibnamefont {Ogoshi}}, \bibinfo {author} {\bibfnamefont {N.~A.}\
  \bibnamefont {Orr}}, \bibinfo {author} {\bibfnamefont {H.}~\bibnamefont
  {Otsu}}, \bibinfo {author} {\bibfnamefont {H.}~\bibnamefont {Sato}}, \bibinfo
  {author} {\bibfnamefont {Y.}~\bibnamefont {Satou}}, \bibinfo {author}
  {\bibfnamefont {Y.}~\bibnamefont {Shimizu}}, \bibinfo {author} {\bibfnamefont
  {H.}~\bibnamefont {Suzuki}}, \bibinfo {author} {\bibfnamefont
  {K.}~\bibnamefont {Takahashi}}, \bibinfo {author} {\bibfnamefont
  {H.}~\bibnamefont {Takeda}}, \bibinfo {author} {\bibfnamefont
  {S.}~\bibnamefont {Takeuchi}}, \bibinfo {author} {\bibfnamefont
  {R.}~\bibnamefont {Tanaka}}, \bibinfo {author} {\bibfnamefont
  {Y.}~\bibnamefont {Togano}}, \bibinfo {author} {\bibfnamefont
  {J.}~\bibnamefont {Tsubota}}, \bibinfo {author} {\bibfnamefont {A.~G.}\
  \bibnamefont {Tuff}}, \bibinfo {author} {\bibfnamefont {M.}~\bibnamefont
  {Vandebrouck}}, \ and\ \bibinfo {author} {\bibfnamefont {K.}~\bibnamefont
  {Yoneda}},\ }\href {\doibase 10.1103/PhysRevLett.124.212503} {\bibfield
  {journal} {\bibinfo  {journal} {Phys. Rev. Lett.}\ }\textbf {\bibinfo
  {volume} {124}},\ \bibinfo {pages} {212503} (\bibinfo {year}
  {2020})}\BibitemShut {NoStop}%
\bibitem [{\citenamefont {Danilin}\ \emph {et~al.}(1998)\citenamefont
  {Danilin}, \citenamefont {Thompson}, \citenamefont {Vaagen},\ and\
  \citenamefont {Zhukov}}]{Danilin98}%
  \BibitemOpen
  \bibfield  {author} {\bibinfo {author} {\bibfnamefont {B.}~\bibnamefont
  {Danilin}}, \bibinfo {author} {\bibfnamefont {I.}~\bibnamefont {Thompson}},
  \bibinfo {author} {\bibfnamefont {J.}~\bibnamefont {Vaagen}}, \ and\ \bibinfo
  {author} {\bibfnamefont {M.}~\bibnamefont {Zhukov}},\ }\href {\doibase
  https://doi.org/10.1016/S0375-9474(98)00002-5} {\bibfield  {journal}
  {\bibinfo  {journal} {Nucl. Phys. A}\ }\textbf {\bibinfo {volume} {632}},\
  \bibinfo {pages} {383 } (\bibinfo {year} {1998})}\BibitemShut {NoStop}%
\bibitem [{\citenamefont {Myo}\ \emph {et~al.}(2001)\citenamefont {Myo},
  \citenamefont {Kat\ifmmode~\bar{o}\else \={o}\fi{}}, \citenamefont {Aoyama},\
  and\ \citenamefont {Ikeda}}]{Myo01}%
  \BibitemOpen
  \bibfield  {author} {\bibinfo {author} {\bibfnamefont {T.}~\bibnamefont
  {Myo}}, \bibinfo {author} {\bibfnamefont {K.}~\bibnamefont
  {Kat\ifmmode~\bar{o}\else \={o}\fi{}}}, \bibinfo {author} {\bibfnamefont
  {S.}~\bibnamefont {Aoyama}}, \ and\ \bibinfo {author} {\bibfnamefont
  {K.}~\bibnamefont {Ikeda}},\ }\href {\doibase 10.1103/PhysRevC.63.054313}
  {\bibfield  {journal} {\bibinfo  {journal} {Phys. Rev. C}\ }\textbf {\bibinfo
  {volume} {63}},\ \bibinfo {pages} {054313} (\bibinfo {year}
  {2001})}\BibitemShut {NoStop}%
\bibitem [{\citenamefont {Rodr\'{\i}guez-Gallardo}\ \emph
  {et~al.}(2005)\citenamefont {Rodr\'{\i}guez-Gallardo}, \citenamefont {Arias},
  \citenamefont {G\'omez-Camacho}, \citenamefont {Moro}, \citenamefont
  {Thompson},\ and\ \citenamefont {Tostevin}}]{MRoGa05}%
  \BibitemOpen
  \bibfield  {author} {\bibinfo {author} {\bibfnamefont {M.}~\bibnamefont
  {Rodr\'{\i}guez-Gallardo}}, \bibinfo {author} {\bibfnamefont {J.~M.}\
  \bibnamefont {Arias}}, \bibinfo {author} {\bibfnamefont {J.}~\bibnamefont
  {G\'omez-Camacho}}, \bibinfo {author} {\bibfnamefont {A.~M.}\ \bibnamefont
  {Moro}}, \bibinfo {author} {\bibfnamefont {I.~J.}\ \bibnamefont {Thompson}},
  \ and\ \bibinfo {author} {\bibfnamefont {J.~A.}\ \bibnamefont {Tostevin}},\
  }\href {\doibase 10.1103/PhysRevC.72.024007} {\bibfield  {journal} {\bibinfo
  {journal} {Phys. Rev. C}\ }\textbf {\bibinfo {volume} {72}},\ \bibinfo
  {pages} {024007} (\bibinfo {year} {2005})}\BibitemShut {NoStop}%
\bibitem [{\citenamefont {Horiuchi}\ and\ \citenamefont
  {Suzuki}(2006)}]{Horiuchi06}%
  \BibitemOpen
  \bibfield  {author} {\bibinfo {author} {\bibfnamefont {W.}~\bibnamefont
  {Horiuchi}}\ and\ \bibinfo {author} {\bibfnamefont {Y.}~\bibnamefont
  {Suzuki}},\ }\href {\doibase 10.1103/PhysRevC.74.034311} {\bibfield
  {journal} {\bibinfo  {journal} {Phys. Rev. C}\ }\textbf {\bibinfo {volume}
  {74}},\ \bibinfo {pages} {034311} (\bibinfo {year} {2006})}\BibitemShut
  {NoStop}%
\bibitem [{\citenamefont {Horiuchi}\ and\ \citenamefont
  {Suzuki}(2007)}]{Horiuchi07}%
  \BibitemOpen
  \bibfield  {author} {\bibinfo {author} {\bibfnamefont {W.}~\bibnamefont
  {Horiuchi}}\ and\ \bibinfo {author} {\bibfnamefont {Y.}~\bibnamefont
  {Suzuki}},\ }\href {\doibase 10.1103/PhysRevC.76.024311} {\bibfield
  {journal} {\bibinfo  {journal} {Phys. Rev. C}\ }\textbf {\bibinfo {volume}
  {76}},\ \bibinfo {pages} {024311} (\bibinfo {year} {2007})}\BibitemShut
  {NoStop}%
\bibitem [{\citenamefont {Hagino}\ \emph {et~al.}(2009)\citenamefont {Hagino},
  \citenamefont {Sagawa}, \citenamefont {Nakamura},\ and\ \citenamefont
  {Shimoura}}]{hagino09}%
  \BibitemOpen
  \bibfield  {author} {\bibinfo {author} {\bibfnamefont {K.}~\bibnamefont
  {Hagino}}, \bibinfo {author} {\bibfnamefont {H.}~\bibnamefont {Sagawa}},
  \bibinfo {author} {\bibfnamefont {T.}~\bibnamefont {Nakamura}}, \ and\
  \bibinfo {author} {\bibfnamefont {S.}~\bibnamefont {Shimoura}},\ }\href
  {\doibase 10.1103/PhysRevC.80.031301} {\bibfield  {journal} {\bibinfo
  {journal} {Phys. Rev. C}\ }\textbf {\bibinfo {volume} {80}},\ \bibinfo
  {pages} {031301} (\bibinfo {year} {2009})}\BibitemShut {NoStop}%
\bibitem [{\citenamefont {de~Diego}\ \emph {et~al.}(2010)\citenamefont
  {de~Diego}, \citenamefont {Garrido}, \citenamefont {Fedorov},\ and\
  \citenamefont {Jensen}}]{RdDiego10}%
  \BibitemOpen
  \bibfield  {author} {\bibinfo {author} {\bibfnamefont {R.}~\bibnamefont
  {de~Diego}}, \bibinfo {author} {\bibfnamefont {E.}~\bibnamefont {Garrido}},
  \bibinfo {author} {\bibfnamefont {D.~V.}\ \bibnamefont {Fedorov}}, \ and\
  \bibinfo {author} {\bibfnamefont {A.~S.}\ \bibnamefont {Jensen}},\ }\href
  {http://stacks.iop.org/0295-5075/90/i=5/a=52001} {\bibfield  {journal}
  {\bibinfo  {journal} {Europhys. Lett.}\ }\textbf {\bibinfo {volume} {90}},\
  \bibinfo {pages} {52001} (\bibinfo {year} {2010})}\BibitemShut {NoStop}%
\bibitem [{\citenamefont {Casal}\ \emph {et~al.}(2013)\citenamefont {Casal},
  \citenamefont {Rodr\'{\i}guez-Gallardo},\ and\ \citenamefont
  {Arias}}]{JCasal13}%
  \BibitemOpen
  \bibfield  {author} {\bibinfo {author} {\bibfnamefont {J.}~\bibnamefont
  {Casal}}, \bibinfo {author} {\bibfnamefont {M.}~\bibnamefont
  {Rodr\'{\i}guez-Gallardo}}, \ and\ \bibinfo {author} {\bibfnamefont {J.~M.}\
  \bibnamefont {Arias}},\ }\href {\doibase 10.1103/PhysRevC.88.014327}
  {\bibfield  {journal} {\bibinfo  {journal} {Phys. Rev. C}\ }\textbf {\bibinfo
  {volume} {88}},\ \bibinfo {pages} {014327} (\bibinfo {year}
  {2013})}\BibitemShut {NoStop}%
\bibitem [{\citenamefont {Fern\'andez-Garc\'{\i}a}\ \emph
  {et~al.}(2013)\citenamefont {Fern\'andez-Garc\'{\i}a}, \citenamefont
  {Cubero}, \citenamefont {Rodr\'{\i}guez-Gallardo}, \citenamefont {Acosta},
  \citenamefont {Alcorta}, \citenamefont {Alvarez}, \citenamefont {Borge},
  \citenamefont {Buchmann}, \citenamefont {Diget}, \citenamefont {Falou},
  \citenamefont {Fulton}, \citenamefont {Fynbo}, \citenamefont {Galaviz},
  \citenamefont {G\'omez-Camacho}, \citenamefont {Kanungo}, \citenamefont
  {Lay}, \citenamefont {Madurga}, \citenamefont {Martel}, \citenamefont {Moro},
  \citenamefont {Mukha}, \citenamefont {Nilsson}, \citenamefont
  {S\'anchez-Ben\'{\i}tez}, \citenamefont {Shotter}, \citenamefont {Tengblad},\
  and\ \citenamefont {Walden}}]{JPFernandezGarcia13}%
  \BibitemOpen
  \bibfield  {author} {\bibinfo {author} {\bibfnamefont {J.~P.}\ \bibnamefont
  {Fern\'andez-Garc\'{\i}a}}, \bibinfo {author} {\bibfnamefont
  {M.}~\bibnamefont {Cubero}}, \bibinfo {author} {\bibfnamefont
  {M.}~\bibnamefont {Rodr\'{\i}guez-Gallardo}}, \bibinfo {author}
  {\bibfnamefont {L.}~\bibnamefont {Acosta}}, \bibinfo {author} {\bibfnamefont
  {M.}~\bibnamefont {Alcorta}}, \bibinfo {author} {\bibfnamefont {M.~A.~G.}\
  \bibnamefont {Alvarez}}, \bibinfo {author} {\bibfnamefont {M.~J.~G.}\
  \bibnamefont {Borge}}, \bibinfo {author} {\bibfnamefont {L.}~\bibnamefont
  {Buchmann}}, \bibinfo {author} {\bibfnamefont {C.~A.}\ \bibnamefont {Diget}},
  \bibinfo {author} {\bibfnamefont {H.~A.}\ \bibnamefont {Falou}}, \bibinfo
  {author} {\bibfnamefont {B.~R.}\ \bibnamefont {Fulton}}, \bibinfo {author}
  {\bibfnamefont {H.~O.~U.}\ \bibnamefont {Fynbo}}, \bibinfo {author}
  {\bibfnamefont {D.}~\bibnamefont {Galaviz}}, \bibinfo {author} {\bibfnamefont
  {J.}~\bibnamefont {G\'omez-Camacho}}, \bibinfo {author} {\bibfnamefont
  {R.}~\bibnamefont {Kanungo}}, \bibinfo {author} {\bibfnamefont {J.~A.}\
  \bibnamefont {Lay}}, \bibinfo {author} {\bibfnamefont {M.}~\bibnamefont
  {Madurga}}, \bibinfo {author} {\bibfnamefont {I.}~\bibnamefont {Martel}},
  \bibinfo {author} {\bibfnamefont {A.~M.}\ \bibnamefont {Moro}}, \bibinfo
  {author} {\bibfnamefont {I.}~\bibnamefont {Mukha}}, \bibinfo {author}
  {\bibfnamefont {T.}~\bibnamefont {Nilsson}}, \bibinfo {author} {\bibfnamefont
  {A.~M.}\ \bibnamefont {S\'anchez-Ben\'{\i}tez}}, \bibinfo {author}
  {\bibfnamefont {A.}~\bibnamefont {Shotter}}, \bibinfo {author} {\bibfnamefont
  {O.}~\bibnamefont {Tengblad}}, \ and\ \bibinfo {author} {\bibfnamefont
  {P.}~\bibnamefont {Walden}},\ }\href {\doibase
  10.1103/PhysRevLett.110.142701} {\bibfield  {journal} {\bibinfo  {journal}
  {Phys. Rev. Lett.}\ }\textbf {\bibinfo {volume} {110}},\ \bibinfo {pages}
  {142701} (\bibinfo {year} {2013})}\BibitemShut {NoStop}%
\bibitem [{\citenamefont {Riisager}(1994)}]{Riisager94}%
  \BibitemOpen
  \bibfield  {author} {\bibinfo {author} {\bibfnamefont {K.}~\bibnamefont
  {Riisager}},\ }\href {\doibase 10.1103/RevModPhys.66.1105} {\bibfield
  {journal} {\bibinfo  {journal} {Rev. Mod. Phys.}\ }\textbf {\bibinfo {volume}
  {66}},\ \bibinfo {pages} {1105} (\bibinfo {year} {1994})}\BibitemShut
  {NoStop}%
\bibitem [{\citenamefont {Jeppesen}\ \emph {et~al.}(2006)\citenamefont
  {Jeppesen} \emph {et~al.}}]{Jep06}%
  \BibitemOpen
  \bibfield  {author} {\bibinfo {author} {\bibfnamefont {H.~B.}\ \bibnamefont
  {Jeppesen}} \emph {et~al.},\ }\href {\doibase 10.1016/j.physletb.2006.09.060}
  {\bibfield  {journal} {\bibinfo  {journal} {Phys. Lett. B}\ }\textbf
  {\bibinfo {volume} {642}},\ \bibinfo {pages} {449} (\bibinfo {year}
  {2006})}\BibitemShut {NoStop}%
\bibitem [{\citenamefont {Sanetullaev}\ \emph {et~al.}(2016)\citenamefont
  {Sanetullaev}, \citenamefont {Kanungo}, \citenamefont {Tanaka}, \citenamefont
  {Alcorta}, \citenamefont {Andreoiu}, \citenamefont {Bender}, \citenamefont
  {Chen}, \citenamefont {Christian}, \citenamefont {Davids}, \citenamefont
  {Fallis}, \citenamefont {Fortin}, \citenamefont {Galinski}, \citenamefont
  {Gallant}, \citenamefont {Garrett}, \citenamefont {Hackman}, \citenamefont
  {Hadinia}, \citenamefont {Ishimoto}, \citenamefont {Keefe}, \citenamefont
  {Krücken}, \citenamefont {Lighthall}, \citenamefont {McNeice}, \citenamefont
  {Miller}, \citenamefont {Purcell}, \citenamefont {Randhawa}, \citenamefont
  {Roger}, \citenamefont {Rojas}, \citenamefont {Savajols}, \citenamefont
  {Shotter}, \citenamefont {Tanihata}, \citenamefont {Thompson}, \citenamefont
  {Unsworth}, \citenamefont {Voss},\ and\ \citenamefont
  {Wang}}]{sanetullaev2016}%
  \BibitemOpen
  \bibfield  {author} {\bibinfo {author} {\bibfnamefont {A.}~\bibnamefont
  {Sanetullaev}}, \bibinfo {author} {\bibfnamefont {R.}~\bibnamefont
  {Kanungo}}, \bibinfo {author} {\bibfnamefont {J.}~\bibnamefont {Tanaka}},
  \bibinfo {author} {\bibfnamefont {M.}~\bibnamefont {Alcorta}}, \bibinfo
  {author} {\bibfnamefont {C.}~\bibnamefont {Andreoiu}}, \bibinfo {author}
  {\bibfnamefont {P.}~\bibnamefont {Bender}}, \bibinfo {author} {\bibfnamefont
  {A.}~\bibnamefont {Chen}}, \bibinfo {author} {\bibfnamefont {G.}~\bibnamefont
  {Christian}}, \bibinfo {author} {\bibfnamefont {B.}~\bibnamefont {Davids}},
  \bibinfo {author} {\bibfnamefont {J.}~\bibnamefont {Fallis}}, \bibinfo
  {author} {\bibfnamefont {J.}~\bibnamefont {Fortin}}, \bibinfo {author}
  {\bibfnamefont {N.}~\bibnamefont {Galinski}}, \bibinfo {author}
  {\bibfnamefont {A.}~\bibnamefont {Gallant}}, \bibinfo {author} {\bibfnamefont
  {P.}~\bibnamefont {Garrett}}, \bibinfo {author} {\bibfnamefont
  {G.}~\bibnamefont {Hackman}}, \bibinfo {author} {\bibfnamefont
  {B.}~\bibnamefont {Hadinia}}, \bibinfo {author} {\bibfnamefont
  {S.}~\bibnamefont {Ishimoto}}, \bibinfo {author} {\bibfnamefont
  {M.}~\bibnamefont {Keefe}}, \bibinfo {author} {\bibfnamefont
  {R.}~\bibnamefont {Krücken}}, \bibinfo {author} {\bibfnamefont
  {J.}~\bibnamefont {Lighthall}}, \bibinfo {author} {\bibfnamefont
  {E.}~\bibnamefont {McNeice}}, \bibinfo {author} {\bibfnamefont
  {D.}~\bibnamefont {Miller}}, \bibinfo {author} {\bibfnamefont
  {J.}~\bibnamefont {Purcell}}, \bibinfo {author} {\bibfnamefont
  {J.}~\bibnamefont {Randhawa}}, \bibinfo {author} {\bibfnamefont
  {T.}~\bibnamefont {Roger}}, \bibinfo {author} {\bibfnamefont
  {A.}~\bibnamefont {Rojas}}, \bibinfo {author} {\bibfnamefont
  {H.}~\bibnamefont {Savajols}}, \bibinfo {author} {\bibfnamefont
  {A.}~\bibnamefont {Shotter}}, \bibinfo {author} {\bibfnamefont
  {I.}~\bibnamefont {Tanihata}}, \bibinfo {author} {\bibfnamefont
  {I.}~\bibnamefont {Thompson}}, \bibinfo {author} {\bibfnamefont
  {C.}~\bibnamefont {Unsworth}}, \bibinfo {author} {\bibfnamefont
  {P.}~\bibnamefont {Voss}}, \ and\ \bibinfo {author} {\bibfnamefont
  {Z.}~\bibnamefont {Wang}},\ }\href {\doibase 10.1016/j.physletb.2016.02.060}
  {\bibfield  {journal} {\bibinfo  {journal} {Phys. Lett. B}\ }\textbf
  {\bibinfo {volume} {755}},\ \bibinfo {pages} {481} (\bibinfo {year}
  {2016})}\BibitemShut {NoStop}%
\bibitem [{\citenamefont {Hamamoto}(2007)}]{Hamamoto07}%
  \BibitemOpen
  \bibfield  {author} {\bibinfo {author} {\bibfnamefont {I.}~\bibnamefont
  {Hamamoto}},\ }\href {\doibase 10.1103/PhysRevC.76.054319} {\bibfield
  {journal} {\bibinfo  {journal} {Phys. Rev. C}\ }\textbf {\bibinfo {volume}
  {76}},\ \bibinfo {pages} {054319} (\bibinfo {year} {2007})}\BibitemShut
  {NoStop}%
\bibitem [{\citenamefont {Warburton}\ \emph {et~al.}(1990)\citenamefont
  {Warburton}, \citenamefont {Becker},\ and\ \citenamefont {Brown}}]{WAR91}%
  \BibitemOpen
  \bibfield  {author} {\bibinfo {author} {\bibfnamefont {E.~K.}\ \bibnamefont
  {Warburton}}, \bibinfo {author} {\bibfnamefont {J.~A.}\ \bibnamefont
  {Becker}}, \ and\ \bibinfo {author} {\bibfnamefont {B.~A.}\ \bibnamefont
  {Brown}},\ }\href {\doibase 10.1103/PhysRevC.41.1147} {\bibfield  {journal}
  {\bibinfo  {journal} {Phys. Rev. C}\ }\textbf {\bibinfo {volume} {41}},\
  \bibinfo {pages} {1147} (\bibinfo {year} {1990})}\BibitemShut {NoStop}%
\bibitem [{\citenamefont {Nakamura}\ \emph {et~al.}(2009)\citenamefont
  {Nakamura}, \citenamefont {Kobayashi}, \citenamefont {Kondo}, \citenamefont
  {Satou}, \citenamefont {Aoi}, \citenamefont {Baba}, \citenamefont {Deguchi},
  \citenamefont {Fukuda}, \citenamefont {Gibelin}, \citenamefont {Inabe},
  \citenamefont {Ishihara}, \citenamefont {Kameda}, \citenamefont {Kawada},
  \citenamefont {Kubo}, \citenamefont {Kusaka}, \citenamefont {Mengoni},
  \citenamefont {Motobayashi}, \citenamefont {Ohnishi}, \citenamefont {Ohtake},
  \citenamefont {Orr}, \citenamefont {Otsu}, \citenamefont {Otsuka},
  \citenamefont {Saito}, \citenamefont {Sakurai}, \citenamefont {Shimoura},
  \citenamefont {Sumikama}, \citenamefont {Takeda}, \citenamefont {Takeshita},
  \citenamefont {Takechi}, \citenamefont {Takeuchi}, \citenamefont {Tanaka},
  \citenamefont {Tanaka}, \citenamefont {Tanaka}, \citenamefont {Togano},
  \citenamefont {Utsuno}, \citenamefont {Yoneda}, \citenamefont {Yoshida},\
  and\ \citenamefont {Yoshida}}]{Nakamura09}%
  \BibitemOpen
  \bibfield  {author} {\bibinfo {author} {\bibfnamefont {T.}~\bibnamefont
  {Nakamura}}, \bibinfo {author} {\bibfnamefont {N.}~\bibnamefont {Kobayashi}},
  \bibinfo {author} {\bibfnamefont {Y.}~\bibnamefont {Kondo}}, \bibinfo
  {author} {\bibfnamefont {Y.}~\bibnamefont {Satou}}, \bibinfo {author}
  {\bibfnamefont {N.}~\bibnamefont {Aoi}}, \bibinfo {author} {\bibfnamefont
  {H.}~\bibnamefont {Baba}}, \bibinfo {author} {\bibfnamefont {S.}~\bibnamefont
  {Deguchi}}, \bibinfo {author} {\bibfnamefont {N.}~\bibnamefont {Fukuda}},
  \bibinfo {author} {\bibfnamefont {J.}~\bibnamefont {Gibelin}}, \bibinfo
  {author} {\bibfnamefont {N.}~\bibnamefont {Inabe}}, \bibinfo {author}
  {\bibfnamefont {M.}~\bibnamefont {Ishihara}}, \bibinfo {author}
  {\bibfnamefont {D.}~\bibnamefont {Kameda}}, \bibinfo {author} {\bibfnamefont
  {Y.}~\bibnamefont {Kawada}}, \bibinfo {author} {\bibfnamefont
  {T.}~\bibnamefont {Kubo}}, \bibinfo {author} {\bibfnamefont {K.}~\bibnamefont
  {Kusaka}}, \bibinfo {author} {\bibfnamefont {A.}~\bibnamefont {Mengoni}},
  \bibinfo {author} {\bibfnamefont {T.}~\bibnamefont {Motobayashi}}, \bibinfo
  {author} {\bibfnamefont {T.}~\bibnamefont {Ohnishi}}, \bibinfo {author}
  {\bibfnamefont {M.}~\bibnamefont {Ohtake}}, \bibinfo {author} {\bibfnamefont
  {N.~A.}\ \bibnamefont {Orr}}, \bibinfo {author} {\bibfnamefont
  {H.}~\bibnamefont {Otsu}}, \bibinfo {author} {\bibfnamefont {T.}~\bibnamefont
  {Otsuka}}, \bibinfo {author} {\bibfnamefont {A.}~\bibnamefont {Saito}},
  \bibinfo {author} {\bibfnamefont {H.}~\bibnamefont {Sakurai}}, \bibinfo
  {author} {\bibfnamefont {S.}~\bibnamefont {Shimoura}}, \bibinfo {author}
  {\bibfnamefont {T.}~\bibnamefont {Sumikama}}, \bibinfo {author}
  {\bibfnamefont {H.}~\bibnamefont {Takeda}}, \bibinfo {author} {\bibfnamefont
  {E.}~\bibnamefont {Takeshita}}, \bibinfo {author} {\bibfnamefont
  {M.}~\bibnamefont {Takechi}}, \bibinfo {author} {\bibfnamefont
  {S.}~\bibnamefont {Takeuchi}}, \bibinfo {author} {\bibfnamefont
  {K.}~\bibnamefont {Tanaka}}, \bibinfo {author} {\bibfnamefont {K.~N.}\
  \bibnamefont {Tanaka}}, \bibinfo {author} {\bibfnamefont {N.}~\bibnamefont
  {Tanaka}}, \bibinfo {author} {\bibfnamefont {Y.}~\bibnamefont {Togano}},
  \bibinfo {author} {\bibfnamefont {Y.}~\bibnamefont {Utsuno}}, \bibinfo
  {author} {\bibfnamefont {K.}~\bibnamefont {Yoneda}}, \bibinfo {author}
  {\bibfnamefont {A.}~\bibnamefont {Yoshida}}, \ and\ \bibinfo {author}
  {\bibfnamefont {K.}~\bibnamefont {Yoshida}},\ }\href {\doibase
  10.1103/PhysRevLett.103.262501} {\bibfield  {journal} {\bibinfo  {journal}
  {Phys. Rev. Lett.}\ }\textbf {\bibinfo {volume} {103}},\ \bibinfo {pages}
  {262501} (\bibinfo {year} {2009})}\BibitemShut {NoStop}%
\bibitem [{\citenamefont {Motobayashi}\ \emph {et~al.}(1995)\citenamefont
  {Motobayashi}, \citenamefont {Ikeda}, \citenamefont {Ieki}, \citenamefont
  {Inoue}, \citenamefont {Iwasa}, \citenamefont {Kikuchi}, \citenamefont
  {Kurokawa}, \citenamefont {Moriya}, \citenamefont {Ogawa}, \citenamefont
  {Murakami}, \citenamefont {Shimoura}, \citenamefont {Yanagisawa},
  \citenamefont {Nakamura}, \citenamefont {Watanabe}, \citenamefont {Ishihara},
  \citenamefont {Teranishi}, \citenamefont {Okuno},\ and\ \citenamefont
  {Casten}}]{Motobayashi1995}%
  \BibitemOpen
  \bibfield  {author} {\bibinfo {author} {\bibfnamefont {T.}~\bibnamefont
  {Motobayashi}}, \bibinfo {author} {\bibfnamefont {Y.}~\bibnamefont {Ikeda}},
  \bibinfo {author} {\bibfnamefont {K.}~\bibnamefont {Ieki}}, \bibinfo {author}
  {\bibfnamefont {M.}~\bibnamefont {Inoue}}, \bibinfo {author} {\bibfnamefont
  {N.}~\bibnamefont {Iwasa}}, \bibinfo {author} {\bibfnamefont
  {T.}~\bibnamefont {Kikuchi}}, \bibinfo {author} {\bibfnamefont
  {M.}~\bibnamefont {Kurokawa}}, \bibinfo {author} {\bibfnamefont
  {S.}~\bibnamefont {Moriya}}, \bibinfo {author} {\bibfnamefont
  {S.}~\bibnamefont {Ogawa}}, \bibinfo {author} {\bibfnamefont
  {H.}~\bibnamefont {Murakami}}, \bibinfo {author} {\bibfnamefont
  {S.}~\bibnamefont {Shimoura}}, \bibinfo {author} {\bibfnamefont
  {Y.}~\bibnamefont {Yanagisawa}}, \bibinfo {author} {\bibfnamefont
  {T.}~\bibnamefont {Nakamura}}, \bibinfo {author} {\bibfnamefont
  {Y.}~\bibnamefont {Watanabe}}, \bibinfo {author} {\bibfnamefont
  {M.}~\bibnamefont {Ishihara}}, \bibinfo {author} {\bibfnamefont
  {T.}~\bibnamefont {Teranishi}}, \bibinfo {author} {\bibfnamefont
  {H.}~\bibnamefont {Okuno}}, \ and\ \bibinfo {author} {\bibfnamefont
  {R.}~\bibnamefont {Casten}},\ }\href {\doibase
  https://doi.org/10.1016/0370-2693(95)00012-A} {\bibfield  {journal} {\bibinfo
   {journal} {Phys. Lett. B}\ }\textbf {\bibinfo {volume} {346}},\ \bibinfo
  {pages} {9 } (\bibinfo {year} {1995})}\BibitemShut {NoStop}%
\bibitem [{\citenamefont {Gaudefroy}\ \emph {et~al.}(2012)\citenamefont
  {Gaudefroy}, \citenamefont {Mittig}, \citenamefont {Orr}, \citenamefont
  {Varet}, \citenamefont {Chartier}, \citenamefont {Roussel-Chomaz},
  \citenamefont {Ebran}, \citenamefont {Fern\'andez-Dom\'{\i}nguez},
  \citenamefont {Fr\'emont}, \citenamefont {Gangnant}, \citenamefont
  {Gillibert}, \citenamefont {Gr\'evy}, \citenamefont {Libin}, \citenamefont
  {Maslov}, \citenamefont {Paschalis}, \citenamefont {Pietras}, \citenamefont
  {Penionzhkevich}, \citenamefont {Spitaels},\ and\ \citenamefont
  {Villari}}]{Gaudefroy2012}%
  \BibitemOpen
  \bibfield  {author} {\bibinfo {author} {\bibfnamefont {L.}~\bibnamefont
  {Gaudefroy}}, \bibinfo {author} {\bibfnamefont {W.}~\bibnamefont {Mittig}},
  \bibinfo {author} {\bibfnamefont {N.~A.}\ \bibnamefont {Orr}}, \bibinfo
  {author} {\bibfnamefont {S.}~\bibnamefont {Varet}}, \bibinfo {author}
  {\bibfnamefont {M.}~\bibnamefont {Chartier}}, \bibinfo {author}
  {\bibfnamefont {P.}~\bibnamefont {Roussel-Chomaz}}, \bibinfo {author}
  {\bibfnamefont {J.~P.}\ \bibnamefont {Ebran}}, \bibinfo {author}
  {\bibfnamefont {B.}~\bibnamefont {Fern\'andez-Dom\'{\i}nguez}}, \bibinfo
  {author} {\bibfnamefont {G.}~\bibnamefont {Fr\'emont}}, \bibinfo {author}
  {\bibfnamefont {P.}~\bibnamefont {Gangnant}}, \bibinfo {author}
  {\bibfnamefont {A.}~\bibnamefont {Gillibert}}, \bibinfo {author}
  {\bibfnamefont {S.}~\bibnamefont {Gr\'evy}}, \bibinfo {author} {\bibfnamefont
  {J.~F.}\ \bibnamefont {Libin}}, \bibinfo {author} {\bibfnamefont {V.~A.}\
  \bibnamefont {Maslov}}, \bibinfo {author} {\bibfnamefont {S.}~\bibnamefont
  {Paschalis}}, \bibinfo {author} {\bibfnamefont {B.}~\bibnamefont {Pietras}},
  \bibinfo {author} {\bibfnamefont {Y.-E.}\ \bibnamefont {Penionzhkevich}},
  \bibinfo {author} {\bibfnamefont {C.}~\bibnamefont {Spitaels}}, \ and\
  \bibinfo {author} {\bibfnamefont {A.~C.~C.}\ \bibnamefont {Villari}},\ }\href
  {\doibase 10.1103/PhysRevLett.109.202503} {\bibfield  {journal} {\bibinfo
  {journal} {Phys. Rev. Lett.}\ }\textbf {\bibinfo {volume} {109}},\ \bibinfo
  {pages} {202503} (\bibinfo {year} {2012})}\BibitemShut {NoStop}%
\bibitem [{\citenamefont {Ahn}\ \emph {et~al.}(2019)\citenamefont {Ahn},
  \citenamefont {Fukuda}, \citenamefont {Geissel}, \citenamefont {Inabe},
  \citenamefont {Iwasa}, \citenamefont {Kubo}, \citenamefont {Kusaka},
  \citenamefont {Morrissey}, \citenamefont {Murai}, \citenamefont {Nakamura},
  \citenamefont {Ohtake}, \citenamefont {Otsu}, \citenamefont {Sato},
  \citenamefont {Sherrill}, \citenamefont {Shimizu}, \citenamefont {Suzuki},
  \citenamefont {Takeda}, \citenamefont {Tarasov}, \citenamefont {Ueno},
  \citenamefont {Yanagisawa},\ and\ \citenamefont {Yoshida}}]{Ahn2019}%
  \BibitemOpen
  \bibfield  {author} {\bibinfo {author} {\bibfnamefont {D.~S.}\ \bibnamefont
  {Ahn}}, \bibinfo {author} {\bibfnamefont {N.}~\bibnamefont {Fukuda}},
  \bibinfo {author} {\bibfnamefont {H.}~\bibnamefont {Geissel}}, \bibinfo
  {author} {\bibfnamefont {N.}~\bibnamefont {Inabe}}, \bibinfo {author}
  {\bibfnamefont {N.}~\bibnamefont {Iwasa}}, \bibinfo {author} {\bibfnamefont
  {T.}~\bibnamefont {Kubo}}, \bibinfo {author} {\bibfnamefont {K.}~\bibnamefont
  {Kusaka}}, \bibinfo {author} {\bibfnamefont {D.~J.}\ \bibnamefont
  {Morrissey}}, \bibinfo {author} {\bibfnamefont {D.}~\bibnamefont {Murai}},
  \bibinfo {author} {\bibfnamefont {T.}~\bibnamefont {Nakamura}}, \bibinfo
  {author} {\bibfnamefont {M.}~\bibnamefont {Ohtake}}, \bibinfo {author}
  {\bibfnamefont {H.}~\bibnamefont {Otsu}}, \bibinfo {author} {\bibfnamefont
  {H.}~\bibnamefont {Sato}}, \bibinfo {author} {\bibfnamefont {B.~M.}\
  \bibnamefont {Sherrill}}, \bibinfo {author} {\bibfnamefont {Y.}~\bibnamefont
  {Shimizu}}, \bibinfo {author} {\bibfnamefont {H.}~\bibnamefont {Suzuki}},
  \bibinfo {author} {\bibfnamefont {H.}~\bibnamefont {Takeda}}, \bibinfo
  {author} {\bibfnamefont {O.~B.}\ \bibnamefont {Tarasov}}, \bibinfo {author}
  {\bibfnamefont {H.}~\bibnamefont {Ueno}}, \bibinfo {author} {\bibfnamefont
  {Y.}~\bibnamefont {Yanagisawa}}, \ and\ \bibinfo {author} {\bibfnamefont
  {K.}~\bibnamefont {Yoshida}},\ }\href {\doibase
  10.1103/PhysRevLett.123.212501} {\bibfield  {journal} {\bibinfo  {journal}
  {Phys. Rev. Lett.}\ }\textbf {\bibinfo {volume} {123}},\ \bibinfo {pages}
  {212501} (\bibinfo {year} {2019})}\BibitemShut {NoStop}%
\bibitem [{\citenamefont {Masui}\ \emph {et~al.}(2020)\citenamefont {Masui},
  \citenamefont {Horiuchi},\ and\ \citenamefont {Kimura}}]{Masui2020}%
  \BibitemOpen
  \bibfield  {author} {\bibinfo {author} {\bibfnamefont {H.}~\bibnamefont
  {Masui}}, \bibinfo {author} {\bibfnamefont {W.}~\bibnamefont {Horiuchi}}, \
  and\ \bibinfo {author} {\bibfnamefont {M.}~\bibnamefont {Kimura}},\ }\href
  {\doibase 10.1103/PhysRevC.101.041303} {\bibfield  {journal} {\bibinfo
  {journal} {Phys. Rev. C}\ }\textbf {\bibinfo {volume} {101}},\ \bibinfo
  {pages} {041303} (\bibinfo {year} {2020})}\BibitemShut {NoStop}%
\bibitem [{\citenamefont {Michel}\ \emph {et~al.}(2020)\citenamefont {Michel},
  \citenamefont {Li}, \citenamefont {Xu},\ and\ \citenamefont
  {Zuo}}]{Michel2020}%
  \BibitemOpen
  \bibfield  {author} {\bibinfo {author} {\bibfnamefont {N.}~\bibnamefont
  {Michel}}, \bibinfo {author} {\bibfnamefont {J.~G.}\ \bibnamefont {Li}},
  \bibinfo {author} {\bibfnamefont {F.~R.}\ \bibnamefont {Xu}}, \ and\ \bibinfo
  {author} {\bibfnamefont {W.}~\bibnamefont {Zuo}},\ }\href {\doibase
  10.1103/PhysRevC.101.031301} {\bibfield  {journal} {\bibinfo  {journal}
  {Phys. Rev. C}\ }\textbf {\bibinfo {volume} {101}},\ \bibinfo {pages}
  {031301} (\bibinfo {year} {2020})}\BibitemShut {NoStop}%
\bibitem [{\citenamefont {Jurado}\ \emph {et~al.}(2007)\citenamefont {Jurado},
  \citenamefont {Savajols}, \citenamefont {Mittig}, \citenamefont {Orr},
  \citenamefont {Roussel-Chomaz}, \citenamefont {Baiborodin}, \citenamefont
  {Catford}, \citenamefont {Chartier}, \citenamefont {Demonchy}, \citenamefont
  {Dlouhý}, \citenamefont {Gillibert}, \citenamefont {Giot}, \citenamefont
  {Khouaja}, \citenamefont {Lépine-Szily}, \citenamefont {Lukyanov},
  \citenamefont {Mrazek}, \citenamefont {Penionzhkevich}, \citenamefont {Pita},
  \citenamefont {Rousseau},\ and\ \citenamefont {Villari}}]{Jurado07}%
  \BibitemOpen
  \bibfield  {author} {\bibinfo {author} {\bibfnamefont {B.}~\bibnamefont
  {Jurado}}, \bibinfo {author} {\bibfnamefont {H.}~\bibnamefont {Savajols}},
  \bibinfo {author} {\bibfnamefont {W.}~\bibnamefont {Mittig}}, \bibinfo
  {author} {\bibfnamefont {N.}~\bibnamefont {Orr}}, \bibinfo {author}
  {\bibfnamefont {P.}~\bibnamefont {Roussel-Chomaz}}, \bibinfo {author}
  {\bibfnamefont {D.}~\bibnamefont {Baiborodin}}, \bibinfo {author}
  {\bibfnamefont {W.}~\bibnamefont {Catford}}, \bibinfo {author} {\bibfnamefont
  {M.}~\bibnamefont {Chartier}}, \bibinfo {author} {\bibfnamefont
  {C.}~\bibnamefont {Demonchy}}, \bibinfo {author} {\bibfnamefont
  {Z.}~\bibnamefont {Dlouhý}}, \bibinfo {author} {\bibfnamefont
  {A.}~\bibnamefont {Gillibert}}, \bibinfo {author} {\bibfnamefont
  {L.}~\bibnamefont {Giot}}, \bibinfo {author} {\bibfnamefont {A.}~\bibnamefont
  {Khouaja}}, \bibinfo {author} {\bibfnamefont {A.}~\bibnamefont
  {Lépine-Szily}}, \bibinfo {author} {\bibfnamefont {S.}~\bibnamefont
  {Lukyanov}}, \bibinfo {author} {\bibfnamefont {J.}~\bibnamefont {Mrazek}},
  \bibinfo {author} {\bibfnamefont {Y.}~\bibnamefont {Penionzhkevich}},
  \bibinfo {author} {\bibfnamefont {S.}~\bibnamefont {Pita}}, \bibinfo {author}
  {\bibfnamefont {M.}~\bibnamefont {Rousseau}}, \ and\ \bibinfo {author}
  {\bibfnamefont {A.}~\bibnamefont {Villari}},\ }\href {\doibase
  https://doi.org/10.1016/j.physletb.2007.04.006} {\bibfield  {journal}
  {\bibinfo  {journal} {Phys. Lett. B}\ }\textbf {\bibinfo {volume} {649}},\
  \bibinfo {pages} {43 } (\bibinfo {year} {2007})}\BibitemShut {NoStop}%
\bibitem [{\citenamefont {Elekes}\ \emph {et~al.}(2004)\citenamefont {Elekes},
  \citenamefont {Dombrádi}, \citenamefont {Saito}, \citenamefont {Aoi},
  \citenamefont {Baba}, \citenamefont {Demichi}, \citenamefont {Fülöp},
  \citenamefont {Gibelin}, \citenamefont {Gomi}, \citenamefont {Hasegawa},
  \citenamefont {Imai}, \citenamefont {Ishihara}, \citenamefont {Iwasaki},
  \citenamefont {Kanno}, \citenamefont {Kawai}, \citenamefont {Kishida},
  \citenamefont {Kubo}, \citenamefont {Kurita}, \citenamefont {Matsuyama},
  \citenamefont {Michimasa}, \citenamefont {Minemura}, \citenamefont
  {Motobayashi}, \citenamefont {Notani}, \citenamefont {Ohnishi}, \citenamefont
  {Ong}, \citenamefont {Ota}, \citenamefont {Ozawa}, \citenamefont {Sakai},
  \citenamefont {Sakurai}, \citenamefont {Shimoura}, \citenamefont {Takeshita},
  \citenamefont {Takeuchi}, \citenamefont {Tamaki}, \citenamefont {Togano},
  \citenamefont {Yamada}, \citenamefont {Yanagisawa},\ and\ \citenamefont
  {Yoneda}}]{Elekes04}%
  \BibitemOpen
  \bibfield  {author} {\bibinfo {author} {\bibfnamefont {Z.}~\bibnamefont
  {Elekes}}, \bibinfo {author} {\bibfnamefont {Z.}~\bibnamefont {Dombrádi}},
  \bibinfo {author} {\bibfnamefont {A.}~\bibnamefont {Saito}}, \bibinfo
  {author} {\bibfnamefont {N.}~\bibnamefont {Aoi}}, \bibinfo {author}
  {\bibfnamefont {H.}~\bibnamefont {Baba}}, \bibinfo {author} {\bibfnamefont
  {K.}~\bibnamefont {Demichi}}, \bibinfo {author} {\bibfnamefont
  {Z.}~\bibnamefont {Fülöp}}, \bibinfo {author} {\bibfnamefont
  {J.}~\bibnamefont {Gibelin}}, \bibinfo {author} {\bibfnamefont
  {T.}~\bibnamefont {Gomi}}, \bibinfo {author} {\bibfnamefont {H.}~\bibnamefont
  {Hasegawa}}, \bibinfo {author} {\bibfnamefont {N.}~\bibnamefont {Imai}},
  \bibinfo {author} {\bibfnamefont {M.}~\bibnamefont {Ishihara}}, \bibinfo
  {author} {\bibfnamefont {H.}~\bibnamefont {Iwasaki}}, \bibinfo {author}
  {\bibfnamefont {S.}~\bibnamefont {Kanno}}, \bibinfo {author} {\bibfnamefont
  {S.}~\bibnamefont {Kawai}}, \bibinfo {author} {\bibfnamefont
  {T.}~\bibnamefont {Kishida}}, \bibinfo {author} {\bibfnamefont
  {T.}~\bibnamefont {Kubo}}, \bibinfo {author} {\bibfnamefont {K.}~\bibnamefont
  {Kurita}}, \bibinfo {author} {\bibfnamefont {Y.}~\bibnamefont {Matsuyama}},
  \bibinfo {author} {\bibfnamefont {S.}~\bibnamefont {Michimasa}}, \bibinfo
  {author} {\bibfnamefont {T.}~\bibnamefont {Minemura}}, \bibinfo {author}
  {\bibfnamefont {T.}~\bibnamefont {Motobayashi}}, \bibinfo {author}
  {\bibfnamefont {M.}~\bibnamefont {Notani}}, \bibinfo {author} {\bibfnamefont
  {T.}~\bibnamefont {Ohnishi}}, \bibinfo {author} {\bibfnamefont
  {H.}~\bibnamefont {Ong}}, \bibinfo {author} {\bibfnamefont {S.}~\bibnamefont
  {Ota}}, \bibinfo {author} {\bibfnamefont {A.}~\bibnamefont {Ozawa}}, \bibinfo
  {author} {\bibfnamefont {H.}~\bibnamefont {Sakai}}, \bibinfo {author}
  {\bibfnamefont {H.}~\bibnamefont {Sakurai}}, \bibinfo {author} {\bibfnamefont
  {S.}~\bibnamefont {Shimoura}}, \bibinfo {author} {\bibfnamefont
  {E.}~\bibnamefont {Takeshita}}, \bibinfo {author} {\bibfnamefont
  {S.}~\bibnamefont {Takeuchi}}, \bibinfo {author} {\bibfnamefont
  {M.}~\bibnamefont {Tamaki}}, \bibinfo {author} {\bibfnamefont
  {Y.}~\bibnamefont {Togano}}, \bibinfo {author} {\bibfnamefont
  {K.}~\bibnamefont {Yamada}}, \bibinfo {author} {\bibfnamefont
  {Y.}~\bibnamefont {Yanagisawa}}, \ and\ \bibinfo {author} {\bibfnamefont
  {K.}~\bibnamefont {Yoneda}},\ }\href {\doibase
  https://doi.org/10.1016/j.physletb.2004.08.028} {\bibfield  {journal}
  {\bibinfo  {journal} {Phys. Lett. B}\ }\textbf {\bibinfo {volume} {599}},\
  \bibinfo {pages} {17 } (\bibinfo {year} {2004})}\BibitemShut {NoStop}%
\bibitem [{\citenamefont {Christian}\ \emph
  {et~al.}(2012{\natexlab{a}})\citenamefont {Christian}, \citenamefont {Frank},
  \citenamefont {Ash}, \citenamefont {Baumann}, \citenamefont {Bazin},
  \citenamefont {Brown}, \citenamefont {DeYoung}, \citenamefont {Finck},
  \citenamefont {Gade}, \citenamefont {Grinyer}, \citenamefont {Grovom},
  \citenamefont {Hinnefeld}, \citenamefont {Lunderberg}, \citenamefont
  {Luther}, \citenamefont {Mosby}, \citenamefont {Mosby}, \citenamefont {Nagi},
  \citenamefont {Peaslee}, \citenamefont {Rogers}, \citenamefont {Smith},
  \citenamefont {Snyder}, \citenamefont {Spyrou}, \citenamefont {Strongman},
  \citenamefont {Thoennessen}, \citenamefont {Warren}, \citenamefont
  {Weisshaar},\ and\ \citenamefont {Wersal}}]{CHRIS1}%
  \BibitemOpen
  \bibfield  {author} {\bibinfo {author} {\bibfnamefont {G.}~\bibnamefont
  {Christian}}, \bibinfo {author} {\bibfnamefont {N.}~\bibnamefont {Frank}},
  \bibinfo {author} {\bibfnamefont {S.}~\bibnamefont {Ash}}, \bibinfo {author}
  {\bibfnamefont {T.}~\bibnamefont {Baumann}}, \bibinfo {author} {\bibfnamefont
  {D.}~\bibnamefont {Bazin}}, \bibinfo {author} {\bibfnamefont
  {J.}~\bibnamefont {Brown}}, \bibinfo {author} {\bibfnamefont {P.~A.}\
  \bibnamefont {DeYoung}}, \bibinfo {author} {\bibfnamefont {J.~E.}\
  \bibnamefont {Finck}}, \bibinfo {author} {\bibfnamefont {A.}~\bibnamefont
  {Gade}}, \bibinfo {author} {\bibfnamefont {G.~F.}\ \bibnamefont {Grinyer}},
  \bibinfo {author} {\bibfnamefont {A.}~\bibnamefont {Grovom}}, \bibinfo
  {author} {\bibfnamefont {J.~D.}\ \bibnamefont {Hinnefeld}}, \bibinfo {author}
  {\bibfnamefont {E.~M.}\ \bibnamefont {Lunderberg}}, \bibinfo {author}
  {\bibfnamefont {B.}~\bibnamefont {Luther}}, \bibinfo {author} {\bibfnamefont
  {M.}~\bibnamefont {Mosby}}, \bibinfo {author} {\bibfnamefont
  {S.}~\bibnamefont {Mosby}}, \bibinfo {author} {\bibfnamefont
  {T.}~\bibnamefont {Nagi}}, \bibinfo {author} {\bibfnamefont {G.~F.}\
  \bibnamefont {Peaslee}}, \bibinfo {author} {\bibfnamefont {W.~F.}\
  \bibnamefont {Rogers}}, \bibinfo {author} {\bibfnamefont {J.~K.}\
  \bibnamefont {Smith}}, \bibinfo {author} {\bibfnamefont {J.}~\bibnamefont
  {Snyder}}, \bibinfo {author} {\bibfnamefont {A.}~\bibnamefont {Spyrou}},
  \bibinfo {author} {\bibfnamefont {M.~J.}\ \bibnamefont {Strongman}}, \bibinfo
  {author} {\bibfnamefont {M.}~\bibnamefont {Thoennessen}}, \bibinfo {author}
  {\bibfnamefont {M.}~\bibnamefont {Warren}}, \bibinfo {author} {\bibfnamefont
  {D.}~\bibnamefont {Weisshaar}}, \ and\ \bibinfo {author} {\bibfnamefont
  {A.}~\bibnamefont {Wersal}},\ }\href {\doibase
  10.1103/PhysRevLett.108.032501} {\bibfield  {journal} {\bibinfo  {journal}
  {Phys. Rev. Lett.}\ }\textbf {\bibinfo {volume} {108}},\ \bibinfo {pages}
  {032501} (\bibinfo {year} {2012}{\natexlab{a}})}\BibitemShut {NoStop}%
\bibitem [{\citenamefont {Christian}\ \emph
  {et~al.}(2012{\natexlab{b}})\citenamefont {Christian}, \citenamefont {Frank},
  \citenamefont {Ash}, \citenamefont {Baumann}, \citenamefont {DeYoung},
  \citenamefont {Finck}, \citenamefont {Gade}, \citenamefont {Grinyer},
  \citenamefont {Luther}, \citenamefont {Mosby}, \citenamefont {Mosby},
  \citenamefont {Smith}, \citenamefont {Snyder}, \citenamefont {Spyrou},
  \citenamefont {Strongman}, \citenamefont {Thoennessen}, \citenamefont
  {Warren}, \citenamefont {Weisshaar},\ and\ \citenamefont {Wersal}}]{CHRIS2}%
  \BibitemOpen
  \bibfield  {author} {\bibinfo {author} {\bibfnamefont {G.}~\bibnamefont
  {Christian}}, \bibinfo {author} {\bibfnamefont {N.}~\bibnamefont {Frank}},
  \bibinfo {author} {\bibfnamefont {S.}~\bibnamefont {Ash}}, \bibinfo {author}
  {\bibfnamefont {T.}~\bibnamefont {Baumann}}, \bibinfo {author} {\bibfnamefont
  {P.~A.}\ \bibnamefont {DeYoung}}, \bibinfo {author} {\bibfnamefont {J.~E.}\
  \bibnamefont {Finck}}, \bibinfo {author} {\bibfnamefont {A.}~\bibnamefont
  {Gade}}, \bibinfo {author} {\bibfnamefont {G.~F.}\ \bibnamefont {Grinyer}},
  \bibinfo {author} {\bibfnamefont {B.}~\bibnamefont {Luther}}, \bibinfo
  {author} {\bibfnamefont {M.}~\bibnamefont {Mosby}}, \bibinfo {author}
  {\bibfnamefont {S.}~\bibnamefont {Mosby}}, \bibinfo {author} {\bibfnamefont
  {J.~K.}\ \bibnamefont {Smith}}, \bibinfo {author} {\bibfnamefont
  {J.}~\bibnamefont {Snyder}}, \bibinfo {author} {\bibfnamefont
  {A.}~\bibnamefont {Spyrou}}, \bibinfo {author} {\bibfnamefont {M.~J.}\
  \bibnamefont {Strongman}}, \bibinfo {author} {\bibfnamefont {M.}~\bibnamefont
  {Thoennessen}}, \bibinfo {author} {\bibfnamefont {M.}~\bibnamefont {Warren}},
  \bibinfo {author} {\bibfnamefont {D.}~\bibnamefont {Weisshaar}}, \ and\
  \bibinfo {author} {\bibfnamefont {A.}~\bibnamefont {Wersal}},\ }\href
  {\doibase 10.1103/PhysRevC.85.034327} {\bibfield  {journal} {\bibinfo
  {journal} {Phys. Rev. C}\ }\textbf {\bibinfo {volume} {85}},\ \bibinfo
  {pages} {034327} (\bibinfo {year} {2012}{\natexlab{b}})}\BibitemShut
  {NoStop}%
\bibitem [{\citenamefont {Doornenbal}\ \emph {et~al.}(2017)\citenamefont
  {Doornenbal}, \citenamefont {Scheit}, \citenamefont {Takeuchi}, \citenamefont
  {Utsuno}, \citenamefont {Aoi}, \citenamefont {Li}, \citenamefont
  {Matsushita}, \citenamefont {Steppenbeck}, \citenamefont {Wang},
  \citenamefont {Baba}, \citenamefont {Ideguchi}, \citenamefont {Kobayashi},
  \citenamefont {Kondo}, \citenamefont {Lee}, \citenamefont {Michimasa},
  \citenamefont {Motobayashi}, \citenamefont {Otsuka}, \citenamefont {Sakurai},
  \citenamefont {Takechi}, \citenamefont {Togano},\ and\ \citenamefont
  {Yoneda}}]{DOOR2017}%
  \BibitemOpen
  \bibfield  {author} {\bibinfo {author} {\bibfnamefont {P.}~\bibnamefont
  {Doornenbal}}, \bibinfo {author} {\bibfnamefont {H.}~\bibnamefont {Scheit}},
  \bibinfo {author} {\bibfnamefont {S.}~\bibnamefont {Takeuchi}}, \bibinfo
  {author} {\bibfnamefont {Y.}~\bibnamefont {Utsuno}}, \bibinfo {author}
  {\bibfnamefont {N.}~\bibnamefont {Aoi}}, \bibinfo {author} {\bibfnamefont
  {K.}~\bibnamefont {Li}}, \bibinfo {author} {\bibfnamefont {M.}~\bibnamefont
  {Matsushita}}, \bibinfo {author} {\bibfnamefont {D.}~\bibnamefont
  {Steppenbeck}}, \bibinfo {author} {\bibfnamefont {H.}~\bibnamefont {Wang}},
  \bibinfo {author} {\bibfnamefont {H.}~\bibnamefont {Baba}}, \bibinfo {author}
  {\bibfnamefont {E.}~\bibnamefont {Ideguchi}}, \bibinfo {author}
  {\bibfnamefont {N.}~\bibnamefont {Kobayashi}}, \bibinfo {author}
  {\bibfnamefont {Y.}~\bibnamefont {Kondo}}, \bibinfo {author} {\bibfnamefont
  {J.}~\bibnamefont {Lee}}, \bibinfo {author} {\bibfnamefont {S.}~\bibnamefont
  {Michimasa}}, \bibinfo {author} {\bibfnamefont {T.}~\bibnamefont
  {Motobayashi}}, \bibinfo {author} {\bibfnamefont {T.}~\bibnamefont {Otsuka}},
  \bibinfo {author} {\bibfnamefont {H.}~\bibnamefont {Sakurai}}, \bibinfo
  {author} {\bibfnamefont {M.}~\bibnamefont {Takechi}}, \bibinfo {author}
  {\bibfnamefont {Y.}~\bibnamefont {Togano}}, \ and\ \bibinfo {author}
  {\bibfnamefont {K.}~\bibnamefont {Yoneda}},\ }\href {\doibase
  10.1103/PhysRevC.95.041301} {\bibfield  {journal} {\bibinfo  {journal} {Phys.
  Rev. C}\ }\textbf {\bibinfo {volume} {95}},\ \bibinfo {pages} {041301}
  (\bibinfo {year} {2017})}\BibitemShut {NoStop}%
\bibitem [{\citenamefont {Singh}\ \emph {et~al.}(2020)\citenamefont {Singh},
  \citenamefont {Casal}, \citenamefont {Horiuchi}, \citenamefont {Fortunato},\
  and\ \citenamefont {Vitturi}}]{Singh2020}%
  \BibitemOpen
  \bibfield  {author} {\bibinfo {author} {\bibfnamefont {J.}~\bibnamefont
  {Singh}}, \bibinfo {author} {\bibfnamefont {J.}~\bibnamefont {Casal}},
  \bibinfo {author} {\bibfnamefont {W.}~\bibnamefont {Horiuchi}}, \bibinfo
  {author} {\bibfnamefont {L.}~\bibnamefont {Fortunato}}, \ and\ \bibinfo
  {author} {\bibfnamefont {A.}~\bibnamefont {Vitturi}},\ }\href {\doibase
  10.1103/PhysRevC.101.024310} {\bibfield  {journal} {\bibinfo  {journal}
  {Phys. Rev. C}\ }\textbf {\bibinfo {volume} {101}},\ \bibinfo {pages}
  {024310} (\bibinfo {year} {2020})}\BibitemShut {NoStop}%
\bibitem [{\citenamefont {Revel}\ \emph {et~al.}(2020)\citenamefont {Revel},
  \citenamefont {Sorlin}, \citenamefont {Marqu\'es}, \citenamefont {Kondo},
  \citenamefont {Kahlbow}, \citenamefont {Nakamura}, \citenamefont {Orr},
  \citenamefont {Nowacki}, \citenamefont {Tostevin}, \citenamefont {Yuan},
  \citenamefont {Achouri}, \citenamefont {Al~Falou}, \citenamefont {Atar},
  \citenamefont {Aumann}, \citenamefont {Baba}, \citenamefont {Boretzky},
  \citenamefont {Caesar}, \citenamefont {Calvet}, \citenamefont {Chae},
  \citenamefont {Chiga}, \citenamefont {Corsi}, \citenamefont {Crawford},
  \citenamefont {Delaunay}, \citenamefont {Delbart}, \citenamefont {Deshayes},
  \citenamefont {Dombr\'adi}, \citenamefont {Douma}, \citenamefont {Elekes},
  \citenamefont {Fallon}, \citenamefont {Ga\ifmmode \check{s}\else
  \v{s}\fi{}pari\ifmmode~\acute{c}\else \'{c}\fi{}}, \citenamefont {Gheller},
  \citenamefont {Gibelin}, \citenamefont {Gillibert}, \citenamefont {Harakeh},
  \citenamefont {He}, \citenamefont {Hirayama}, \citenamefont {Hoffman},
  \citenamefont {Holl}, \citenamefont {Horvat}, \citenamefont {Horv\'ath},
  \citenamefont {Hwang}, \citenamefont {Isobe}, \citenamefont
  {Kalantar-Nayestanaki}, \citenamefont {Kawase}, \citenamefont {Kim},
  \citenamefont {Kisamori}, \citenamefont {Kobayashi}, \citenamefont
  {K\"orper}, \citenamefont {Koyama}, \citenamefont {Kuti}, \citenamefont
  {Lapoux}, \citenamefont {Lindberg}, \citenamefont {Masuoka}, \citenamefont
  {Mayer}, \citenamefont {Miki}, \citenamefont {Murakami}, \citenamefont
  {Najafi}, \citenamefont {Nakano}, \citenamefont {Nakatsuka}, \citenamefont
  {Nilsson}, \citenamefont {Obertelli}, \citenamefont {de~Oliveira~Santos},
  \citenamefont {Otsu}, \citenamefont {Ozaki}, \citenamefont {Panin},
  \citenamefont {Paschalis}, \citenamefont {Rossi}, \citenamefont {Saito},
  \citenamefont {Saito}, \citenamefont {Sasano}, \citenamefont {Sato},
  \citenamefont {Satou}, \citenamefont {Scheit}, \citenamefont {Schindler},
  \citenamefont {Schrock}, \citenamefont {Shikata}, \citenamefont {Shimizu},
  \citenamefont {Simon}, \citenamefont {Sohler}, \citenamefont {Stuhl},
  \citenamefont {Takeuchi}, \citenamefont {Tanaka}, \citenamefont
  {Thoennessen}, \citenamefont {T\"ornqvist}, \citenamefont {Togano},
  \citenamefont {Tomai}, \citenamefont {Tscheuschner}, \citenamefont {Tsubota},
  \citenamefont {Uesaka}, \citenamefont {Yang}, \citenamefont {Yasuda},\ and\
  \citenamefont {Yoneda}}]{Revel2020}%
  \BibitemOpen
  \bibfield  {author} {\bibinfo {author} {\bibfnamefont {A.}~\bibnamefont
  {Revel}}, \bibinfo {author} {\bibfnamefont {O.}~\bibnamefont {Sorlin}},
  \bibinfo {author} {\bibfnamefont {F.~M.}\ \bibnamefont {Marqu\'es}}, \bibinfo
  {author} {\bibfnamefont {Y.}~\bibnamefont {Kondo}}, \bibinfo {author}
  {\bibfnamefont {J.}~\bibnamefont {Kahlbow}}, \bibinfo {author} {\bibfnamefont
  {T.}~\bibnamefont {Nakamura}}, \bibinfo {author} {\bibfnamefont {N.~A.}\
  \bibnamefont {Orr}}, \bibinfo {author} {\bibfnamefont {F.}~\bibnamefont
  {Nowacki}}, \bibinfo {author} {\bibfnamefont {J.~A.}\ \bibnamefont
  {Tostevin}}, \bibinfo {author} {\bibfnamefont {C.~X.}\ \bibnamefont {Yuan}},
  \bibinfo {author} {\bibfnamefont {N.~L.}\ \bibnamefont {Achouri}}, \bibinfo
  {author} {\bibfnamefont {H.}~\bibnamefont {Al~Falou}}, \bibinfo {author}
  {\bibfnamefont {L.}~\bibnamefont {Atar}}, \bibinfo {author} {\bibfnamefont
  {T.}~\bibnamefont {Aumann}}, \bibinfo {author} {\bibfnamefont
  {H.}~\bibnamefont {Baba}}, \bibinfo {author} {\bibfnamefont {K.}~\bibnamefont
  {Boretzky}}, \bibinfo {author} {\bibfnamefont {C.}~\bibnamefont {Caesar}},
  \bibinfo {author} {\bibfnamefont {D.}~\bibnamefont {Calvet}}, \bibinfo
  {author} {\bibfnamefont {H.}~\bibnamefont {Chae}}, \bibinfo {author}
  {\bibfnamefont {N.}~\bibnamefont {Chiga}}, \bibinfo {author} {\bibfnamefont
  {A.}~\bibnamefont {Corsi}}, \bibinfo {author} {\bibfnamefont {H.~L.}\
  \bibnamefont {Crawford}}, \bibinfo {author} {\bibfnamefont {F.}~\bibnamefont
  {Delaunay}}, \bibinfo {author} {\bibfnamefont {A.}~\bibnamefont {Delbart}},
  \bibinfo {author} {\bibfnamefont {Q.}~\bibnamefont {Deshayes}}, \bibinfo
  {author} {\bibfnamefont {Z.}~\bibnamefont {Dombr\'adi}}, \bibinfo {author}
  {\bibfnamefont {C.~A.}\ \bibnamefont {Douma}}, \bibinfo {author}
  {\bibfnamefont {Z.}~\bibnamefont {Elekes}}, \bibinfo {author} {\bibfnamefont
  {P.}~\bibnamefont {Fallon}}, \bibinfo {author} {\bibfnamefont
  {I.}~\bibnamefont {Ga\ifmmode \check{s}\else
  \v{s}\fi{}pari\ifmmode~\acute{c}\else \'{c}\fi{}}}, \bibinfo {author}
  {\bibfnamefont {J.-M.}\ \bibnamefont {Gheller}}, \bibinfo {author}
  {\bibfnamefont {J.}~\bibnamefont {Gibelin}}, \bibinfo {author} {\bibfnamefont
  {A.}~\bibnamefont {Gillibert}}, \bibinfo {author} {\bibfnamefont {M.~N.}\
  \bibnamefont {Harakeh}}, \bibinfo {author} {\bibfnamefont {W.}~\bibnamefont
  {He}}, \bibinfo {author} {\bibfnamefont {A.}~\bibnamefont {Hirayama}},
  \bibinfo {author} {\bibfnamefont {C.~R.}\ \bibnamefont {Hoffman}}, \bibinfo
  {author} {\bibfnamefont {M.}~\bibnamefont {Holl}}, \bibinfo {author}
  {\bibfnamefont {A.}~\bibnamefont {Horvat}}, \bibinfo {author} {\bibfnamefont
  {A.}~\bibnamefont {Horv\'ath}}, \bibinfo {author} {\bibfnamefont {J.~W.}\
  \bibnamefont {Hwang}}, \bibinfo {author} {\bibfnamefont {T.}~\bibnamefont
  {Isobe}}, \bibinfo {author} {\bibfnamefont {N.}~\bibnamefont
  {Kalantar-Nayestanaki}}, \bibinfo {author} {\bibfnamefont {S.}~\bibnamefont
  {Kawase}}, \bibinfo {author} {\bibfnamefont {S.}~\bibnamefont {Kim}},
  \bibinfo {author} {\bibfnamefont {K.}~\bibnamefont {Kisamori}}, \bibinfo
  {author} {\bibfnamefont {T.}~\bibnamefont {Kobayashi}}, \bibinfo {author}
  {\bibfnamefont {D.}~\bibnamefont {K\"orper}}, \bibinfo {author}
  {\bibfnamefont {S.}~\bibnamefont {Koyama}}, \bibinfo {author} {\bibfnamefont
  {I.}~\bibnamefont {Kuti}}, \bibinfo {author} {\bibfnamefont {V.}~\bibnamefont
  {Lapoux}}, \bibinfo {author} {\bibfnamefont {S.}~\bibnamefont {Lindberg}},
  \bibinfo {author} {\bibfnamefont {S.}~\bibnamefont {Masuoka}}, \bibinfo
  {author} {\bibfnamefont {J.}~\bibnamefont {Mayer}}, \bibinfo {author}
  {\bibfnamefont {K.}~\bibnamefont {Miki}}, \bibinfo {author} {\bibfnamefont
  {T.}~\bibnamefont {Murakami}}, \bibinfo {author} {\bibfnamefont
  {M.}~\bibnamefont {Najafi}}, \bibinfo {author} {\bibfnamefont
  {K.}~\bibnamefont {Nakano}}, \bibinfo {author} {\bibfnamefont
  {N.}~\bibnamefont {Nakatsuka}}, \bibinfo {author} {\bibfnamefont
  {T.}~\bibnamefont {Nilsson}}, \bibinfo {author} {\bibfnamefont
  {A.}~\bibnamefont {Obertelli}}, \bibinfo {author} {\bibfnamefont
  {F.}~\bibnamefont {de~Oliveira~Santos}}, \bibinfo {author} {\bibfnamefont
  {H.}~\bibnamefont {Otsu}}, \bibinfo {author} {\bibfnamefont {T.}~\bibnamefont
  {Ozaki}}, \bibinfo {author} {\bibfnamefont {V.}~\bibnamefont {Panin}},
  \bibinfo {author} {\bibfnamefont {S.}~\bibnamefont {Paschalis}}, \bibinfo
  {author} {\bibfnamefont {D.}~\bibnamefont {Rossi}}, \bibinfo {author}
  {\bibfnamefont {A.~T.}\ \bibnamefont {Saito}}, \bibinfo {author}
  {\bibfnamefont {T.}~\bibnamefont {Saito}}, \bibinfo {author} {\bibfnamefont
  {M.}~\bibnamefont {Sasano}}, \bibinfo {author} {\bibfnamefont
  {H.}~\bibnamefont {Sato}}, \bibinfo {author} {\bibfnamefont {Y.}~\bibnamefont
  {Satou}}, \bibinfo {author} {\bibfnamefont {H.}~\bibnamefont {Scheit}},
  \bibinfo {author} {\bibfnamefont {F.}~\bibnamefont {Schindler}}, \bibinfo
  {author} {\bibfnamefont {P.}~\bibnamefont {Schrock}}, \bibinfo {author}
  {\bibfnamefont {M.}~\bibnamefont {Shikata}}, \bibinfo {author} {\bibfnamefont
  {Y.}~\bibnamefont {Shimizu}}, \bibinfo {author} {\bibfnamefont
  {H.}~\bibnamefont {Simon}}, \bibinfo {author} {\bibfnamefont
  {D.}~\bibnamefont {Sohler}}, \bibinfo {author} {\bibfnamefont
  {L.}~\bibnamefont {Stuhl}}, \bibinfo {author} {\bibfnamefont
  {S.}~\bibnamefont {Takeuchi}}, \bibinfo {author} {\bibfnamefont
  {M.}~\bibnamefont {Tanaka}}, \bibinfo {author} {\bibfnamefont
  {M.}~\bibnamefont {Thoennessen}}, \bibinfo {author} {\bibfnamefont
  {H.}~\bibnamefont {T\"ornqvist}}, \bibinfo {author} {\bibfnamefont
  {Y.}~\bibnamefont {Togano}}, \bibinfo {author} {\bibfnamefont
  {T.}~\bibnamefont {Tomai}}, \bibinfo {author} {\bibfnamefont
  {J.}~\bibnamefont {Tscheuschner}}, \bibinfo {author} {\bibfnamefont
  {J.}~\bibnamefont {Tsubota}}, \bibinfo {author} {\bibfnamefont
  {T.}~\bibnamefont {Uesaka}}, \bibinfo {author} {\bibfnamefont
  {Z.}~\bibnamefont {Yang}}, \bibinfo {author} {\bibfnamefont {M.}~\bibnamefont
  {Yasuda}}, \ and\ \bibinfo {author} {\bibfnamefont {K.}~\bibnamefont
  {Yoneda}} (\bibinfo {collaboration} {SAMURAI21 collaboration}),\ }\href
  {\doibase 10.1103/PhysRevLett.124.152502} {\bibfield  {journal} {\bibinfo
  {journal} {Phys. Rev. Lett.}\ }\textbf {\bibinfo {volume} {124}},\ \bibinfo
  {pages} {152502} (\bibinfo {year} {2020})}\BibitemShut {NoStop}%
\bibitem [{\citenamefont {Bagchi}\ \emph {et~al.}(2020)\citenamefont {Bagchi},
  \citenamefont {Kanungo}, \citenamefont {Tanaka}, \citenamefont {Geissel},
  \citenamefont {Doornenbal}, \citenamefont {Horiuchi}, \citenamefont {Hagen},
  \citenamefont {Suzuki}, \citenamefont {Tsunoda}, \citenamefont {Ahn},
  \citenamefont {Baba}, \citenamefont {Behr}, \citenamefont {Browne},
  \citenamefont {Chen}, \citenamefont {Cort\'es}, \citenamefont {Estrad\'e},
  \citenamefont {Fukuda}, \citenamefont {Holl}, \citenamefont {Itahashi},
  \citenamefont {Iwasa}, \citenamefont {Jansen}, \citenamefont {Jiang},
  \citenamefont {Kaur}, \citenamefont {Macchiavelli}, \citenamefont
  {Matsumoto}, \citenamefont {Momiyama}, \citenamefont {Murray}, \citenamefont
  {Nakamura}, \citenamefont {Novario}, \citenamefont {Ong}, \citenamefont
  {Otsuka}, \citenamefont {Papenbrock}, \citenamefont {Paschalis},
  \citenamefont {Prochazka}, \citenamefont {Scheidenberger}, \citenamefont
  {Schrock}, \citenamefont {Shimizu}, \citenamefont {Steppenbeck},
  \citenamefont {Sakurai}, \citenamefont {Suzuki}, \citenamefont {Suzuki},
  \citenamefont {Takechi}, \citenamefont {Takeda}, \citenamefont {Takeuchi},
  \citenamefont {Taniuchi}, \citenamefont {Wimmer},\ and\ \citenamefont
  {Yoshida}}]{Bagchi2020}%
  \BibitemOpen
  \bibfield  {author} {\bibinfo {author} {\bibfnamefont {S.}~\bibnamefont
  {Bagchi}}, \bibinfo {author} {\bibfnamefont {R.}~\bibnamefont {Kanungo}},
  \bibinfo {author} {\bibfnamefont {Y.~K.}\ \bibnamefont {Tanaka}}, \bibinfo
  {author} {\bibfnamefont {H.}~\bibnamefont {Geissel}}, \bibinfo {author}
  {\bibfnamefont {P.}~\bibnamefont {Doornenbal}}, \bibinfo {author}
  {\bibfnamefont {W.}~\bibnamefont {Horiuchi}}, \bibinfo {author}
  {\bibfnamefont {G.}~\bibnamefont {Hagen}}, \bibinfo {author} {\bibfnamefont
  {T.}~\bibnamefont {Suzuki}}, \bibinfo {author} {\bibfnamefont
  {N.}~\bibnamefont {Tsunoda}}, \bibinfo {author} {\bibfnamefont {D.~S.}\
  \bibnamefont {Ahn}}, \bibinfo {author} {\bibfnamefont {H.}~\bibnamefont
  {Baba}}, \bibinfo {author} {\bibfnamefont {K.}~\bibnamefont {Behr}}, \bibinfo
  {author} {\bibfnamefont {F.}~\bibnamefont {Browne}}, \bibinfo {author}
  {\bibfnamefont {S.}~\bibnamefont {Chen}}, \bibinfo {author} {\bibfnamefont
  {M.~L.}\ \bibnamefont {Cort\'es}}, \bibinfo {author} {\bibfnamefont
  {A.}~\bibnamefont {Estrad\'e}}, \bibinfo {author} {\bibfnamefont
  {N.}~\bibnamefont {Fukuda}}, \bibinfo {author} {\bibfnamefont
  {M.}~\bibnamefont {Holl}}, \bibinfo {author} {\bibfnamefont {K.}~\bibnamefont
  {Itahashi}}, \bibinfo {author} {\bibfnamefont {N.}~\bibnamefont {Iwasa}},
  \bibinfo {author} {\bibfnamefont {G.~R.}\ \bibnamefont {Jansen}}, \bibinfo
  {author} {\bibfnamefont {W.~G.}\ \bibnamefont {Jiang}}, \bibinfo {author}
  {\bibfnamefont {S.}~\bibnamefont {Kaur}}, \bibinfo {author} {\bibfnamefont
  {A.~O.}\ \bibnamefont {Macchiavelli}}, \bibinfo {author} {\bibfnamefont
  {S.~Y.}\ \bibnamefont {Matsumoto}}, \bibinfo {author} {\bibfnamefont
  {S.}~\bibnamefont {Momiyama}}, \bibinfo {author} {\bibfnamefont
  {I.}~\bibnamefont {Murray}}, \bibinfo {author} {\bibfnamefont
  {T.}~\bibnamefont {Nakamura}}, \bibinfo {author} {\bibfnamefont {S.~J.}\
  \bibnamefont {Novario}}, \bibinfo {author} {\bibfnamefont {H.~J.}\
  \bibnamefont {Ong}}, \bibinfo {author} {\bibfnamefont {T.}~\bibnamefont
  {Otsuka}}, \bibinfo {author} {\bibfnamefont {T.}~\bibnamefont {Papenbrock}},
  \bibinfo {author} {\bibfnamefont {S.}~\bibnamefont {Paschalis}}, \bibinfo
  {author} {\bibfnamefont {A.}~\bibnamefont {Prochazka}}, \bibinfo {author}
  {\bibfnamefont {C.}~\bibnamefont {Scheidenberger}}, \bibinfo {author}
  {\bibfnamefont {P.}~\bibnamefont {Schrock}}, \bibinfo {author} {\bibfnamefont
  {Y.}~\bibnamefont {Shimizu}}, \bibinfo {author} {\bibfnamefont
  {D.}~\bibnamefont {Steppenbeck}}, \bibinfo {author} {\bibfnamefont
  {H.}~\bibnamefont {Sakurai}}, \bibinfo {author} {\bibfnamefont
  {D.}~\bibnamefont {Suzuki}}, \bibinfo {author} {\bibfnamefont
  {H.}~\bibnamefont {Suzuki}}, \bibinfo {author} {\bibfnamefont
  {M.}~\bibnamefont {Takechi}}, \bibinfo {author} {\bibfnamefont
  {H.}~\bibnamefont {Takeda}}, \bibinfo {author} {\bibfnamefont
  {S.}~\bibnamefont {Takeuchi}}, \bibinfo {author} {\bibfnamefont
  {R.}~\bibnamefont {Taniuchi}}, \bibinfo {author} {\bibfnamefont
  {K.}~\bibnamefont {Wimmer}}, \ and\ \bibinfo {author} {\bibfnamefont
  {K.}~\bibnamefont {Yoshida}},\ }\href {\doibase
  10.1103/PhysRevLett.124.222504} {\bibfield  {journal} {\bibinfo  {journal}
  {Phys. Rev. Lett.}\ }\textbf {\bibinfo {volume} {124}},\ \bibinfo {pages}
  {222504} (\bibinfo {year} {2020})}\BibitemShut {NoStop}%
\bibitem [{\citenamefont {Rodr\'{\i}guez-Gallardo}\ \emph
  {et~al.}(2008)\citenamefont {Rodr\'{\i}guez-Gallardo}, \citenamefont {Arias},
  \citenamefont {G\'omez-Camacho}, \citenamefont {Johnson}, \citenamefont
  {Moro}, \citenamefont {Thompson},\ and\ \citenamefont {Tostevin}}]{MRoGa08}%
  \BibitemOpen
  \bibfield  {author} {\bibinfo {author} {\bibfnamefont {M.}~\bibnamefont
  {Rodr\'{\i}guez-Gallardo}}, \bibinfo {author} {\bibfnamefont {J.~M.}\
  \bibnamefont {Arias}}, \bibinfo {author} {\bibfnamefont {J.}~\bibnamefont
  {G\'omez-Camacho}}, \bibinfo {author} {\bibfnamefont {R.~C.}\ \bibnamefont
  {Johnson}}, \bibinfo {author} {\bibfnamefont {A.~M.}\ \bibnamefont {Moro}},
  \bibinfo {author} {\bibfnamefont {I.~J.}\ \bibnamefont {Thompson}}, \ and\
  \bibinfo {author} {\bibfnamefont {J.~A.}\ \bibnamefont {Tostevin}},\ }\href
  {\doibase 10.1103/PhysRevC.77.064609} {\bibfield  {journal} {\bibinfo
  {journal} {Phys. Rev. C}\ }\textbf {\bibinfo {volume} {77}},\ \bibinfo
  {pages} {064609} (\bibinfo {year} {2008})}\BibitemShut {NoStop}%
\bibitem [{\citenamefont {Cubero}\ \emph {et~al.}(2012)\citenamefont {Cubero},
  \citenamefont {Fern\'andez-Garc\'{\i}a}, \citenamefont
  {Rodr\'{\i}guez-Gallardo}, \citenamefont {Acosta}, \citenamefont {Alcorta},
  \citenamefont {Alvarez}, \citenamefont {Borge}, \citenamefont {Buchmann},
  \citenamefont {Diget}, \citenamefont {Falou}, \citenamefont {Fulton},
  \citenamefont {Fynbo}, \citenamefont {Galaviz}, \citenamefont
  {G\'omez-Camacho}, \citenamefont {Kanungo}, \citenamefont {Lay},
  \citenamefont {Madurga}, \citenamefont {Martel}, \citenamefont {Moro},
  \citenamefont {Mukha}, \citenamefont {Nilsson}, \citenamefont
  {S\'anchez-Ben\'{\i}tez}, \citenamefont {Shotter}, \citenamefont {Tengblad},\
  and\ \citenamefont {Walden}}]{Cubero12}%
  \BibitemOpen
  \bibfield  {author} {\bibinfo {author} {\bibfnamefont {M.}~\bibnamefont
  {Cubero}}, \bibinfo {author} {\bibfnamefont {J.~P.}\ \bibnamefont
  {Fern\'andez-Garc\'{\i}a}}, \bibinfo {author} {\bibfnamefont
  {M.}~\bibnamefont {Rodr\'{\i}guez-Gallardo}}, \bibinfo {author}
  {\bibfnamefont {L.}~\bibnamefont {Acosta}}, \bibinfo {author} {\bibfnamefont
  {M.}~\bibnamefont {Alcorta}}, \bibinfo {author} {\bibfnamefont {M.~A.~G.}\
  \bibnamefont {Alvarez}}, \bibinfo {author} {\bibfnamefont {M.~J.~G.}\
  \bibnamefont {Borge}}, \bibinfo {author} {\bibfnamefont {L.}~\bibnamefont
  {Buchmann}}, \bibinfo {author} {\bibfnamefont {C.~A.}\ \bibnamefont {Diget}},
  \bibinfo {author} {\bibfnamefont {H.~A.}\ \bibnamefont {Falou}}, \bibinfo
  {author} {\bibfnamefont {B.~R.}\ \bibnamefont {Fulton}}, \bibinfo {author}
  {\bibfnamefont {H.~O.~U.}\ \bibnamefont {Fynbo}}, \bibinfo {author}
  {\bibfnamefont {D.}~\bibnamefont {Galaviz}}, \bibinfo {author} {\bibfnamefont
  {J.}~\bibnamefont {G\'omez-Camacho}}, \bibinfo {author} {\bibfnamefont
  {R.}~\bibnamefont {Kanungo}}, \bibinfo {author} {\bibfnamefont {J.~A.}\
  \bibnamefont {Lay}}, \bibinfo {author} {\bibfnamefont {M.}~\bibnamefont
  {Madurga}}, \bibinfo {author} {\bibfnamefont {I.}~\bibnamefont {Martel}},
  \bibinfo {author} {\bibfnamefont {A.~M.}\ \bibnamefont {Moro}}, \bibinfo
  {author} {\bibfnamefont {I.}~\bibnamefont {Mukha}}, \bibinfo {author}
  {\bibfnamefont {T.}~\bibnamefont {Nilsson}}, \bibinfo {author} {\bibfnamefont
  {A.~M.}\ \bibnamefont {S\'anchez-Ben\'{\i}tez}}, \bibinfo {author}
  {\bibfnamefont {A.}~\bibnamefont {Shotter}}, \bibinfo {author} {\bibfnamefont
  {O.}~\bibnamefont {Tengblad}}, \ and\ \bibinfo {author} {\bibfnamefont
  {P.}~\bibnamefont {Walden}},\ }\href {\doibase
  10.1103/PhysRevLett.109.262701} {\bibfield  {journal} {\bibinfo  {journal}
  {Phys. Rev. Lett.}\ }\textbf {\bibinfo {volume} {109}},\ \bibinfo {pages}
  {262701} (\bibinfo {year} {2012})}\BibitemShut {NoStop}%
\bibitem [{\citenamefont {Fortunato}\ \emph {et~al.}(2020)\citenamefont
  {Fortunato}, \citenamefont {Casal}, \citenamefont {Horiuchi}, \citenamefont
  {Singh},\ and\ \citenamefont {Vitturi}}]{Fortunato2020}%
  \BibitemOpen
  \bibfield  {author} {\bibinfo {author} {\bibfnamefont {L.}~\bibnamefont
  {Fortunato}}, \bibinfo {author} {\bibfnamefont {J.}~\bibnamefont {Casal}},
  \bibinfo {author} {\bibfnamefont {W.}~\bibnamefont {Horiuchi}}, \bibinfo
  {author} {\bibfnamefont {J.}~\bibnamefont {Singh}}, \ and\ \bibinfo {author}
  {\bibfnamefont {A.}~\bibnamefont {Vitturi}},\ }\href {\doibase
  10.1038/s42005-020-00402-5} {\bibfield  {journal} {\bibinfo  {journal}
  {Commun. Phys.}\ }\textbf {\bibinfo {volume} {3}},\ \bibinfo {pages} {132}
  (\bibinfo {year} {2020})}\BibitemShut {NoStop}%
\bibitem [{\citenamefont {Nielsen}\ \emph {et~al.}(2001)\citenamefont
  {Nielsen}, \citenamefont {Fedorov}, \citenamefont {Jensen},\ and\
  \citenamefont {Garrido}}]{Nielsen01}%
  \BibitemOpen
  \bibfield  {author} {\bibinfo {author} {\bibfnamefont {E.}~\bibnamefont
  {Nielsen}}, \bibinfo {author} {\bibfnamefont {D.}~\bibnamefont {Fedorov}},
  \bibinfo {author} {\bibfnamefont {A.}~\bibnamefont {Jensen}}, \ and\ \bibinfo
  {author} {\bibfnamefont {E.}~\bibnamefont {Garrido}},\ }\href {\doibase
  https://doi.org/10.1016/S0370-1573(00)00107-1} {\bibfield  {journal}
  {\bibinfo  {journal} {Phys. Rep.}\ }\textbf {\bibinfo {volume} {347}},\
  \bibinfo {pages} {373 } (\bibinfo {year} {2001})}\BibitemShut {NoStop}%
\bibitem [{\citenamefont {Thompson}\ \emph {et~al.}(2004)\citenamefont
  {Thompson}, \citenamefont {Nunes},\ and\ \citenamefont
  {Danilin}}]{IJThompson04}%
  \BibitemOpen
  \bibfield  {author} {\bibinfo {author} {\bibfnamefont {I.}~\bibnamefont
  {Thompson}}, \bibinfo {author} {\bibfnamefont {F.}~\bibnamefont {Nunes}}, \
  and\ \bibinfo {author} {\bibfnamefont {B.}~\bibnamefont {Danilin}},\ }\href
  {\doibase https://doi.org/10.1016/j.cpc.2004.03.007} {\bibfield  {journal}
  {\bibinfo  {journal} {Comput. Phys. Commun.}\ }\textbf {\bibinfo {volume}
  {161}},\ \bibinfo {pages} {87 } (\bibinfo {year} {2004})}\BibitemShut
  {NoStop}%
\bibitem [{\citenamefont {Matsumoto}\ \emph {et~al.}(2019)\citenamefont
  {Matsumoto}, \citenamefont {Tanaka},\ and\ \citenamefont
  {Ogata}}]{Matsumoto19}%
  \BibitemOpen
  \bibfield  {author} {\bibinfo {author} {\bibfnamefont {T.}~\bibnamefont
  {Matsumoto}}, \bibinfo {author} {\bibfnamefont {J.}~\bibnamefont {Tanaka}}, \
  and\ \bibinfo {author} {\bibfnamefont {K.}~\bibnamefont {Ogata}},\ }\href
  {\doibase 10.1093/ptep/ptz126} {\bibfield  {journal} {\bibinfo  {journal}
  {Prog. Theor. Exp. Phys.}\ }\textbf {\bibinfo {volume} {2019}},\ \bibinfo
  {pages} {123D02} (\bibinfo {year} {2019})}\BibitemShut {NoStop}%
\bibitem [{\citenamefont {Tolstikhin}\ \emph {et~al.}(1997)\citenamefont
  {Tolstikhin}, \citenamefont {Ostrovsky},\ and\ \citenamefont
  {Nakamura}}]{Tolstikhin97}%
  \BibitemOpen
  \bibfield  {author} {\bibinfo {author} {\bibfnamefont {O.~I.}\ \bibnamefont
  {Tolstikhin}}, \bibinfo {author} {\bibfnamefont {V.~N.}\ \bibnamefont
  {Ostrovsky}}, \ and\ \bibinfo {author} {\bibfnamefont {H.}~\bibnamefont
  {Nakamura}},\ }\href {\doibase 10.1103/PhysRevLett.79.2026} {\bibfield
  {journal} {\bibinfo  {journal} {Phys. Rev. Lett.}\ }\textbf {\bibinfo
  {volume} {79}},\ \bibinfo {pages} {2026} (\bibinfo {year}
  {1997})}\BibitemShut {NoStop}%
\bibitem [{\citenamefont {Descouvemont}\ \emph {et~al.}(2003)\citenamefont
  {Descouvemont}, \citenamefont {Daniel},\ and\ \citenamefont {Baye}}]{Desc03}%
  \BibitemOpen
  \bibfield  {author} {\bibinfo {author} {\bibfnamefont {P.}~\bibnamefont
  {Descouvemont}}, \bibinfo {author} {\bibfnamefont {C.}~\bibnamefont
  {Daniel}}, \ and\ \bibinfo {author} {\bibfnamefont {D.}~\bibnamefont
  {Baye}},\ }\href {\doibase 10.1103/PhysRevC.67.044309} {\bibfield  {journal}
  {\bibinfo  {journal} {Phys. Rev. C}\ }\textbf {\bibinfo {volume} {67}},\
  \bibinfo {pages} {044309} (\bibinfo {year} {2003})}\BibitemShut {NoStop}%
\bibitem [{\citenamefont {Matsumoto}\ \emph
  {et~al.}(2004{\natexlab{a}})\citenamefont {Matsumoto}, \citenamefont
  {Hiyama}, \citenamefont {Yahiro}, \citenamefont {K.Ogata}, \citenamefont
  {Iseri},\ and\ \citenamefont {Kamimura}}]{Matsumoto04}%
  \BibitemOpen
  \bibfield  {author} {\bibinfo {author} {\bibfnamefont {T.}~\bibnamefont
  {Matsumoto}}, \bibinfo {author} {\bibfnamefont {E.}~\bibnamefont {Hiyama}},
  \bibinfo {author} {\bibfnamefont {M.}~\bibnamefont {Yahiro}}, \bibinfo
  {author} {\bibnamefont {K.Ogata}}, \bibinfo {author} {\bibfnamefont
  {Y.}~\bibnamefont {Iseri}}, \ and\ \bibinfo {author} {\bibfnamefont
  {M.}~\bibnamefont {Kamimura}},\ }\href {\doibase
  10.1016/j.nuclphysa.2004.04.089} {\bibfield  {journal} {\bibinfo  {journal}
  {Nucl. Phys. A}\ }\textbf {\bibinfo {volume} {738}},\ \bibinfo {pages} {471}
  (\bibinfo {year} {2004}{\natexlab{a}})}\BibitemShut {NoStop}%
\bibitem [{\citenamefont {Karataglidis}\ \emph {et~al.}(2005)\citenamefont
  {Karataglidis}, \citenamefont {Amos},\ and\ \citenamefont
  {Giraud}}]{Karataglidis}%
  \BibitemOpen
  \bibfield  {author} {\bibinfo {author} {\bibfnamefont {S.}~\bibnamefont
  {Karataglidis}}, \bibinfo {author} {\bibfnamefont {K.}~\bibnamefont {Amos}},
  \ and\ \bibinfo {author} {\bibfnamefont {B.~G.}\ \bibnamefont {Giraud}},\
  }\href {\doibase 10.1103/PhysRevC.71.064601} {\bibfield  {journal} {\bibinfo
  {journal} {Phys. Rev. C}\ }\textbf {\bibinfo {volume} {71}},\ \bibinfo
  {pages} {064601} (\bibinfo {year} {2005})}\BibitemShut {NoStop}%
\bibitem [{\citenamefont {Casal}(2018)}]{JCasal18}%
  \BibitemOpen
  \bibfield  {author} {\bibinfo {author} {\bibfnamefont {J.}~\bibnamefont
  {Casal}},\ }\href {\doibase 10.1103/PhysRevC.97.034613} {\bibfield  {journal}
  {\bibinfo  {journal} {Phys. Rev. C}\ }\textbf {\bibinfo {volume} {97}},\
  \bibinfo {pages} {034613} (\bibinfo {year} {2018})}\BibitemShut {NoStop}%
\bibitem [{\citenamefont {Gogny}\ \emph {et~al.}(1970)\citenamefont {Gogny},
  \citenamefont {Pires},\ and\ \citenamefont {Tourreil}}]{GPT}%
  \BibitemOpen
  \bibfield  {author} {\bibinfo {author} {\bibfnamefont {D.}~\bibnamefont
  {Gogny}}, \bibinfo {author} {\bibfnamefont {P.}~\bibnamefont {Pires}}, \ and\
  \bibinfo {author} {\bibfnamefont {R.~D.}\ \bibnamefont {Tourreil}},\ }\href
  {\doibase https://doi.org/10.1016/0370-2693(70)90552-6} {\bibfield  {journal}
  {\bibinfo  {journal} {Phys. Lett. B}\ }\textbf {\bibinfo {volume} {32}},\
  \bibinfo {pages} {591 } (\bibinfo {year} {1970})}\BibitemShut {NoStop}%
\bibitem [{\citenamefont {Garrido}\ \emph {et~al.}(2004)\citenamefont
  {Garrido}, \citenamefont {Fedorov},\ and\ \citenamefont
  {Jensen}}]{Garrido04}%
  \BibitemOpen
  \bibfield  {author} {\bibinfo {author} {\bibfnamefont {E.}~\bibnamefont
  {Garrido}}, \bibinfo {author} {\bibfnamefont {D.~V.}\ \bibnamefont
  {Fedorov}}, \ and\ \bibinfo {author} {\bibfnamefont {A.~S.}\ \bibnamefont
  {Jensen}},\ }\href {\doibase 10.1016/j.nuclphysa.2003.12.016} {\bibfield
  {journal} {\bibinfo  {journal} {Nucl. Phys. A}\ }\textbf {\bibinfo {volume}
  {733}},\ \bibinfo {pages} {85} (\bibinfo {year} {2004})}\BibitemShut
  {NoStop}%
\bibitem [{\citenamefont {Wiringa}\ \emph {et~al.}(1995)\citenamefont
  {Wiringa}, \citenamefont {Stoks},\ and\ \citenamefont {Schiavilla}}]{av18}%
  \BibitemOpen
  \bibfield  {author} {\bibinfo {author} {\bibfnamefont {R.~B.}\ \bibnamefont
  {Wiringa}}, \bibinfo {author} {\bibfnamefont {V.~G.~J.}\ \bibnamefont
  {Stoks}}, \ and\ \bibinfo {author} {\bibfnamefont {R.}~\bibnamefont
  {Schiavilla}},\ }\href {\doibase 10.1103/PhysRevC.51.38} {\bibfield
  {journal} {\bibinfo  {journal} {Phys. Rev. C}\ }\textbf {\bibinfo {volume}
  {51}},\ \bibinfo {pages} {38} (\bibinfo {year} {1995})}\BibitemShut {NoStop}%
\bibitem [{\citenamefont {Horiuchi}\ \emph {et~al.}(2010)\citenamefont
  {Horiuchi}, \citenamefont {Suzuki}, \citenamefont {Capel},\ and\
  \citenamefont {Baye}}]{Horiuchi10}%
  \BibitemOpen
  \bibfield  {author} {\bibinfo {author} {\bibfnamefont {W.}~\bibnamefont
  {Horiuchi}}, \bibinfo {author} {\bibfnamefont {Y.}~\bibnamefont {Suzuki}},
  \bibinfo {author} {\bibfnamefont {P.}~\bibnamefont {Capel}}, \ and\ \bibinfo
  {author} {\bibfnamefont {D.}~\bibnamefont {Baye}},\ }\href {\doibase
  10.1103/PhysRevC.81.024606} {\bibfield  {journal} {\bibinfo  {journal} {Phys.
  Rev. C}\ }\textbf {\bibinfo {volume} {81}},\ \bibinfo {pages} {024606}
  (\bibinfo {year} {2010})}\BibitemShut {NoStop}%
\bibitem [{\citenamefont {Bohr}\ and\ \citenamefont {Mottelson}(1969)}]{BOHR}%
  \BibitemOpen
  \bibfield  {author} {\bibinfo {author} {\bibfnamefont {A.}~\bibnamefont
  {Bohr}}\ and\ \bibinfo {author} {\bibfnamefont {B.~R.}\ \bibnamefont
  {Mottelson}},\ }\href@noop {} {\emph {\bibinfo {title} {Nuclear Structure}}}\
  (\bibinfo  {publisher} {Benjamin, New York},\ \bibinfo {year}
  {1969})\BibitemShut {NoStop}%
\bibitem [{\citenamefont {Baye}(1987)}]{Baye287}%
  \BibitemOpen
  \bibfield  {author} {\bibinfo {author} {\bibfnamefont {D.}~\bibnamefont
  {Baye}},\ }\href {\doibase 10.1088/0305-4470/20/16/027} {\bibfield  {journal}
  {\bibinfo  {journal} {Jour. Phys. A}\ }\textbf {\bibinfo {volume} {20}},\
  \bibinfo {pages} {5529} (\bibinfo {year} {1987})}\BibitemShut {NoStop}%
\bibitem [{\citenamefont {Singh}\ \emph {et~al.}(2016)\citenamefont {Singh},
  \citenamefont {Fortunato}, \citenamefont {Vitturi},\ and\ \citenamefont
  {Chatterjee}}]{Singh16}%
  \BibitemOpen
  \bibfield  {author} {\bibinfo {author} {\bibfnamefont {J.}~\bibnamefont
  {Singh}}, \bibinfo {author} {\bibfnamefont {L.}~\bibnamefont {Fortunato}},
  \bibinfo {author} {\bibfnamefont {A.}~\bibnamefont {Vitturi}}, \ and\
  \bibinfo {author} {\bibfnamefont {R.}~\bibnamefont {Chatterjee}},\ }\href
  {\doibase 10.1140/epja/i2016-16209-8} {\bibfield  {journal} {\bibinfo
  {journal} {Eur. Phys. J. A}\ }\textbf {\bibinfo {volume} {52}},\ \bibinfo
  {pages} {209} (\bibinfo {year} {2016})}\BibitemShut {NoStop}%
\bibitem [{\citenamefont {Glauber}(1959)}]{Glauber}%
  \BibitemOpen
  \bibfield  {author} {\bibinfo {author} {\bibfnamefont {R.~J.}\ \bibnamefont
  {Glauber}},\ }\href@noop {} {\emph {\bibinfo {title} {Lectures in Theoretical
  Physics}}},\ edited by\ \bibinfo {editor} {\bibfnamefont {W.~E.}\
  \bibnamefont {Brittin}}\ and\ \bibinfo {editor} {\bibfnamefont {L.~G.}\
  \bibnamefont {Dunham}},\ Vol.~\bibinfo {volume} {1}\ (\bibinfo  {publisher}
  {Interscience},\ \bibinfo {address} {New York},\ \bibinfo {year} {1959})\ p.\
  \bibinfo {pages} {315}\BibitemShut {NoStop}%
\bibitem [{\citenamefont {Abu-Ibrahim}\ and\ \citenamefont
  {Suzuki}(2000)}]{AbuIbrahim00}%
  \BibitemOpen
  \bibfield  {author} {\bibinfo {author} {\bibfnamefont {B.}~\bibnamefont
  {Abu-Ibrahim}}\ and\ \bibinfo {author} {\bibfnamefont {Y.}~\bibnamefont
  {Suzuki}},\ }\href {\doibase 10.1103/PhysRevC.61.051601} {\bibfield
  {journal} {\bibinfo  {journal} {Phys. Rev. C}\ }\textbf {\bibinfo {volume}
  {61}},\ \bibinfo {pages} {051601} (\bibinfo {year} {2000})}\BibitemShut
  {NoStop}%
\bibitem [{\citenamefont {Abu-Ibrahim}\ \emph {et~al.}(2008)\citenamefont
  {Abu-Ibrahim}, \citenamefont {Horiuchi}, \citenamefont {Kohama},\ and\
  \citenamefont {Suzuki}}]{AbuIbrahim08}%
  \BibitemOpen
  \bibfield  {author} {\bibinfo {author} {\bibfnamefont {B.}~\bibnamefont
  {Abu-Ibrahim}}, \bibinfo {author} {\bibfnamefont {W.}~\bibnamefont
  {Horiuchi}}, \bibinfo {author} {\bibfnamefont {A.}~\bibnamefont {Kohama}}, \
  and\ \bibinfo {author} {\bibfnamefont {Y.}~\bibnamefont {Suzuki}},\ }\href
  {\doibase 10.1103/PhysRevC.77.034607} {\bibfield  {journal} {\bibinfo
  {journal} {Phys. Rev. C}\ }\textbf {\bibinfo {volume} {77}},\ \bibinfo
  {pages} {034607} (\bibinfo {year} {2008})}\BibitemShut {NoStop}%
\bibitem [{\citenamefont {Abu-Ibrahim}\ \emph {et~al.}(2009)\citenamefont
  {Abu-Ibrahim}, \citenamefont {Horiuchi}, \citenamefont {Kohama},\ and\
  \citenamefont {Suzuki}}]{AbuIbrahim09E}%
  \BibitemOpen
  \bibfield  {author} {\bibinfo {author} {\bibfnamefont {B.}~\bibnamefont
  {Abu-Ibrahim}}, \bibinfo {author} {\bibfnamefont {W.}~\bibnamefont
  {Horiuchi}}, \bibinfo {author} {\bibfnamefont {A.}~\bibnamefont {Kohama}}, \
  and\ \bibinfo {author} {\bibfnamefont {Y.}~\bibnamefont {Suzuki}},\ }\href
  {\doibase 10.1103/PhysRevC.80.029903} {\bibfield  {journal} {\bibinfo
  {journal} {Phys. Rev. C}\ }\textbf {\bibinfo {volume} {80}},\ \bibinfo
  {pages} {029903} (\bibinfo {year} {2009})}\BibitemShut {NoStop}%
\bibitem [{\citenamefont {Abu-Ibrahim}\ \emph {et~al.}(2010)\citenamefont
  {Abu-Ibrahim}, \citenamefont {Horiuchi}, \citenamefont {Kohama},\ and\
  \citenamefont {Suzuki}}]{AbuIbrahim10E}%
  \BibitemOpen
  \bibfield  {author} {\bibinfo {author} {\bibfnamefont {B.}~\bibnamefont
  {Abu-Ibrahim}}, \bibinfo {author} {\bibfnamefont {W.}~\bibnamefont
  {Horiuchi}}, \bibinfo {author} {\bibfnamefont {A.}~\bibnamefont {Kohama}}, \
  and\ \bibinfo {author} {\bibfnamefont {Y.}~\bibnamefont {Suzuki}},\ }\href
  {\doibase 10.1103/PhysRevC.81.019901} {\bibfield  {journal} {\bibinfo
  {journal} {Phys. Rev. C}\ }\textbf {\bibinfo {volume} {81}},\ \bibinfo
  {pages} {019901} (\bibinfo {year} {2010})}\BibitemShut {NoStop}%
\bibitem [{\citenamefont {Horiuchi}\ \emph {et~al.}(2012)\citenamefont
  {Horiuchi}, \citenamefont {Inakura}, \citenamefont {Nakatsukasa},\ and\
  \citenamefont {Suzuki}}]{Horiuchi12}%
  \BibitemOpen
  \bibfield  {author} {\bibinfo {author} {\bibfnamefont {W.}~\bibnamefont
  {Horiuchi}}, \bibinfo {author} {\bibfnamefont {T.}~\bibnamefont {Inakura}},
  \bibinfo {author} {\bibfnamefont {T.}~\bibnamefont {Nakatsukasa}}, \ and\
  \bibinfo {author} {\bibfnamefont {Y.}~\bibnamefont {Suzuki}},\ }\href
  {\doibase 10.1103/PhysRevC.86.024614} {\bibfield  {journal} {\bibinfo
  {journal} {Phys. Rev. C}\ }\textbf {\bibinfo {volume} {86}},\ \bibinfo
  {pages} {024614} (\bibinfo {year} {2012})}\BibitemShut {NoStop}%
\bibitem [{\citenamefont {Horiuchi}\ \emph {et~al.}(2015)\citenamefont
  {Horiuchi}, \citenamefont {Inakura}, \citenamefont {Nakatsukasa},\ and\
  \citenamefont {Suzuki}}]{Horiuchi15jps}%
  \BibitemOpen
  \bibfield  {author} {\bibinfo {author} {\bibfnamefont {W.}~\bibnamefont
  {Horiuchi}}, \bibinfo {author} {\bibfnamefont {T.}~\bibnamefont {Inakura}},
  \bibinfo {author} {\bibfnamefont {T.}~\bibnamefont {Nakatsukasa}}, \ and\
  \bibinfo {author} {\bibfnamefont {Y.}~\bibnamefont {Suzuki}},\ }\href
  {\doibase 10.7566/JPSCP.6.030079} {\bibfield  {journal} {\bibinfo  {journal}
  {JPS Conf. Proc.}\ }\textbf {\bibinfo {volume} {6}},\ \bibinfo {pages}
  {030079} (\bibinfo {year} {2015})}\BibitemShut {NoStop}%
\bibitem [{\citenamefont {Catara}\ \emph {et~al.}(1984)\citenamefont {Catara},
  \citenamefont {Insolia}, \citenamefont {Maglione},\ and\ \citenamefont
  {Vitturi}}]{Catara84}%
  \BibitemOpen
  \bibfield  {author} {\bibinfo {author} {\bibfnamefont {F.}~\bibnamefont
  {Catara}}, \bibinfo {author} {\bibfnamefont {A.}~\bibnamefont {Insolia}},
  \bibinfo {author} {\bibfnamefont {E.}~\bibnamefont {Maglione}}, \ and\
  \bibinfo {author} {\bibfnamefont {A.}~\bibnamefont {Vitturi}},\ }\href
  {\doibase 10.1103/PhysRevC.29.1091} {\bibfield  {journal} {\bibinfo
  {journal} {Phys. Rev. C}\ }\textbf {\bibinfo {volume} {29}},\ \bibinfo
  {pages} {1091} (\bibinfo {year} {1984})}\BibitemShut {NoStop}%
\bibitem [{\citenamefont {Raynal}\ and\ \citenamefont {Revai}(1970)}]{RR70}%
  \BibitemOpen
  \bibfield  {author} {\bibinfo {author} {\bibfnamefont {J.}~\bibnamefont
  {Raynal}}\ and\ \bibinfo {author} {\bibfnamefont {J.}~\bibnamefont {Revai}},\
  }\href {\doibase 10.1007/BF02756127} {\bibfield  {journal} {\bibinfo
  {journal} {Nuovo Cim.}\ }\textbf {\bibinfo {volume} {68A}},\ \bibinfo {pages}
  {612} (\bibinfo {year} {1970})}\BibitemShut {NoStop}%
\bibitem [{\citenamefont {Pinilla}\ \emph {et~al.}(2011)\citenamefont
  {Pinilla}, \citenamefont {Baye}, \citenamefont {Descouvemont}, \citenamefont
  {Horiuchi},\ and\ \citenamefont {Suzuki}}]{Pinilla2010}%
  \BibitemOpen
  \bibfield  {author} {\bibinfo {author} {\bibfnamefont {E.}~\bibnamefont
  {Pinilla}}, \bibinfo {author} {\bibfnamefont {D.}~\bibnamefont {Baye}},
  \bibinfo {author} {\bibfnamefont {P.}~\bibnamefont {Descouvemont}}, \bibinfo
  {author} {\bibfnamefont {W.}~\bibnamefont {Horiuchi}}, \ and\ \bibinfo
  {author} {\bibfnamefont {Y.}~\bibnamefont {Suzuki}},\ }\href {\doibase
  https://doi.org/10.1016/j.nuclphysa.2011.06.030} {\bibfield  {journal}
  {\bibinfo  {journal} {Nucl. Phys. A}\ }\textbf {\bibinfo {volume} {865}},\
  \bibinfo {pages} {43 } (\bibinfo {year} {2011})}\BibitemShut {NoStop}%
\bibitem [{\citenamefont {Casal}\ \emph {et~al.}(2014)\citenamefont {Casal},
  \citenamefont {Rodr\'{\i}guez-Gallardo}, \citenamefont {Arias},\ and\
  \citenamefont {Thompson}}]{JCasal14}%
  \BibitemOpen
  \bibfield  {author} {\bibinfo {author} {\bibfnamefont {J.}~\bibnamefont
  {Casal}}, \bibinfo {author} {\bibfnamefont {M.}~\bibnamefont
  {Rodr\'{\i}guez-Gallardo}}, \bibinfo {author} {\bibfnamefont {J.~M.}\
  \bibnamefont {Arias}}, \ and\ \bibinfo {author} {\bibfnamefont {I.~J.}\
  \bibnamefont {Thompson}},\ }\href {\doibase 10.1103/PhysRevC.90.044304}
  {\bibfield  {journal} {\bibinfo  {journal} {Phys. Rev. C}\ }\textbf {\bibinfo
  {volume} {90}},\ \bibinfo {pages} {044304} (\bibinfo {year}
  {2014})}\BibitemShut {NoStop}%
\bibitem [{\citenamefont {Descouvemont}(2020)}]{Descouvemont20}%
  \BibitemOpen
  \bibfield  {author} {\bibinfo {author} {\bibfnamefont {P.}~\bibnamefont
  {Descouvemont}},\ }\href {\doibase 10.1103/PhysRevC.101.064611} {\bibfield
  {journal} {\bibinfo  {journal} {Phys. Rev. C}\ }\textbf {\bibinfo {volume}
  {101}},\ \bibinfo {pages} {064611} (\bibinfo {year} {2020})}\BibitemShut
  {NoStop}%
\bibitem [{\citenamefont {de~Diego}\ \emph {et~al.}(2008)\citenamefont
  {de~Diego}, \citenamefont {Garrido}, \citenamefont {Jensen},\ and\
  \citenamefont {Fedorov}}]{RdDiego08}%
  \BibitemOpen
  \bibfield  {author} {\bibinfo {author} {\bibfnamefont {R.}~\bibnamefont
  {de~Diego}}, \bibinfo {author} {\bibfnamefont {E.}~\bibnamefont {Garrido}},
  \bibinfo {author} {\bibfnamefont {A.~S.}\ \bibnamefont {Jensen}}, \ and\
  \bibinfo {author} {\bibfnamefont {D.~V.}\ \bibnamefont {Fedorov}},\ }\href
  {\doibase 10.1103/PhysRevC.77.024001} {\bibfield  {journal} {\bibinfo
  {journal} {Phys. Rev. C}\ }\textbf {\bibinfo {volume} {77}},\ \bibinfo
  {pages} {024001} (\bibinfo {year} {2008})}\BibitemShut {NoStop}%
\bibitem [{\citenamefont {Nakamura}\ \emph {et~al.}(1994)\citenamefont
  {Nakamura}, \citenamefont {Shimoura}, \citenamefont {Kobayashi},
  \citenamefont {Teranishi}, \citenamefont {Abe}, \citenamefont {Aoi},
  \citenamefont {Doki}, \citenamefont {Fujimaki}, \citenamefont {Inabe},
  \citenamefont {Iwasa}, \citenamefont {Katori}, \citenamefont {Kubo},
  \citenamefont {Okuno}, \citenamefont {Suzuki}, \citenamefont {Tanihata},
  \citenamefont {Watanabe}, \citenamefont {Yoshida},\ and\ \citenamefont
  {Ishihara}}]{Nakamura94}%
  \BibitemOpen
  \bibfield  {author} {\bibinfo {author} {\bibfnamefont {T.}~\bibnamefont
  {Nakamura}}, \bibinfo {author} {\bibfnamefont {S.}~\bibnamefont {Shimoura}},
  \bibinfo {author} {\bibfnamefont {T.}~\bibnamefont {Kobayashi}}, \bibinfo
  {author} {\bibfnamefont {T.}~\bibnamefont {Teranishi}}, \bibinfo {author}
  {\bibfnamefont {K.}~\bibnamefont {Abe}}, \bibinfo {author} {\bibfnamefont
  {N.}~\bibnamefont {Aoi}}, \bibinfo {author} {\bibfnamefont {Y.}~\bibnamefont
  {Doki}}, \bibinfo {author} {\bibfnamefont {M.}~\bibnamefont {Fujimaki}},
  \bibinfo {author} {\bibfnamefont {N.}~\bibnamefont {Inabe}}, \bibinfo
  {author} {\bibfnamefont {N.}~\bibnamefont {Iwasa}}, \bibinfo {author}
  {\bibfnamefont {K.}~\bibnamefont {Katori}}, \bibinfo {author} {\bibfnamefont
  {T.}~\bibnamefont {Kubo}}, \bibinfo {author} {\bibfnamefont {H.}~\bibnamefont
  {Okuno}}, \bibinfo {author} {\bibfnamefont {T.}~\bibnamefont {Suzuki}},
  \bibinfo {author} {\bibfnamefont {I.}~\bibnamefont {Tanihata}}, \bibinfo
  {author} {\bibfnamefont {Y.}~\bibnamefont {Watanabe}}, \bibinfo {author}
  {\bibfnamefont {A.}~\bibnamefont {Yoshida}}, \ and\ \bibinfo {author}
  {\bibfnamefont {M.}~\bibnamefont {Ishihara}},\ }\href {\doibase
  https://doi.org/10.1016/0370-2693(94)91055-3} {\bibfield  {journal} {\bibinfo
   {journal} {Phys. Lett. B}\ }\textbf {\bibinfo {volume} {331}},\ \bibinfo
  {pages} {296 } (\bibinfo {year} {1994})}\BibitemShut {NoStop}%
\bibitem [{\citenamefont {Nagarajan}\ \emph {et~al.}(2005)\citenamefont
  {Nagarajan}, \citenamefont {Lenzi},\ and\ \citenamefont
  {Vitturi}}]{Nagarajan05}%
  \BibitemOpen
  \bibfield  {author} {\bibinfo {author} {\bibfnamefont {M.~A.}\ \bibnamefont
  {Nagarajan}}, \bibinfo {author} {\bibfnamefont {S.~M.}\ \bibnamefont
  {Lenzi}}, \ and\ \bibinfo {author} {\bibfnamefont {A.}~\bibnamefont
  {Vitturi}},\ }\href {\doibase 10.1140/epja/i2004-10129-2} {\bibfield
  {journal} {\bibinfo  {journal} {Eur. Phys. J. A}\ }\textbf {\bibinfo {volume}
  {24}},\ \bibinfo {pages} {63} (\bibinfo {year} {2005})}\BibitemShut {NoStop}%
\bibitem [{\citenamefont {Casal}\ and\ \citenamefont
  {G\'omez-Camacho}(2019)}]{JCasal19}%
  \BibitemOpen
  \bibfield  {author} {\bibinfo {author} {\bibfnamefont {J.}~\bibnamefont
  {Casal}}\ and\ \bibinfo {author} {\bibfnamefont {J.}~\bibnamefont
  {G\'omez-Camacho}},\ }\href {\doibase 10.1103/PhysRevC.99.014604} {\bibfield
  {journal} {\bibinfo  {journal} {Phys. Rev. C}\ }\textbf {\bibinfo {volume}
  {99}},\ \bibinfo {pages} {014604} (\bibinfo {year} {2019})}\BibitemShut
  {NoStop}%
\bibitem [{\citenamefont {Thompson}()}]{sturmxx}%
  \BibitemOpen
  \bibfield  {author} {\bibinfo {author} {\bibfnamefont {I.~J.}\ \bibnamefont
  {Thompson}},\ }\href@noop {} {\bibinfo  {journal} {Program {\tt sturmxx},
  {\it Sturmian Bound-state and Scattering program}, available from:
  http://www.fresco.org.uk/programs/sturmxx/index.html}\ }\BibitemShut
  {NoStop}%
\bibitem [{\citenamefont {Winther}\ and\ \citenamefont
  {Alder}(1979)}]{WintherAdler79}%
  \BibitemOpen
\bibfield  {journal} {  }\bibfield  {author} {\bibinfo {author} {\bibfnamefont
  {A.}~\bibnamefont {Winther}}\ and\ \bibinfo {author} {\bibfnamefont
  {K.}~\bibnamefont {Alder}},\ }\href {\doibase
  https://doi.org/10.1016/0375-9474(79)90528-1} {\bibfield  {journal} {\bibinfo
   {journal} {Nucl. Phys. A}\ }\textbf {\bibinfo {volume} {319}},\ \bibinfo
  {pages} {518 } (\bibinfo {year} {1979})}\BibitemShut {NoStop}%
\bibitem [{\citenamefont {Bertulani}\ and\ \citenamefont
  {Baur}(1988)}]{BertulaniEPM}%
  \BibitemOpen
  \bibfield  {author} {\bibinfo {author} {\bibfnamefont {C.~A.}\ \bibnamefont
  {Bertulani}}\ and\ \bibinfo {author} {\bibfnamefont {G.}~\bibnamefont
  {Baur}},\ }\href {\doibase https://doi.org/10.1016/0370-1573(88)90142-1}
  {\bibfield  {journal} {\bibinfo  {journal} {Phys. Rep.}\ }\textbf {\bibinfo
  {volume} {163}},\ \bibinfo {pages} {299 } (\bibinfo {year}
  {1988})}\BibitemShut {NoStop}%
\bibitem [{\citenamefont {Austern}\ \emph {et~al.}(1987)\citenamefont
  {Austern}, \citenamefont {Iseri}, \citenamefont {Kamimura}, \citenamefont
  {Kawai}, \citenamefont {Rawitscher},\ and\ \citenamefont
  {Yahiro}}]{Austern87}%
  \BibitemOpen
  \bibfield  {author} {\bibinfo {author} {\bibfnamefont {N.}~\bibnamefont
  {Austern}}, \bibinfo {author} {\bibfnamefont {Y.}~\bibnamefont {Iseri}},
  \bibinfo {author} {\bibfnamefont {M.}~\bibnamefont {Kamimura}}, \bibinfo
  {author} {\bibfnamefont {M.}~\bibnamefont {Kawai}}, \bibinfo {author}
  {\bibfnamefont {G.}~\bibnamefont {Rawitscher}}, \ and\ \bibinfo {author}
  {\bibfnamefont {M.}~\bibnamefont {Yahiro}},\ }\href {\doibase
  https://doi.org/10.1016/0370-1573(87)90094-9} {\bibfield  {journal} {\bibinfo
   {journal} {Phys. Rep.}\ }\textbf {\bibinfo {volume} {154}},\ \bibinfo
  {pages} {125 } (\bibinfo {year} {1987})}\BibitemShut {NoStop}%
\bibitem [{\citenamefont {Yahiro}\ \emph {et~al.}(1986)\citenamefont {Yahiro},
  \citenamefont {Iseri}, \citenamefont {Kameyama}, \citenamefont {Kamimura},\
  and\ \citenamefont {Kawai}}]{Yahiro86}%
  \BibitemOpen
  \bibfield  {author} {\bibinfo {author} {\bibfnamefont {M.}~\bibnamefont
  {Yahiro}}, \bibinfo {author} {\bibfnamefont {Y.}~\bibnamefont {Iseri}},
  \bibinfo {author} {\bibfnamefont {H.}~\bibnamefont {Kameyama}}, \bibinfo
  {author} {\bibfnamefont {M.}~\bibnamefont {Kamimura}}, \ and\ \bibinfo
  {author} {\bibfnamefont {M.}~\bibnamefont {Kawai}},\ }\href@noop {}
  {\bibfield  {journal} {\bibinfo  {journal} {Prog. Theor. Phys. Suppl.}\
  }\textbf {\bibinfo {volume} {89}},\ \bibinfo {pages} {32} (\bibinfo {year}
  {1986})}\BibitemShut {NoStop}%
\bibitem [{\citenamefont {Matsumoto}\ \emph
  {et~al.}(2004{\natexlab{b}})\citenamefont {Matsumoto}, \citenamefont
  {Hiyama}, \citenamefont {Ogata}, \citenamefont {Iseri}, \citenamefont
  {Kamimura}, \citenamefont {Chiba},\ and\ \citenamefont
  {Yahiro}}]{Matsumoto04-2}%
  \BibitemOpen
  \bibfield  {author} {\bibinfo {author} {\bibfnamefont {T.}~\bibnamefont
  {Matsumoto}}, \bibinfo {author} {\bibfnamefont {E.}~\bibnamefont {Hiyama}},
  \bibinfo {author} {\bibfnamefont {K.}~\bibnamefont {Ogata}}, \bibinfo
  {author} {\bibfnamefont {Y.}~\bibnamefont {Iseri}}, \bibinfo {author}
  {\bibfnamefont {M.}~\bibnamefont {Kamimura}}, \bibinfo {author}
  {\bibfnamefont {S.}~\bibnamefont {Chiba}}, \ and\ \bibinfo {author}
  {\bibfnamefont {M.}~\bibnamefont {Yahiro}},\ }\href {\doibase
  10.1103/PhysRevC.70.061601} {\bibfield  {journal} {\bibinfo  {journal} {Phys.
  Rev. C}\ }\textbf {\bibinfo {volume} {70}},\ \bibinfo {pages} {061601}
  (\bibinfo {year} {2004}{\natexlab{b}})}\BibitemShut {NoStop}%
\bibitem [{\citenamefont {Casal}(2016)}]{CasalTh}%
  \BibitemOpen
  \bibfield  {author} {\bibinfo {author} {\bibfnamefont {J.}~\bibnamefont
  {Casal}},\ }\href {https://idus.us.es/xmlui/handle/11441/41814} {\enquote
  {\bibinfo {title} {Weakly-bound three-body nuclear systems: Structure,
  reactions and astrophysical implications},}\ }\bibinfo {howpublished} {Ph.D.
  thesis, Universidad de Sevilla} (\bibinfo {year} {2016})\BibitemShut
  {NoStop}%
\bibitem [{\citenamefont {Thompson}(1988)}]{fresco}%
  \BibitemOpen
  \bibfield  {author} {\bibinfo {author} {\bibfnamefont {I.~J.}\ \bibnamefont
  {Thompson}},\ }\href {\doibase 10.1016/0167-7977(88)90005-6} {\bibfield
  {journal} {\bibinfo  {journal} {Comput. Phys. Rep.}\ }\textbf {\bibinfo
  {volume} {7}},\ \bibinfo {pages} {167 } (\bibinfo {year} {1988})}\BibitemShut
  {NoStop}%
\bibitem [{\citenamefont {Koning}\ and\ \citenamefont {Delaroche}(2008)}]{KD}%
  \BibitemOpen
  \bibfield  {author} {\bibinfo {author} {\bibfnamefont {A.~J.}\ \bibnamefont
  {Koning}}\ and\ \bibinfo {author} {\bibfnamefont {J.~P.}\ \bibnamefont
  {Delaroche}},\ }\href@noop {} {\bibfield  {journal} {\bibinfo  {journal}
  {Nucl. Phys. A}\ }\textbf {\bibinfo {volume} {803}},\ \bibinfo {pages} {30}
  (\bibinfo {year} {2008})}\BibitemShut {NoStop}%
\bibitem [{\citenamefont {Chamon}\ \emph {et~al.}(2002)\citenamefont {Chamon},
  \citenamefont {Carlson}, \citenamefont {Gasques}, \citenamefont {Pereira},
  \citenamefont {De~Conti}, \citenamefont {Alvarez}, \citenamefont {Hussein},
  \citenamefont {C\^andido~Ribeiro}, \citenamefont {Rossi},\ and\ \citenamefont
  {Silva}}]{SPP}%
  \BibitemOpen
  \bibfield  {author} {\bibinfo {author} {\bibfnamefont {L.~C.}\ \bibnamefont
  {Chamon}}, \bibinfo {author} {\bibfnamefont {B.~V.}\ \bibnamefont {Carlson}},
  \bibinfo {author} {\bibfnamefont {L.~R.}\ \bibnamefont {Gasques}}, \bibinfo
  {author} {\bibfnamefont {D.}~\bibnamefont {Pereira}}, \bibinfo {author}
  {\bibfnamefont {C.}~\bibnamefont {De~Conti}}, \bibinfo {author}
  {\bibfnamefont {M.~A.~G.}\ \bibnamefont {Alvarez}}, \bibinfo {author}
  {\bibfnamefont {M.~S.}\ \bibnamefont {Hussein}}, \bibinfo {author}
  {\bibfnamefont {M.~A.}\ \bibnamefont {C\^andido~Ribeiro}}, \bibinfo {author}
  {\bibfnamefont {E.~S.}\ \bibnamefont {Rossi}}, \ and\ \bibinfo {author}
  {\bibfnamefont {C.~P.}\ \bibnamefont {Silva}},\ }\href {\doibase
  10.1103/PhysRevC.66.014610} {\bibfield  {journal} {\bibinfo  {journal} {Phys.
  Rev. C}\ }\textbf {\bibinfo {volume} {66}},\ \bibinfo {pages} {014610}
  (\bibinfo {year} {2002})}\BibitemShut {NoStop}%
\bibitem [{\citenamefont {Acosta}\ \emph {et~al.}(2011)\citenamefont {Acosta},
  \citenamefont {S\'anchez-Ben\'{\i}tez}, \citenamefont {G\'omez},
  \citenamefont {Martel}, \citenamefont {P\'erez-Bernal}, \citenamefont
  {Pizarro}, \citenamefont {Rodr\'{\i}guez-Quintero}, \citenamefont {Rusek},
  \citenamefont {Alvarez}, \citenamefont {Andr\'es}, \citenamefont {Espino},
  \citenamefont {Fern\'andez-Garc\'{\i}a}, \citenamefont {G\'omez-Camacho},
  \citenamefont {Moro}, \citenamefont {Angulo}, \citenamefont {Cabrera},
  \citenamefont {Casarejos}, \citenamefont {Demaret}, \citenamefont {Borge},
  \citenamefont {Escrig}, \citenamefont {Tengblad}, \citenamefont {Cherubini},
  \citenamefont {Figuera}, \citenamefont {Gulino}, \citenamefont {Freer},
  \citenamefont {Metelko}, \citenamefont {Ziman}, \citenamefont {Raabe},
  \citenamefont {Mukha}, \citenamefont {Smirnov}, \citenamefont {Kakuee},\ and\
  \citenamefont {Rahighi}}]{Acosta11}%
  \BibitemOpen
  \bibfield  {author} {\bibinfo {author} {\bibfnamefont {L.}~\bibnamefont
  {Acosta}}, \bibinfo {author} {\bibfnamefont {A.~M.}\ \bibnamefont
  {S\'anchez-Ben\'{\i}tez}}, \bibinfo {author} {\bibfnamefont {M.~E.}\
  \bibnamefont {G\'omez}}, \bibinfo {author} {\bibfnamefont {I.}~\bibnamefont
  {Martel}}, \bibinfo {author} {\bibfnamefont {F.}~\bibnamefont
  {P\'erez-Bernal}}, \bibinfo {author} {\bibfnamefont {F.}~\bibnamefont
  {Pizarro}}, \bibinfo {author} {\bibfnamefont {J.}~\bibnamefont
  {Rodr\'{\i}guez-Quintero}}, \bibinfo {author} {\bibfnamefont
  {K.}~\bibnamefont {Rusek}}, \bibinfo {author} {\bibfnamefont {M.~A.~G.}\
  \bibnamefont {Alvarez}}, \bibinfo {author} {\bibfnamefont {M.~V.}\
  \bibnamefont {Andr\'es}}, \bibinfo {author} {\bibfnamefont {J.~M.}\
  \bibnamefont {Espino}}, \bibinfo {author} {\bibfnamefont {J.~P.}\
  \bibnamefont {Fern\'andez-Garc\'{\i}a}}, \bibinfo {author} {\bibfnamefont
  {J.}~\bibnamefont {G\'omez-Camacho}}, \bibinfo {author} {\bibfnamefont
  {A.~M.}\ \bibnamefont {Moro}}, \bibinfo {author} {\bibfnamefont
  {C.}~\bibnamefont {Angulo}}, \bibinfo {author} {\bibfnamefont
  {J.}~\bibnamefont {Cabrera}}, \bibinfo {author} {\bibfnamefont
  {E.}~\bibnamefont {Casarejos}}, \bibinfo {author} {\bibfnamefont
  {P.}~\bibnamefont {Demaret}}, \bibinfo {author} {\bibfnamefont {M.~J.~G.}\
  \bibnamefont {Borge}}, \bibinfo {author} {\bibfnamefont {D.}~\bibnamefont
  {Escrig}}, \bibinfo {author} {\bibfnamefont {O.}~\bibnamefont {Tengblad}},
  \bibinfo {author} {\bibfnamefont {S.}~\bibnamefont {Cherubini}}, \bibinfo
  {author} {\bibfnamefont {P.}~\bibnamefont {Figuera}}, \bibinfo {author}
  {\bibfnamefont {M.}~\bibnamefont {Gulino}}, \bibinfo {author} {\bibfnamefont
  {M.}~\bibnamefont {Freer}}, \bibinfo {author} {\bibfnamefont
  {C.}~\bibnamefont {Metelko}}, \bibinfo {author} {\bibfnamefont
  {V.}~\bibnamefont {Ziman}}, \bibinfo {author} {\bibfnamefont
  {R.}~\bibnamefont {Raabe}}, \bibinfo {author} {\bibfnamefont
  {I.}~\bibnamefont {Mukha}}, \bibinfo {author} {\bibfnamefont
  {D.}~\bibnamefont {Smirnov}}, \bibinfo {author} {\bibfnamefont {O.~R.}\
  \bibnamefont {Kakuee}}, \ and\ \bibinfo {author} {\bibfnamefont
  {J.}~\bibnamefont {Rahighi}},\ }\href {\doibase 10.1103/PhysRevC.84.044604}
  {\bibfield  {journal} {\bibinfo  {journal} {Phys. Rev. C}\ }\textbf {\bibinfo
  {volume} {84}},\ \bibinfo {pages} {044604} (\bibinfo {year}
  {2011})}\BibitemShut {NoStop}%
\bibitem [{\citenamefont {Di~Pietro}\ \emph {et~al.}(2004)\citenamefont
  {Di~Pietro}, \citenamefont {Figuera}, \citenamefont {Amorini}, \citenamefont
  {Angulo}, \citenamefont {Cardella}, \citenamefont {Cherubini}, \citenamefont
  {Davinson}, \citenamefont {Leanza}, \citenamefont {Lu}, \citenamefont
  {Mahmud}, \citenamefont {Milin}, \citenamefont {Musumarra}, \citenamefont
  {Ninane}, \citenamefont {Papa}, \citenamefont {Pellegriti}, \citenamefont
  {Raabe}, \citenamefont {Rizzo}, \citenamefont {Ruiz}, \citenamefont
  {Shotter}, \citenamefont {Soi\ifmmode~\acute{c}\else \'{c}\fi{}},
  \citenamefont {Tudisco},\ and\ \citenamefont {Weissman}}]{AdiPietro04}%
  \BibitemOpen
  \bibfield  {author} {\bibinfo {author} {\bibfnamefont {A.}~\bibnamefont
  {Di~Pietro}}, \bibinfo {author} {\bibfnamefont {P.}~\bibnamefont {Figuera}},
  \bibinfo {author} {\bibfnamefont {F.}~\bibnamefont {Amorini}}, \bibinfo
  {author} {\bibfnamefont {C.}~\bibnamefont {Angulo}}, \bibinfo {author}
  {\bibfnamefont {G.}~\bibnamefont {Cardella}}, \bibinfo {author}
  {\bibfnamefont {S.}~\bibnamefont {Cherubini}}, \bibinfo {author}
  {\bibfnamefont {T.}~\bibnamefont {Davinson}}, \bibinfo {author}
  {\bibfnamefont {D.}~\bibnamefont {Leanza}}, \bibinfo {author} {\bibfnamefont
  {J.}~\bibnamefont {Lu}}, \bibinfo {author} {\bibfnamefont {H.}~\bibnamefont
  {Mahmud}}, \bibinfo {author} {\bibfnamefont {M.}~\bibnamefont {Milin}},
  \bibinfo {author} {\bibfnamefont {A.}~\bibnamefont {Musumarra}}, \bibinfo
  {author} {\bibfnamefont {A.}~\bibnamefont {Ninane}}, \bibinfo {author}
  {\bibfnamefont {M.}~\bibnamefont {Papa}}, \bibinfo {author} {\bibfnamefont
  {M.~G.}\ \bibnamefont {Pellegriti}}, \bibinfo {author} {\bibfnamefont
  {R.}~\bibnamefont {Raabe}}, \bibinfo {author} {\bibfnamefont
  {F.}~\bibnamefont {Rizzo}}, \bibinfo {author} {\bibfnamefont
  {C.}~\bibnamefont {Ruiz}}, \bibinfo {author} {\bibfnamefont {A.~C.}\
  \bibnamefont {Shotter}}, \bibinfo {author} {\bibfnamefont {N.}~\bibnamefont
  {Soi\ifmmode~\acute{c}\else \'{c}\fi{}}}, \bibinfo {author} {\bibfnamefont
  {S.}~\bibnamefont {Tudisco}}, \ and\ \bibinfo {author} {\bibfnamefont
  {L.}~\bibnamefont {Weissman}},\ }\href {\doibase 10.1103/PhysRevC.69.044613}
  {\bibfield  {journal} {\bibinfo  {journal} {Phys. Rev. C}\ }\textbf {\bibinfo
  {volume} {69}},\ \bibinfo {pages} {044613} (\bibinfo {year}
  {2004})}\BibitemShut {NoStop}%
\end{thebibliography}%

\end{document}